\documentclass[twocolumn]{aastex63}
\usepackage{lineno}

\usepackage{graphicx}
\usepackage{wrapfig}
\usepackage{natbib}
\usepackage{color}
\graphicspath{{./figures/}} 
\usepackage{mathtools}
\usepackage{epstopdf}
\usepackage[autostyle]{csquotes}
\usepackage{hyperref} 
\def    \apjl  		{\rm {ApJL}}

\def    \apjl  		{\rm {ApJL}}

\def	\cm		{\,{\rm {cm}}}
\def	\K		{\,{\rm K}}
\def	\g		{\,{\rm {g}}}
\def	\mum	{\,{\mu \rm{m}}}
\def	\erg    		{\,{\rm {erg}}}
\def	\s		{\,{\rm {s}}}
\def	\H		{\,{\rm {H}}}

\def    \P     {\,\boldsymbol{P}\,}
\def    \B     {\,\boldsymbol{B}\,}
\def    \k     {\,\boldsymbol{k}\,}
\def    \J     {\,\boldsymbol{J}\,}

\def \ahat {\,{\hat{\boldsymbol{a}}}}
\def \bea {\begin{eqnarray}}
\def \ena {\end{eqnarray}}                 

\begin{document}
\shorttitle{ }
\title{Synthetic Modelling of Polarized Dust Emission in Intermediate-Mass YSOs: I: Constraining the Role of Iron Inclusions and Inelastic Relaxation on Grain Alignment with ALMA Polarization}

\author{Nguyen Chau Giang}
\affil{Korea Astronomy and Space Science Institute, Daejeon 34055, Republic of Korea}
\email{chaugiang@kasi.re.kr}
\affil{Department of Astronomy and Space Science, University of Science and Technology, 217 Gajeong-ro, Yuseong-gu, Daejeon, 34113, Republic of Korea}

\author{V. J. M. Le Gouellec}
\email{valentinlegouellec@gmail.com}
\affil{NASA Ames Research Center, Space Science and Astrobiology Division M.S. 245-6 Moffett Field, CA 94035, US}
\affil{NASA Postdoctoral Program Fellow}

\author{Thiem Hoang}
\affil{Korea Astronomy and Space Science Institute, Daejeon 34055, Republic of Korea}
\email{thiemhoang@kasi.re.kr}
\affil{Department of Astronomy and Space Science, University of Science and Technology, 217 Gajeong-ro, Yuseong-gu, Daejeon, 34113, Republic of Korea}

\author{A. J. Maury}
\affil{Institute of Space Sciences (ICE), CSIC, Campus UAB, Barcelona, Spain}
\affil{ICREA, Pg. Lluís Companys 23, Barcelona, Spain}

\author{P. Hennebelle}
\affil{AIM, CEA, CNRS, Université Paris-Saclay, Université Paris Diderot, Sorbonne Paris Cité, F-91191 Gif-sur-Yvette, France} 

\begin{abstract}
Iron inclusions embedded inside dust grains play a crucial role in both internal alignment (IA) via Barnett relaxation and external alignment via the MAgnetically Enhanced RAdiative Torque (MRAT) mechanism. Moreover, inelastic relaxation is predicted to dominate over Barnett relaxation in driving the IA of micron-sized and very large grains above $10\mum$ (VLGs). Yet, a detailed modeling of polarized thermal dust emission from Class 0/I Young Stellar Objects (YSOs) taking into account these effects and their observational constraints is still lacking. In this paper, we update the POLARIS code and use it to perform synthetic dust polarization modeling for MHD simulations of an intermediate-mass YSO. Results will be post-processed with CASA to confront ALMA polarimetric observations. We found that to reproduce the high polarization degree of $p \sim 5-30\%$ observed in protostellar envelopes by ALMA, micron-sized and VLGs must contain iron inclusions with $N_{\rm cl} \sim 5 - 10^{3}$ iron atoms per cluster, assuming $30\%$ of iron abundance locked inside dust grains under the cluster form. Inside the inner $\sim 500$ au region, inelastic relaxation must participate in driving the grain internal alignment, and grains must contain larger iron inclusions of $N_{\rm cl} \sim 10^{2}-10^{4}$ and grow beyond $\geq 10\mum$ to reproduce $\sim 3-10\%$ of dust polarization observed by ALMA. But given such a combination, the internal alignment and MRAT efficiency acting on VLGs still decrease toward the center, inducing the decrease of $p(\%)$ with increasing gas density, reaching $p \sim 1\%$ inside the disk.

\end{abstract}

 \section{Introduction}\label{sec:intro}
 Magnetic fields ($\B$) are thought to play a crucial role in regulating the star and disk formation inside low- and intermediate-mass protostellar cores (\citealt{Shu_1987}, \citealt{Yusuke_2022}). Magnetic fields can guide the infalling gas flow which shapes the core structure, and control the growth of the protostar (\citealt{Galli_2006}, \citealt{Li_2014}) and the disk size (\citealt{Melon_2008}, \citealt{Machida_2010}) via the magnetic braking effect (\citealt{Mestel_Spitzer_1956}, \citealt{Melon_2008}). Magnetic fields are also crucial for the formation and collimation of protostellar jets and outflows (\citealt{Konigl_2000}, \citealt{Frank_2014}). They also help to track the stellar feedback in their parent core (\citealt{Seifried_2015}, \citealt{Offner_2017}). Therefore, measuring $\mathbf{B}$ fields throughout the entire protostellar core at different evolution stages is essential to understanding how they control the core and star evolution and how stellar feedback alters the natal $\B$ fields inside the core.
 
The most popular technique to map $\B$ fields toward low/intermediate-mass Class 0/I Young Stellar Objects (YSOs) is to use polarized thermal emission from dust grains that are aligned with magnetic fields (\citealt{Hull_2019}, \citealt{Tram_Hoang_2022}). Submillimeter/millimeter observations toward the protostellar core scale of 0.1 pc down to the disk scale of 100 au are widely conducted by both single-dish and interferometric telescopes such as James Clerk Maxwell Telescope (JCMT), Jansky Very Large Array (JVLA), SubMillimeter Array (SMA), and Atacama Large Millimeter Array (ALMA). The significant improvement in the spatial resolution reveals in detail the complicated evolution of $\B$ fields following the collapse/fragmentation of the core and the formation/evolution of YSOs (\citealt{Stephens_2013}, \citealt{Hull_2017b}, \citealt{Cox_2018}, \citealt{Maury_2018}, \citealt{Sadavoy_2019}, \citealt{Takahashi_2019}, \citealt{Kwon_2019}, \citealt{Liu_2021L}). However, the alignment of dust grains with magnetic fields is expected to significantly decrease inside protostellar environments due to the efficient gaseous damping (\citealt{Lazarian_2007}, \citealt{Hoang+2021}). It thus raises the question about the reliability of $\B$ field maps obtained by this technique within thousands au scale around the protostar. 

Besides the issue related to the origin of dust polarization observed in protostellar cores, ALMA observations frequently report the enhancement in polarized dust intensity along the dense outflow cavity wall (\citealt{Hull_2017b}, \citealt{Maury_2018}, \citealt{Valentin_2019}, \citealt{Kwon_2019}, \citealt{Hull_Valentin_2020}, \citealt{Valentin_2023a}). The achieve polarization degree can be high up to $p \sim 10-30\%$ in the envelope scale (\citealt{Cox_2018}, \citealt{Maury_2018}, \citealt{Kwon_2019},  \citealt{Valentin_2023a}), but it rapidly decreases toward the center, reaching $p \sim 1\%$ within $\sim 100$ au around the protostar (\citealt{Hull_2014}, \citealt{Cox_2018},  \citealt{Valentin_2023a}). Note that such a depolarization phenomenon is also observed in the cloud and filament scales by JCMT (\citealt{Pattle_2019}, \citealt{Lyo_2021}, \citealt{Ngoc_2021}), down to the core scale by CARMA (\citealt{Hull_2014}), SMA (\citealt{Girart_2006}), JVLA (\citealt{Rao_2009}, \citealt{Liu_2018}). The exact origin behind the depolarization in each scale remains unclear. But different from JCMT which can only resolve the cloud to sub-parsec scale, or SMA, CARMA, JVLA that can resolve the core to a few hundred au, ALMA can resolve YSOs up to a few tens of au. The detailed resolved structure of $\B$ fields by ALMA may weaken the impact of the beam size in producing low polarization degrees in the central region. Thus, it is likely that either the grain alignment efficiency, the low intrinsic polarization degree (i.e., grain elongation, \citealt{Hoang_Bao_2024}), or the disorganization of B fields by turbulence producing low polarization degrees near the stellar source. However, the exact relative importance of these effects is not yet quantified, which requires the detailed modeling of grain magnetic alignment to accurately interpret observed properties of dust polarization by ALMA in YSO environments.
  
Synthetic modeling of polarized dust emission based on the realistic model of grain alignment is critical to accurately interpret the observed dust polarization and constrain dust properties and dust alignment physics inside low/intermediate Class 0/I YSOs. Theoretically, the grain alignment undergoes two major processes: 1) the internal alignment (IA) which drives the alignment between the grain angular momentum $\J$ and the principal axis, i.e., axis with maximum inertia moment $\ahat_{1}$, and 2) the external alignment which brings $\J$ to align with magnetic fields (see \citealt{Anderson_2015}, \citealt{Lazarian_2015} for the reviews). For paramagnetic (PM) grains (namely grains containing diffusely distributed iron atoms),  they can dissipate their rotational energy and return to the lowest energy state with $\J \parallel \ahat_{1}$ through Barnett relaxation (\citealt{Purcell_1979}), nuclear relaxation (\citealt{Lazarian_Draine_1999}), and inelastic relaxation (\citealt{Lazarian_Efroisky_1999}, \citealt{Efroimsky_Lazarian_2000}). Two former mechanisms dissipate the grain rotational energy based on the precession of the Barnett-induced grain magnetic moment around the principal axis within the grain inertia frame (\citealt{Barnett_1915}). In contrast, inelastic relaxation dissipates the rotational energy via the deformation of inelastic grains spinning around the non-principal axis induced by centrifugal acceleration. As shown in \cite{Lazarian_Hoang_2019}, \cite{Hoang+2022}, sub-micron grains tend to have internal alignment by Barnett relaxation, while micron-sized and very large grains beyond $\geq 10\mu m$ (VLGs) experience this process by nuclear relaxation and inelastic relaxation. For the external alignment, the interaction of the grain magnetic moment with the ambient magnetic fields induces the Larmor precession which drives the coupling of dust grains with $\B$ fields. Dust grains then can have the magnetic alignment by paramagnetic relaxation mechanism \citep{David_1951} (aka Davis-Greeenstein mechanism). However, the Davis-Greenstein mechanism is shown to be inefficient \citep{Hoang_Lazarian_2016}, and the leading external alignment mechanism is based on Radiative Torques (RATs) (\citealt{Dolginov_1976}, \citealt{Lazarian_Hoang_2007a}, \citealt{Hoang_Lazarian_2008}) produced by differential scattering and absorption of an anisotropic radiation field with irregular dust grains. The RAT can stabilize the grain magnetic coupling by inducing suprathermal grain rotation and drive the alignment between $\J$ and $\B$ by radiative alignment torques. However, only a fraction of grains can have the magnetic alignment at their suprathermal rotation, or high-\textit{J} attractors. The rest of grains will experience spin-down by radiative torques and be aligned with $\B$ at thermal rotation, or low-\textit{J} attractors (see \citealt{Hoang+2022}). The fraction of grains at high-\textit{J} attractors, denoted by $f_{\rm high-J}$, varies with grain properties (shape, size, and composition) and the angle between the radiation direction and magnetic fields (\citealt{Lazarian_2007}, \citealt{Hoang_Lazarian_2008}, \citealt{Herranen_2021}). Typically, about $\sim 25\%$ of PM grains can have the magnetic alignment at high-\textit{J} attractors by RATs \citep{Hoang_Lazarian_2008,Herranen_2021}. However, for grains containing embedded iron clusters (i.e., superparamagnetic (SPM) grains), \cite{Hoang_Lazarian_2016_mrat} demonstrated through numerical simulations that SPM grains can achieve a high degree of alignment with $f_{\rm high-J}\approx 1$, regardless of grain shapes, due to the joint effect between RATs and enhanced magnetic relaxation. This mechanism is named Magnetically Enhanced RAdiative Torque (MRAT) alignment.

The recent publication of POLArized RadIative Simulator (POLARIS) by \cite{Reissl_2014} and \cite{Reissl_2016} establishes a reliable platform for connecting dust properties and RAT alignment theorem with observations of polarized dust emission. Using the standard RAT model for PM grains, \cite{Valdivia_2019} found that the polarization degree above $5\%$ detected in the envelope of low-mass Class 0/I YSOs can be produced by the emission of aligned VLGs beyond $\geq 10\mum$. However, ALMA polarization data suggest that VLGs to have perfect magnetic alignment in protostellar environments (\citealt{Valentin_2020}). This anomalous feature can be explained by the ideal RAT alignment model with $f_{\rm high-J}= 1$ (\citealt{Valentin_2023b}). However, \cite{Hoang_2022} and \cite{Hoang+2022} showed that in protostellar environments, micron-size grains cannot be aligned with $\B$ if they are PM owing to their slow Larmor precession compared to the fast gas damping timescale. The RAT alignment, which majorly drives the magnetic alignment for PM grains, thus becomes ineffective there. However, if protostellar grains are SPM, micron-sized and VLGs can gain fast internal relaxation and be able to couple with magnetic fields (\citealt{Hoang_2022}, \citealt{Hoang+2022}). The enhanced magnetic susceptibility of SPM grains operates MRAT mechanism to drive the magnetic alignment of SPM grains (\citealt{Lazarian_Hoang_2008}, \citealt{Hoang_Lazarian_2016_mrat}), allowing protostellar grains up to $\sim 50-100\mum$ to achieve perfect magnetic alignment with $f_{\rm high-J} \sim 1$ (\citealt{Hoang+2022}, \citealt{Giang_2023a},  \citealt{Giang_2023b}). This finding can explain why the ideal RAT model with $f_{\rm high-J}=1$ in \cite{Valentin_2023b} could reproduce ALMA data. Given that grains tend to be SPM in protostellar environments, we confirm the finding in \cite{Valdivia_2022} that dust polarization is the reliable tracer of magnetic fields in the protostellar envelope and inner envelope scale. However, within $\sim 100$ au around the protostar, the origin of dust polarization remains unclear because almost SPM grains are poorly aligned with $\B$ regardless of their sizes owing to the strong gaseous damping there (\citealt{Yang_2021}, \citealt{Lam_2021}, \citealt{Giang_2023b}).

Although \cite{Giang_2023a} incorporated the detailed MRAT mechanism and (super-)Barnett relaxation into the updated POLARIS, they disregarded the effect of inelastic relaxation which was expected to dominate over the (super-)Barnett relaxation in driving the internal alignment of micron-sized and VLGs in protostellar environments (\citealt{Lazarian_Hoang_2019}, \citealt{Hoang+2022}). Consequently, the internal alignment degree predicted for VLGs in \cite{Giang_2023a} and \cite{Giang_2023b} may be underestimating their realistic situation in YSOs environments. Moreover, our previous studies indicated that the perfect magnetic alignment of very large SPM grains driven by MRAT mechanism in the envelope can produce $p \sim 30\%$ observed by ALMA at this scale. However, \cite{Valentin_2020} and \cite{Valentin_2023b} found that the maximum level of dust polarization in the outermost region may be biased by interferometric filtering, which questions the realistic origin for such ALMA feature. Another missing point is that \cite{Giang_2023b} only focuses on studying the dust polarization in the protostellar envelope and disks of low-mass Class 0 YSOs, yet properties of dust polarization from the protostellar outflow remain unclear. Even though this area has a lower density than the protostellar envelope and protostellar disk, it is still much higher than the diffuse interstellar medium (ISM) and molecular clouds (MCs) where the RAT model is always valid for grains below $1\mum$ (see \citealt{Tram_Hoang_2022} for a review). Therefore, to achieve reliable constraints on the grain alignment physics and dust properties (i.e., iron inclusions and grain growth) in protostellar environments, in this paper (Paper I), we will perform synthetic modeling of polarized dust emission toward an intermediate Class 0 YSO with the clear outflow feature, using all well-known alignment mechanisms incorporated in POLARIS code. Synthetic results will be post-processed with Common Astronomy Software Application (CASA) to confront ALMA observations (\citealt{Valentin_2020}).

Additionally, \cite{Valentin_2023b} found that with the high bolometric luminosity beyond $\geq 20L_{\odot}$ required to reproduce the grain alignment degree inferred from ALMA dust polarization, RAdiative Torque Disruption (RATD) proposed by \cite{Hoang_2019} can destroy very-fast spinning aggregate$-$type VLGs up to thousands au scale inside the outflow cavity. Consequently, VLGs must have composite or compact structures to avoid being destructed by RATD and to reproduce the observed dust polarization fraction (\citealt{Valentin_2023b}), yet how it can happen remains unclear. We will study in detail the impact of RATD on dust population and thermal dust emission polarization in our follow-up paper (Paper II).

The structure of this paper is as follows. We first describe the MHD simulation of the self-collapsing protostellar core and our synthetic modeling of dust polarization in Sections \ref{sec:MHD} and \ref{sec:post_processing}. We show the spatial distribution of the critical alignment size in Section \ref{sec:Distribution}. The effect of iron inclusions and inelastic relaxation on the polarization degree map will be shown in Section \ref{sec:Iron_Inelastic_properties}, followed by Section \ref{sec:ALMA_observations} where we constrain the grain alignment mechanism and grain physical properties from ALMA observations. Further discussion and summary of our study will be carried out in Sections \ref{sec:discussion} and \ref{sec:summary}, respectively.

 \section{RAMSES MHD protostellar core}\label{sec:MHD}
We use a non-ideal MHD simulation of the self-collapsing core from \cite{Valentin_2023b} to study the effect of iron inclusions and inelastic relaxation on thermal dust polarization. The simulation is performed by the RAMSES code (\citealt{Teyssier_2002}), starting with the collapse of a strong magnetized intermediate prestellar core with a mass of $30 M_{\odot}$ and a dust$-$to$-$gas mass ratio $\mu = 5$. Ambipolar diffusion is considered to account for the decoupling between magnetic fields and infall gas (\citealt{Masson_2012}). The core has no initial turbulence and contains the uniform magnetic field along the rotation z$-$axis. They use the Adaptive Mesh Refinement (AMR) method to capture the star and disk formation at the smallest resolution of 5 au. The sink particle is used to track the evolution of the newly formed stellar object (\citealt{Krumholz_2004}, \citealt{Bleuler_2014}). The jet is implemented by hand right after the sink particle is merged into the simulation (\citealt{Verliat_2022}). 
 
For our study, we take the snapshot when the core evolves 38.52 kyr (Class 0 protostar). The sink particle has an age of 14.8 kyr, a mass of $1.2M_{\odot}$, a radius of $1.58R_{\odot}$, and a stellar luminosity of $L_{\rm star} = 0.58L_{\odot}$. We show in the upper left panel of Figure \ref{fig:distribution_mcrt} the spatial distribution of the gas volume density $n_{\rm H_{2}}$ on the slice containing the sink particle. The observed direction is along the edge$-$on direction, and the gas velocity field is overplotted in the panel. Without initial turbulence, the core collapses isotropically and forms a symmetric density structure around the vertical z$-$direction. The flattened disk with high $n_{\rm H_{2}} \sim 10^{10}\cm^{-3}$ and radius of $\sim 100$ au is formed on the equatorial midplane. The material above and below the disk is swept up outward following the jet propagation, creating the outflow cavity with an opening angle of $\sim 30^{\circ}$ and low density $n_{\rm H_{2}} \sim 10^{4}-10^{6}\cm^{-3}$. The outflow cavity wall with higher $n_{\rm H_{2}} \sim 10^{7}\cm^{-3}$ is formed from the compression between infalling and outflowing gas. The protostar, protostellar disk, and outflow/jet are embedded inside the envelope with $n_{\rm H_{2}} \sim 10^{6}\cm^{-3}$.

\section{Synthetic modelling of dust polarization with POLARIS}\label{sec:post_processing}
\subsection{Radiation fields and dust model} \label{sec:modelling_setup}
Considering the RAT alignment mechanism, \cite{Valentin_2023b} found that the central of low/intermediate-mass protostellar cores must radiate at least $\geq 20L_{\odot}$ to reproduce the grain alignment efficiency inferred by ALMA, assuming the maximum grain size of $10\mum$. Given the low stellar luminosity $L_{\rm star} = 0.58L_{\odot}$  (Section \ref{sec:MHD}), such required high luminosity above tentatively released from the accretion of supersonic gas from the accretion disk onto the protostar photosphere. Given the complex accretion activities and since we do not aim to monitor the time-varying dust polarization properties during the accretion burst event (\citealt{Valentin_2023b}), we simply consider our sink particle (Section \ref{sec:MHD}) to be a black body object and vary the temperature $T_{\rm star}$ to get the chosen bolometric luminosity. We adopt the high central luminosity of $100L_{\odot}$ (corresponds to $T_{\rm star} = 16457$ K) in the rest of our study to better quantify the grain alignment dynamic and dust polarization properties inside YSO environments. Extent discussion about the impact of bolometric luminosity and their spectral energy distribution on dust polarization will be raised in Section \ref{sec:discuss_Lstar}. We do not consider the interstellar radiation field in our post-processing due to their negligible effect on dust heating and grain alignment within thousands au scale around the protostar. The radiation spectrum extends from the upper bound of photoionization regism with $\lambda_{\rm min} = 0.1\mum$ until the millimeter range with $\lambda = 3$ mm, where the dust-radiation interaction becomes negligible.

\begin{table}
\centering
\caption{Setup for the radiation field and dust model in POLARIS}
\begin{tabular}{lll}
\hline
Quantity & Symbol & Value \\   
\hline
\multicolumn{3}{c}{\textbf{ Radiation sources}}\\
\hline 
Stellar radius              & $R_{\rm star}$       & $1.23R_{\odot}$ \\ 
Effective temperature       & $T_{\rm star}$       & 7800, 11000, 16457 K  \\
Central luminosity          & $L_{\rm center}$       & $5, 20, 100 L_{\odot}$ \\

\hline
\multicolumn{3}{c}{\textbf{Dust model}}\\
\hline
Grain axial ratio           & $s$                    & 0.5\\
Dust-to-gas mass ratio      & $\eta$               & 0.01\\
Initial size distribution   & $\rm dn/da$              & C$ a^{-3.5}$\\
Minimum grain size          & $a_{\rm min}$        & 5\AA  \\ 
Maximum grain size          & $a_{\rm max}$        & $1, 5, 10, 50, 100\mu m$  \\
Fraction of silicate        &                      & $67.5\%$  \\
Fraction of graphite        &                      & $32.5\%$  \\

\hline
\multicolumn{3}{c}{\textbf{Grain magnetic properties}}\\
\hline
Iron fraction & $f_{\rm p}$ & 0.1 \\
Iron atom/cluster & $N_{\rm cl}$ & 5, $ 10^{2}, 10^{3}, 10^{4}$ \\
Volume filling factor  & $\phi_{\rm sp}$ & 0.1 \\
of iron clusters & & \\

\hline 
\multicolumn{3}{c}{\textbf{Elasticity of grains}}\\
\hline
Elasticity & $\mu Q$ & $3e9, 3e10, 3e12 \erg\cm^{-3}$ \\

 \hline
\multicolumn{3}{c}{\textbf{Internal alignment degree }}\\
\hline
\multicolumn{3}{c}{Grains with fast internal relaxation}\\
\hline
High-$J$ attractors & $Q_{\rm X}^{\rm high-J}$ & 1 \\
Low-$J$ attractors & $Q_{\rm X}^{\rm low-J}$ & TE Boltzmann distribution\\
\hline
\multicolumn{3}{c}{Grains with slow internal relaxation}\\
\hline
High-$J$ attractors & $Q_{\rm X}^{\rm high-J}$ & 0.15 \\
Low-$J$ attractors & $Q_{\rm X}^{\rm low-J}$ & 0.05\\

 \hline
\multicolumn{3}{c}{\textbf{External alignment degree}}\\
\hline
High-$J$ attractors & $Q_{\rm J}^{\rm high-J}$ &  1 \\
Low-$J$ attractors & $Q_{\rm J}^{\rm low-J}$ & 1 \\

  \hline
    \label{tab:parameter}
    \end{tabular}   
\end{table} 

\begin{table*}
\centering
\caption{Notations used in the paper}
\begin{tabular}{ll}
\hline
Meaning & Symbol  \\   
  \hline
Minimum size for external alignment by RATs & $a_{\rm align}$ \\ 
Maximum size for external alignment by Larmor precession & $a_{\rm max,JB}^{\rm Lar}$ \\ 
Alignment range & $[a_{\rm align}-a_{\rm max,JB}^{\rm Lar}]$\\
Fraction of grains aligned with $\B$ at high-\textit{J} attractors & $f_{\rm high-J}$ \\ 
Size range for internal alignment by fast Barnett relaxation at high-\textit{J} attractors & $a_{\rm min,aJ}^{\rm Bar,highJ} - a_{\rm max,aJ}^{\rm Bar,highJ}$ \\

Size range for internal alignment by fast Barnett relaxation at low-\textit{J} attractors & $a_{\rm min,aJ}^{\rm Bar,lowJ} - a_{\rm max,aJ}^{\rm Bar,lowJ}$ \\

Size range for internal alignment by fast inelastic relaxation at high-\textit{J} attractors & $a_{\rm min,aJ}^{\rm iNER,highJ} - a_{\rm max,aJ}^{\rm iNER,highJ}$ \\

Maximum size for internal alignment by fast inelastic relaxation at low-\textit{J} attractors & $a_{\rm max,aJ}^{\rm iNER,lowJ}$ \\

Maximum size having $f_{\rm high-J} = 0.5$ by MRAT alignment & $a_{\rm max,JB}^{\rm DG,0.5}$ \\

Maximum size having $f_{\rm high-J} = 1$ by MRAT alignment & $a_{\rm max,JB}^{\rm DG,1}$ \\
Rayleigh reduction factor (degree of grain alignment) & $R$ \\

Observed angle to the z-direction & $\Theta$ \\

(Dimensionless) Radiation field strength & $U_{\rm rad}$ \\

Stokes parameter & [I Q U V] \\

Polarization fraction & $p(\%)$ \\

Magnetic moment of iron atom & $p$ \\

    \label{tab:denotion}
    \end{tabular}   
\end{table*} 

For the dust model, we adopt a constant dust-to-gas mass ratio of $\eta = 0.01$ found in ISM to describe the distribution of dust grains inside the protostellar core. Dust grains are assumed to have a composite structure, including $67.25\%$ of silicate and $33.75\%$ of graphite, with both silicate and graphite contributing to polarized dust emission. To study the effect of iron inclusions on dust polarization, we consider five types of dust grains: paramagnetic (PM) grains with iron atoms occupying $f_{\rm p} = 0.1$ the grain volume, superparamagnetic (SPM) grains with iron clusters occupying $\phi_{\rm sp} = 0.1$ the grain volume (corresponding to $30\%$ of iron depleting from the gas phase to be embedded inside grains under cluster form), with each cluster containing $N_{\rm cl} = 5, 10^{2}, 10^{3}$, and $10^{4}$ iron atoms \footnote{The composite dust model containing embedded iron inclusions is the leading dust model for the diffuse ISM and molecular clouds, but the exact size of iron clusters is unknown \citep{Jones_2013, Jones_2017,Draine_Hensley_2021a,Ysard_2024}}. 
Finally, to understand the impact of inelastic relaxation on the magnetic alignment process, we describe the inelasticity of dust grains via the product of $\mu$ the shear modulus, and $Q$ the quality factor. The $\mu$ characterizes the deformation level of material when applying force, which can vary among $\mu \sim 3.6\times 10^{7} - 3.46\times 10^{9}\erg\cm^{-3}$ (\citealt{Knapmeyer_2018}). While the $Q$ factor characterizes how fast grains dissipate rotational energy via grain deformation, which can change from $Q = 100$ for silicate rocks (\citealt{Efroimsky_Lazarian_2000}) to $Q \sim 400-2000$ for vitreous silicate (\citealt{Purcell_1979}). To roughly describe the inelasticity of protostellar grains, we use two values of $\mu Q  = 3\times 10^{9}$ and $3\times 10^{10}\erg\cm^{-3}$ to characterize material with strong inelasticity, and $\mu Q = 3\times 10^{12}\erg\cm^{-3}$ to characterize high elastic material. We assume dust grains to follow the standard grain distribution with $dn/da \propto a^{-3.5}$ (\citealt{Mathis_1977}), with the minimum size $a_{\rm min} = 5$ \AA. To quantify the effect of grain growth on dust polarization, we consider different maximum grain sizes from $a_{\rm max} = 1\mum$ and $a_{\rm max} = 100\mum$. We summarize all parameters used in our post-processing in Table \ref{tab:parameter}.

\subsection{Radiative transfer and Dust temperature}\label{sec:modelling_mcrt}
Given the stellar radiation source and the dust model, we first perform the Monte-Carlo three-dimensional (3D) radiative transfer of photons, accounting for scattering, absorption, and spontaneous thermal dust emission following the technique introduced by \cite{Lucy_1999}. The grain temperature of each grain size $T_{\rm d}(a)$ is calculated after the energy density distribution is determined, using the energy conservation between dust absorption and thermal dust emission. The average dust temperature inside the cell $T_{\rm d,rt}$ is found by integrating $T_{\rm d}(a)$ over the grain size distribution. In our simulation, we assume the effective energy exchange between dust grains and protostellar gas over time. Therefore, we sum $T_{\rm d,rt}$ with the gas temperature $T_{\rm g,mhd}$ inside the MHD simulation, giving the final dust and gas temperature of $T_{\rm d} = T_{\rm g} = T_{\rm d,rt} + T_{\rm g,mhd}$. Note that in the MHD simulation, $T_{\rm g,mhd}$ is a result of the stellar heating ($L_{\rm star}$), accretion disk heating (by the viscosity heating and energy releasing when material falls into the disk and to the protostar photosphere), and iron-neutral gas drift and gas compression inside the jet. In POLARIS, $T_{\rm d,rt}$ is determined by the radiation heating released from the protostar and the accretion burst (represented by $L_{\rm center}$ in Section \ref{sec:modelling_mcrt}). As the stellar luminosity $L_{\rm star} = 0.58L_{\odot}$ is subdominant to the accretion luminosity and mechanical heating inside jets, duplicating the contribution from the photosphere does not affect the dust and gas temperature. Thus, it is safe to take the sum of $T_{\rm g,mhd}$ and $T_{\rm d,rt}$ to get $T_{\rm d}$ and $T_{\rm g}$ in our post-processing.

\subsection{Modelling of grain alignment}\label{sec:modelling_alignment}
We first determine the minimum and maximum size that grains can be aligned with $\B$, $a_{\rm align}$ and $a_{\rm max,JB}^{\rm Lar}$. The minimum alignment size is determined by the suprathermal rotation condition of RATs (\citealt{Hoang_Lazarian_2014}, \citealt{Hoang_Lazarian_2016}), while the maximum alignment size is determined when the Larmor precession is slower than the gas randomization a factor of 10 (see \citealt{Giang_2023a} for detailed calculations) \footnote{The updated POLARIS is available in \cite{POLARIS_link}}. Depending on the grain magnetic susceptibility, some values of gas density could lead all dust grains to be misaligned with $\B$ due to their slow Larmor precession, i.e., $a_{\rm max,JB}^{\rm Lar} < a_{\rm align}$. In such cases, grains will have random orientation in space even if they can get suprathermal by RATs.

For grains that can be coupled with magnetic fields, the external alignment mechanism (RATs or MRAT alignment) will be determined based on the magnetic relaxation ratio $\delta_{\rm m}$, which characterizes how fast the magnetic relaxation compared with the gas damping (\citealt{Lazarian_Hoang_2008}, \citealt{Hoang_Lazarian_2016_mrat}). We illustrate the external alignment mechanism accompanied with the fraction of grains at high-\textit{J} attractors $f_{\rm high-J}$ as a function of grain sizes in Appendix \ref{sec:appen_align_fhighJ} (see original figure in \citealt{Giang_2023a}). For a quick summary, RAT will be the alignment mechanism for grains having weak magnetic relaxation with $\delta_{\rm m} < 1$, with the typical value of $f_{\rm high-J} = 0.25$. In contrast, MRAT will lead the external alignment of grains, with $f_{\rm high-J} = 0.5$ for grains having $1 \leq \delta_{\rm m} \leq 10$, and $f_{\rm high-J} = 1$ for grains with $\delta_{\rm m}\geq 10$. The minimum and maximum size for $50\%$ and $100\%$ of grains being aligned with $\B$ at high-\textit{J} attractors are labeled as $a_{\rm min,JB}^{\rm DG,0.5}$, $a_{\rm max, JB}^{\rm DG, 0.5}$, and $a_{\rm min,JB}^{\rm DG,1}$, $a_{\rm max,JB}^{\rm DG,1}$, respectively. 

For the internal alignment, we consider that grains can only have fast internal relaxation if the internal alignment timescale driven by either (super-)Barnett relaxation or inelastic relaxation is shorter than the gas damping timescale. The typical range of size with fast Barnett relaxation is denoted as $a_{\rm min,aJ}^{\rm Bar,lowJ}$ to $a_{\rm max,aJ}^{\rm Bar,lowJ}$ for grains aligning with $\B$ at low-\textit{J} attractors, and $a_{\rm min,aJ}^{\rm Bar,highJ}$ to $a_{\rm min,aJ}^{\rm Bar,highJ}$ for grain aligning with $\B$ at high-\textit{J} attractors. The range for grains having fast inelastic relaxation is denoted as $a_{\rm max,aJ}^{\rm iNER, lowJ}$ for grains at low-\textit{J}, and $a_{\rm min,aJ}^{\rm iNER, highJ}$ and $a_{\rm max,aJ}^{\rm iNER, highJ}$ for grains at high-\textit{J} attractors. The summary of all notations describing the alignment state used in our paper is listed in Table \ref{tab:denotion}. The detailed calculation for the critical size of Barnett and inelastic relaxation is described in detail in \cite{Giang_2023a} and Appendix \ref{sec:inelastic_theory}, respectively.

The net alignment degree of grains of size $a$ with magnetic fields is described by the Rayleigh reduction factor $R$ (\citealt{Greenberg_1968}). According to the RAT mechanism, the value of $R$ can be described as \citep{Hoang_Lazarian_2014}:
\begin{equation}
R = f_{\rm high-J} Q_{\rm X}^{\rm highJ} Q_{\rm J}^{\rm highJ} + (1-f_{\rm high-J})Q_{\rm x}^{\rm lowJ}Q_{\rm J}^{\rm lowJ}, \label{eq:Rayleigh}
\end{equation}
where $Q_{\rm X}$ and $Q_{\rm J}$ describe the internal and external alignment degree, respectively. Grains beyond the alignment range determined by $a_{\rm align}-a_{max,JB}^{\rm Lar}$ will have the random orientation with $R = 0$. For grains within the alignment range, we consider the perfect external alignment with B-fields for grains at both high and low-\textit{J} attractors, i.e., $Q_{\rm J,lowJ} = Q_{\rm J,highJ} = 1$. For the internal alignment, grains having either the fast Barnett relaxation or fast inelastic relaxation at high-\textit{J} will achieve perfect internal alignment with $Q_{\rm X,highJ} = 1$. Grains with fast internal relaxation at low-\textit{J} will have imperfect IA due to internal thermal fluctuation. Their alignment state can be described by the local thermal equilibrium Boltzmann distribution (\citealt{Lazarian_Roberge_1997}, \citealt{Hoang_Lazarian_2014}) due to the efficient exchange between the rotation-vibration energy inside dust grains. Grains beyond both the fast Barnett and fast inelastic relaxation range in high and low-\textit{J} attractors will be considered to have inefficient IA by slow internal relaxation, whose alignment direction and alignment degree are not yet well constrained (\citealt{Hoang_Lazarian_2009}). In the rest of our study, we assume grains in this regime have the right IA \footnote{Indeed, \cite{Hoang_Lazarian_2009} found that grains with slow internal relaxation can also have the wrong IA (grains rotate around their longest axis) if they rotate thermally. However, we do not consider the wrong alignment scenario in our study because of its minor effect on the net polarization degree (\citealt{Giang_2023a}, \citealt{Giang_2023b}). Since our study does not focus on the dust polarization pattern, considering only the case of right alignment is enough for exploring the impact of iron inclusions and inelastic relaxation on dust polarization properties.} with $Q_{\rm X,highJ} = 0.15$ and $Q_{\rm X,lowJ} = 0.05$. \footnote{Comparing to the modeling of radiative alignment for thermal grains by \cite{Tazaki_2017}, we use higher external alignment degree ($Q_{\rm J,lowJ} = 1$) than values adopted in \cite{Tazaki_2017} ($Q_{\rm J,lowJ} = 0.1-0.2$). However, for the internal alignment degree, we adopt lower $Q_{\rm X,lowJ} \sim 0.3$ and $Q_{\rm X,lowJ} = 0.05$ for thermal grains having fast and slow internal relaxation, while \cite{Tazaki_2017} considered $Q_{\rm X,lowJ} = 0.5$ regardless of the internal relaxation rate of thermal grains. For grains at high-\textit{J} attractors, while \cite{Tazaki_2017} considered $Q_{\rm X,highJ} = Q_{\rm J,highJ} = 1$ regardless of the internal relaxation timescale, we consider $Q_{\rm X,highJ} = 1$ only for grains with fast internal relaxation, and $Q_{\rm X,highJ} = 0.15$ when they have slow internal relaxation. Combining $Q_{\rm X,lowJ}, Q_{\rm J,lowJ}$, $Q_{\rm X,highJ}$, $Q_{\rm J,highJ}$ with any values of $f_{\rm high-J}$, we get similar net alignment degrees with \cite{Tazaki_2017} study, that neglects the bias of MRAT mechanism over RATs in modeling dust polarization by choosing too high values of $Q_{\rm X}$ and $Q_{\rm J}$.}.

 \begin{table*}
  \centering
         \caption{Model setup}
  \begin{tabular} {ccccccc}
  \hline 
\textbf{Model name} & \textbf{Internal alignment mechanism} & \textbf{Alignment range} & \textbf{Slow IR$^{a}$} & $f_{\rm high-J}$  \\
  \hline 
PA (Perfect Alignment)& -- & $a_{\rm align} - a_{\rm max}$ & No & 1  \\

RATA (RAdiative Torque Alignment)   & Barnett relaxation & $a_{\rm align} - a_{\rm max,JB}^{\rm Lar}$ & Yes & $\delta_{\rm m}$$^{b}$ \\ 

RATA$-$INELASTIC & Barnett + inelastic relaxation & $a_{\rm align} - a_{\rm max,JB}^{\rm Lar}$ & Yes & $\delta_{\rm m}$$^{b}$ \\ 
   \hline 
    \label{tab:model}
    \end{tabular}\\
    \footnotesize{($^{a}$): consider the slow internal relaxation for grains with both Barnett relaxation and inelastic relaxation timescale being longer than the gas damping timescale.} \\
     \footnotesize{($^b$): Fraction of grain aligning with $\B$ at high-\textit{J} attractors is determined by the magnetic relaxation ratio $\delta_{\rm m}$ which depends on the grain magnetic properties and the gas damping timescale.}\\ 
\end{table*}
  
\subsection{Polarized radiative transfer of Stokes parameters and CASA post-processing}\label{sec:modelling_stokes}
Knowing the grain alignment efficiency of each grain size, we finally perform the synthetic observation of dust polarization by solving the polarized radiative transfer of Stokes parameters. We place the plane detector at 300 pc from the object to observe the 8000 au scale of the protostellar core. The plane detector contains 1000$\times$1000 pixels, which resolves the core at 8 au. We observe the core at 1.3mm, which corresponds to Band 6 of ALMA. The observed direction in the rest of our study is along the edge$-$on direction passing through the equatorial midplane. 
 
\begin{figure*}
\centering
    \includegraphics[width=\textwidth,height=\textheight,keepaspectratio]{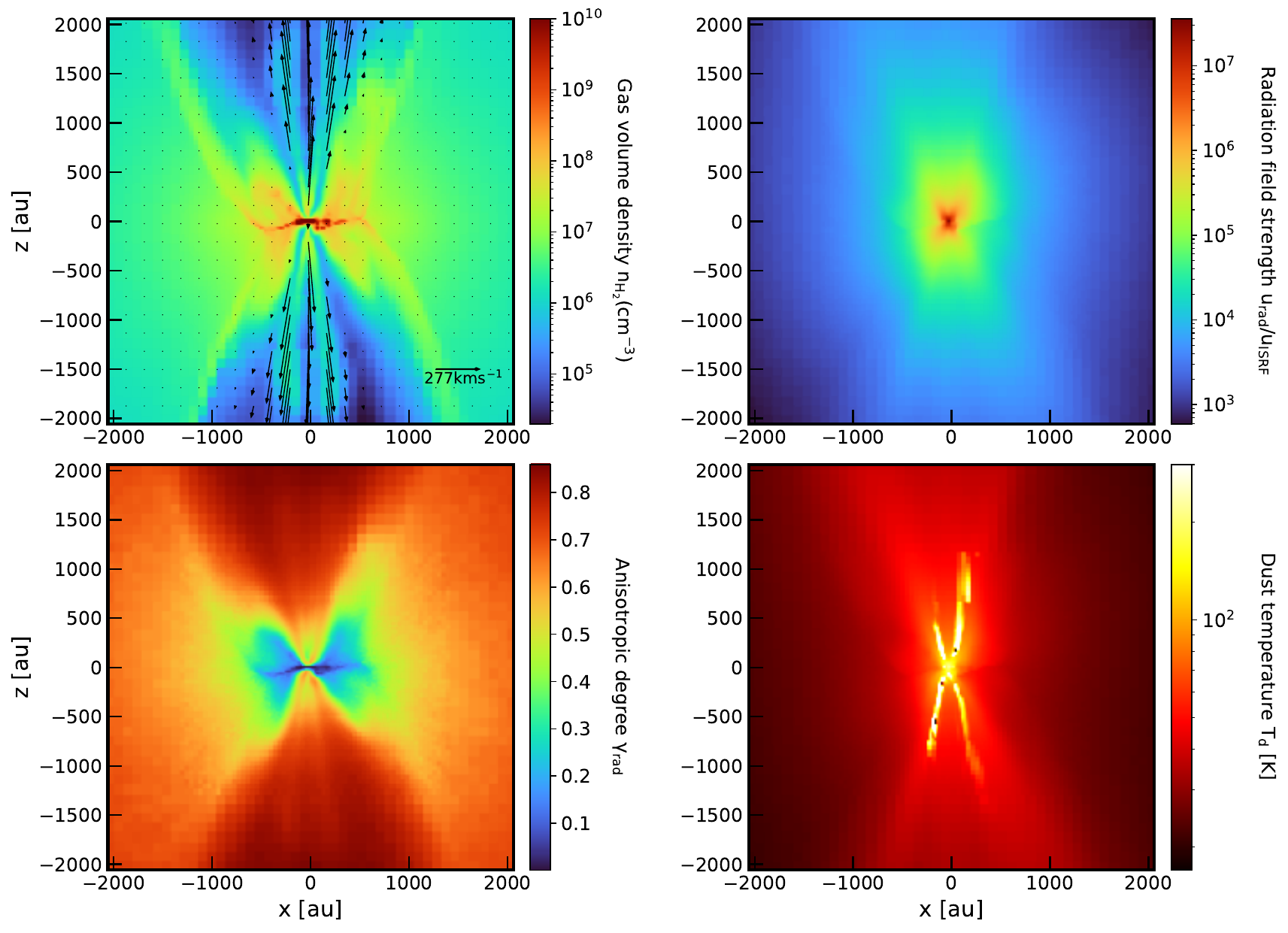}
    \caption{Upper left panel: spatial distribution of the gas volume density $n_{\rm H_{2}}$ overplotted with the gas velocity field on the slice containing the sink particle. The object is seen with edge$-$o direction. Upper right panel and lower left panel: spatial distribution of the radiation field strength $U_{\rm rad} = u_{\rm rad}/u_{\rm ISRF}$ and the anisotropic degree $\gamma_{\rm rad}$ of the radiation field within 4000 au. The corresponding distribution of the dust temperature $T_{\rm d}$ is plotted in the lower right panel. The radiation field strength decreases outward but it will become more anisotropic as reducing dust extinction in the envelope. Grains can be heated to $\sim 100$ K in the center, and gain lower $T_{\rm d} \sim 20-50$K beyond 1000 au} 
     \label{fig:distribution_mcrt}
\end{figure*}
 
In order to understand the effect of interferometric filtering on observations, we post-process the synthetic Stokes parameter maps obtained from each model with CASA (version 5.8) to confront our synthetic results with ALMA observational data. We use the \textit{simobserve} task to observe Stokes I, Q, U maps of each dust model at 1.3mm with three ALMA configurations C$-$2, C$-$4, and C$-$6 during three hours/configuration. We then use \textit{simanalyze} to feature the maps obtained from three ALMA configurations together and use \textit{tclean} to clean the dirty map and find the noise level of Stokes I, Q, and U maps, denoted as $\sigma_{\rm I}, \sigma_{\rm Q},$ and $\sigma_{\rm U}$, respectively. The observed field of view is 20 arcsec, which covers 8000 au around the protostellar core, and the cell size is 0.02 arcsec, which resolves the object at 8 au as setups in the synthetic modeling with POLARIS. The final synthetic ALMA beam size is 140 au. 

The synthetic Stokes I, Q, U maps before filtering with CASA will be smoothed with the Gaussian kernel to reduce the contrast in resolutions between the center and outer regions induced by the AMR technique. To quantify the tangling of $\B$ fields by turbulence in the interferometric map, we regrid both the synthetic maps before and after filtering with CASA into 4 pixels per synthetic beam size to satisfy the Nyquist sampling. The resulting polarized intensity will be $I_{\rm pol} = \sqrt{Q^{2} + U^{2}}$ \footnote{We use $I_{\rm pol}$ instead of $P$ (as the convention in observation) to denote the polarized intensity throughout our paper}. The polarization degree (in the unit of percentage) is calculated by:

\bea 
p(\%) = \frac{I_{\rm pol}}{I} = \frac{\sqrt{Q^{2} + U^{2}}}{I} \times 100\%,
\ena 

and the polarization angle dispersion $S$ for pixel $i$ is calculated by using (\citealt{Valentin_2020}):
\bea 
S(\delta)_{\rm i} = \sqrt{\frac{1}{n} \sum_{\rm j = 1}^{\rm n} \Bigg[\frac{1}{2} \arctan \frac{Q(j)U(i)- U(j)Q(i)}{Q(j)Q(i) + U(j)U(i)}\Bigg]^{2}},
\ena
with $n = 8$ nearest neighborhood cells around pixel $i$.

\section{Grain alignment around YSOs}\label{sec:Distribution}
\subsection{Dust temperature and minimum alignment size distribution}\label{sec:distribution_mc}
The upper right and lower left panels of Figure \ref{fig:distribution_mcrt} show the spatial distribution of the radiation field strength $U_{\rm rad}$ and its anisotropic degree $\gamma_{\rm rad}$ within 2000 au around the protostar. Here, the radiation field strength is defined as $U_{\rm rad} = u_{\rm rad}/u_{\rm ISRF}$, with $u_{\rm rad} = \int u_{\lambda} d\lambda$ presents the total energy density inside the core, and $u_{\rm ISRF} = 8.64\times 10^{-13} \erg\cm^{-3}$ presents the total energy density found in ISM (\citealt{Mathis_1983}). Within 500 au around the protostar, the radiation field strength is very high, reaching $U_{\rm rad} \sim 10^{7}$. This field is highly isotropic with $\gamma_{\rm rad} \sim 0.1$ owing to the strong dust scattering in this very dense central region (see the density distribution in the upper left panel of Figure \ref{fig:distribution_mcrt}). Moving toward the envelope, $U_{\rm rad}$ decreases to $\sim 10^{3}$ as increasing dust extinction, but radiation fields become more highly anisotropic, i.e., $\gamma_{\rm rad} \sim 0.8$, due to the reduced scattering between thermal dust emission radiating from the center and envelope grains. We show the corresponding dust temperature in the lower right panel of Figure \ref{fig:distribution_mcrt}. Warm dust grains with $T_{\rm d} \sim 100$ K are presented in the inner 500 au region, and cold dust with low $T_{\rm d} \sim 60$ K and $T_{\rm d} \sim 30$ K are located inside the outflow cavity and the envelope, respectively. Along the jet lobe of height 1000 au, the dust temperature can exceed $\geq 200$ K, given that grains are coupled with hot outflowing gas heated by the neutral-iron gas friction and gas compressing (\citealt{Valentin_2023b}).
 
\begin{figure*}
\centering
 \hspace*{-13cm}
    \includegraphics[width=0.28\textwidth,height=0.28\textheight,keepaspectratio]{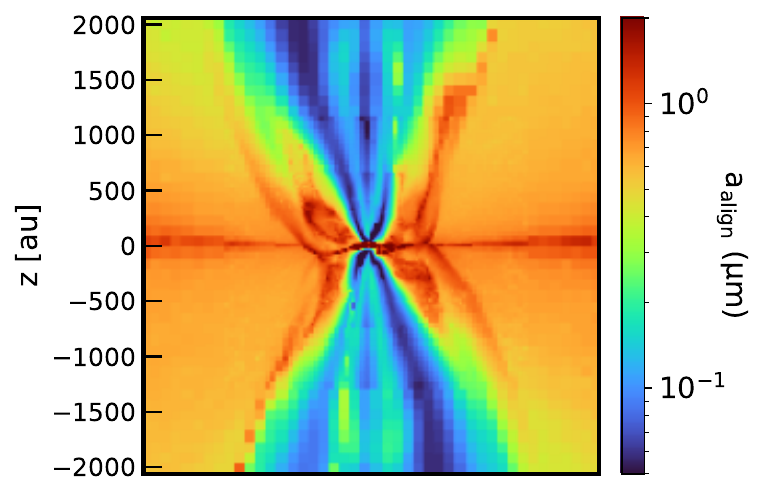}
\centering
 
    \includegraphics[width=\textwidth,height=\textheight,keepaspectratio]{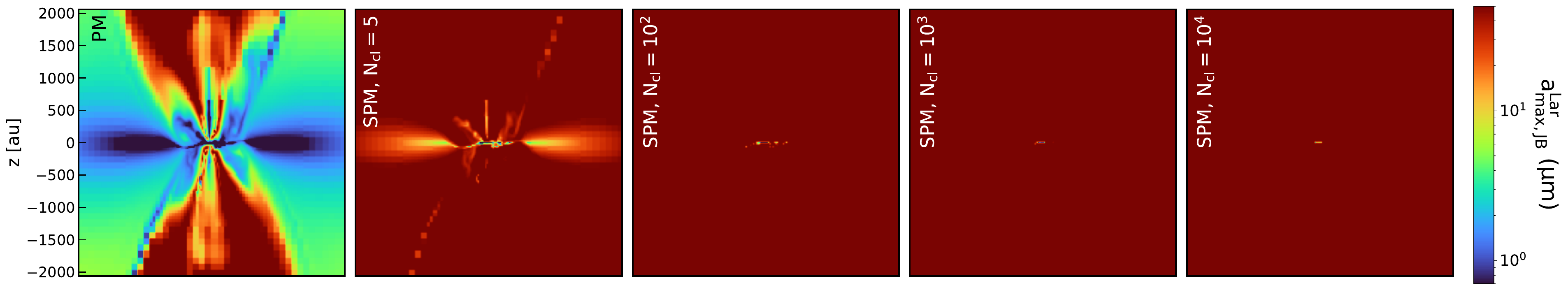}
    \includegraphics[width=\textwidth,height=\textheight,keepaspectratio]{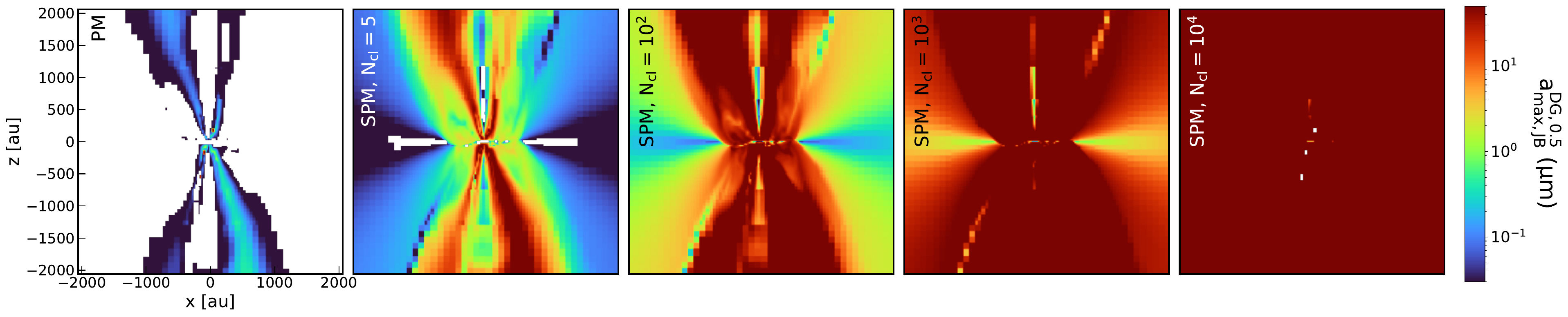}
    \caption{Upper and middle row: Spatial distribution of the minimum alignment size $a_{\rm align}$ and maximum alignment size $a_{\rm max,JB}^{\rm Lar}$ inside the midplane containing the sink particle. We observe the core along the edge$-$on direction and show the map within 4000 au around the protostar. Lower row: distribution of the maximum size being aligned with $\B$ by MRAT mechanism with $f_{\rm high-J} = 0.5$. The first column shows results for PM grains, and the second to fifth columns correspond to SPM grains with low $N_{\rm cl} = 5$, moderate $N_{\rm cl} = 10^{2}$, and high $N_{\rm cl} = 10^{3}$ and $10^{4}$, respectively. Empty cells in the lower row correspond to the region without MRAT effect, i.e., all dust grains are aligned with $\B$ by RATs only. Iron inclusions can help more large micron-sized grains to be aligned with $\B$ by MRAT mechanism owing to the enhanced Larmor precession and magnetic relaxation. For example, almost SPM grains with $N_{\rm cl} \geq 10^{3}$ above $a_{\rm align}$ (upper row) can achieve $f_{\rm high-J} = 0.5$ in the entire protostellar core thanks to MRAT alignment. }
     \label{fig:distribution_alignment}
\end{figure*}
 
\subsection{External Alignment: Larmor Precession and MRAT Alignment}
We show in Figure \ref{fig:distribution_alignment} the distribution of the alignment range and the external alignment mechanism within 2000 au around the protostar, assuming $a_{\rm max} = 50\mum$. The upper row shows the minimum alignment size $a_{\rm align}$ in the unit of $\mum$, the middle row shows the maximum alignment size $a_{\rm max,JB}^{\rm Lar}$. And the lower row shows the maximum size for grains having $f_{\rm high-J} = 0.5$ by MRAT mechanism $a_{\rm max,JB}^{\rm DG,0.5}$. In the middle and lower rows, the first column shows results obtained for PM grains, and the second to fifth columns present results for SPM grains with $N_{\rm cl} = 5, 10^{2}, 10^{3}$, and $10^{4}$, respectively. The colorbar scale in each row is fixed to better visualize the effect of iron inclusions on grain alignment. Empty cells in the lower row show the location where all dust grains are aligned with $\B$ by RATs only.

Inside the outflow cavity, sub-micron grains of $a_{\rm align} \sim 0.06 - 0.1\mum$ can achieve suprathermal rotation by RATs owing to the strong radiation field with $L_{\rm center} = 100L_{\odot}$ (Figure \ref{fig:distribution_mcrt}, \citealt{Valentin_2023b}). However, for PM grains (middle row, first panel), the maximum alignment size is limited to $a_{\rm max,JB}^{\rm Lar} \sim 10\mum$ inside the jet lobe with height 2000 au. Beyond this lobe, $a_{\rm max,JB}^{\rm Lar}$ increases to $50\mum$ as reducing gas density (Figure \ref{fig:distribution_mcrt}). The alignment range is narrower to $\sim 0.6\mum - 8\mum$ in the envelope, and it is more extreme with $a_{\rm align} \sim 1\mum$ and $a_{\rm max,JB}^{\rm Lar} \sim 2\mum$ inside the outflow cavity wall owing to the strong shield of radiation fields and strong gas randomization there. PM grains are almost not aligned with $\B$ on the equatorial midplane, i.e., $a_{\rm align} \sim a_{\rm max,JB}^{\rm Lar} \sim 1\mum$, due to the combined effects of strong gas randomization and the inefficiency of RATs in regions where radiation fields are orthogonal to magnetic fields (\citealt{Lazarian_Hoang_2007a}, \citealt{Hoang_Lazarian_2008}). However, if grains are SPM with low $N_{\rm cl} = 5$ (second panel), all grains from $a_{\rm align}$ to $a_{\rm max} \sim 50\mu m$ within 2000 au (except grains on the equatorial midplane) can be aligned with $\B$ owing to the enhanced Larmor precession by iron inclusions. VLGs on the equatorial midplane can have the magnetic alignment if they contain higher number of $N_{\rm cl} = 10^{2}-10^{4}$ (third to fifth panels). 

\begin{figure*}
\centering
    \includegraphics[width=\textwidth,height=\textheight,keepaspectratio]{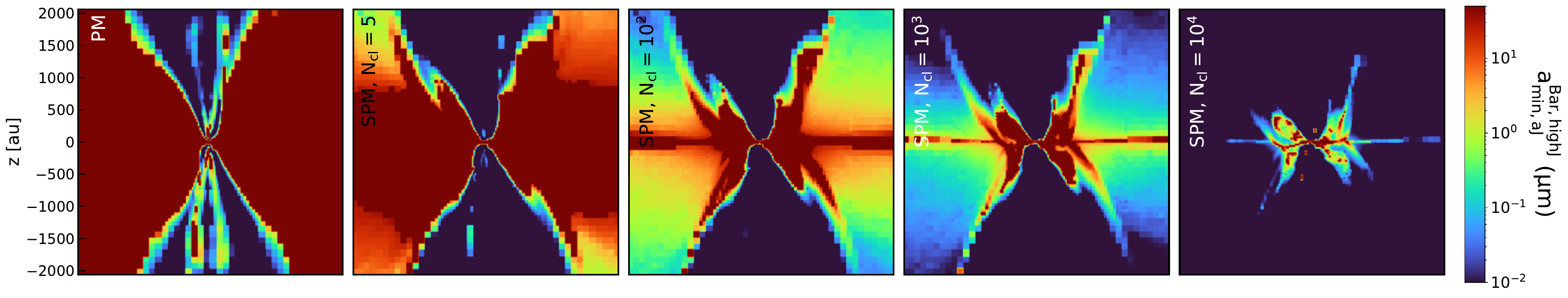}
    \includegraphics[width=\textwidth,height=\textheight,keepaspectratio]{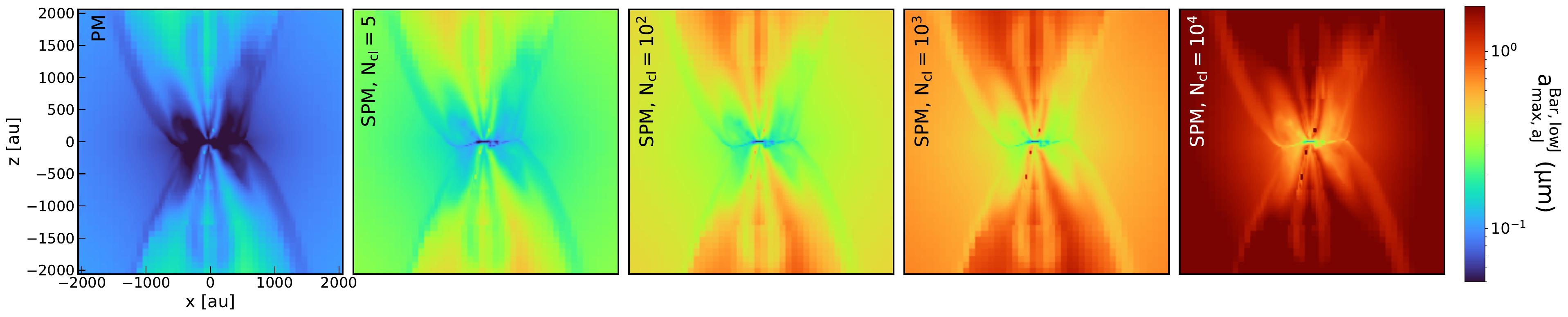}
    \caption{first row: distribution of the minimum size for fast Barnett relaxation at high-\textit{J} attractors $a_{\rm min,aJ}^{\rm Bar, highJ}$. Lower row: distribution of the maximum size for fast Barnett relaxation at low-\textit{J} attractors $a_{\rm max,aJ}^{\rm Bar, low-J}$. The first column shows results for PM grains, and the second to fifth rows correspond to SPM grains with $N_{\rm cl} = 5, N_{\rm cl} = 10^{2}, 10^{3}$, high $N_{\rm cl} = 10^{3}$, respectively. Similar to Figure \ref{fig:distribution_alignment}, iron inclusions help more large grains to have fast internal relaxation at both high and low-\textit{J} attractors owing to the enhanced Barnett relaxation efficiency. However, Barnett relaxation cannot help micron-sized and VLGs with $N_{\rm cl} = 10^{4}$ inside the outflow cavity wall and on the equatorial midplane to have fast internal relaxation at high-\textit{J} attractors. Similarly, grains above $1\mum$ at low-\textit{J} always have slow internal relaxation regardless of values of $N_{\rm cl}$.}
     \label{fig:Barnett}
 
\centering
    \includegraphics[width=\textwidth,height=\textheight,keepaspectratio]{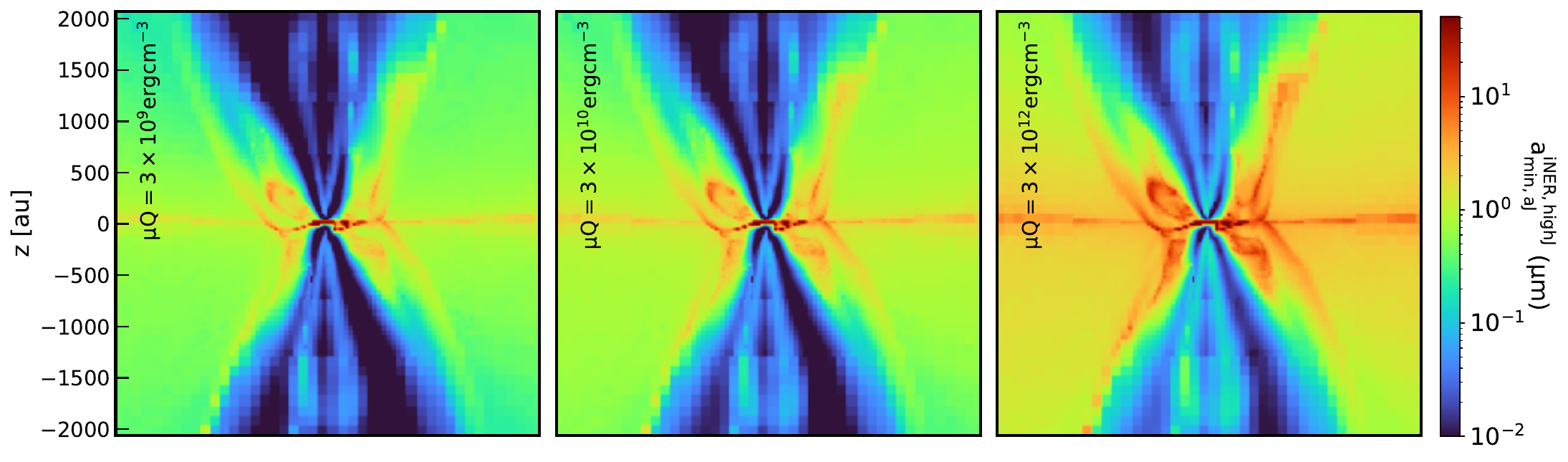}
    \includegraphics[width=\textwidth,height=\textheight,keepaspectratio]{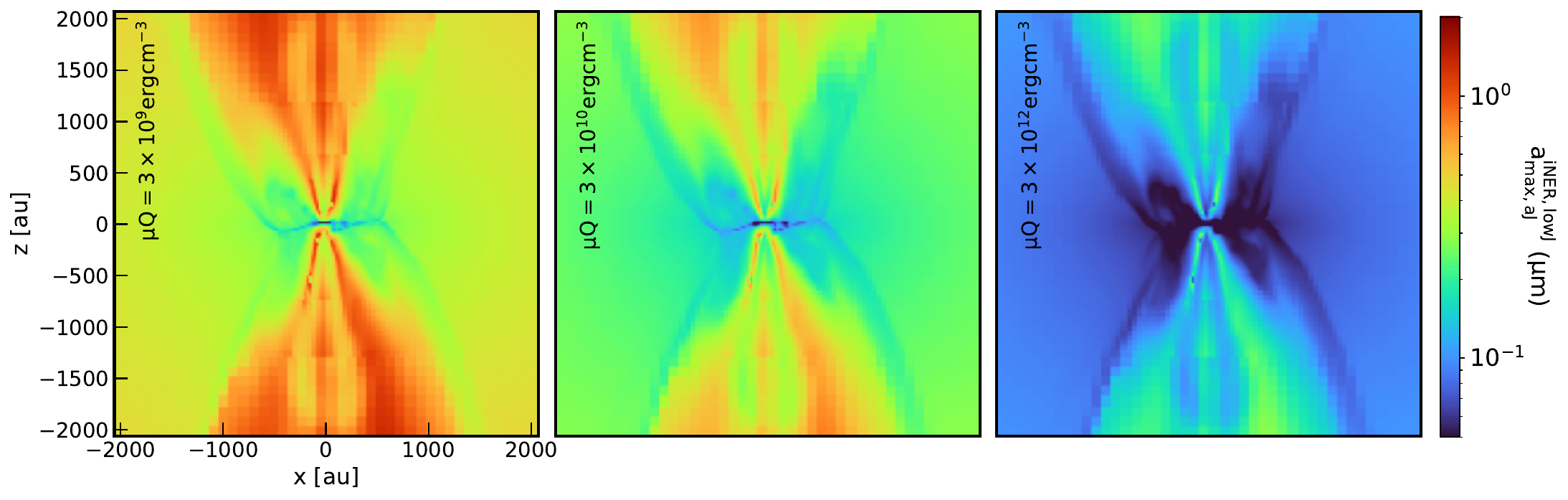}
    \caption{upper row: spatial distribution of the minimum size for efficient inelastic relaxation at high-\textit{J} attractors, $a_{\rm min,aJ}^{\rm iNER, highJ}$, considering high to low inelasticity of grains $\mu Q = 3\times 10^{9}, 3\times 10^{10}$ and $3\times 10^{12}\erg\cm^{-3}$, from left to right, respectively. Lower row: similar results to the upper row but for the maximum size with efficient inelastic relaxation at low-\textit{J} attractors, $a_{\rm max,aJ}^{\rm iNER,lowJ}$. The range for grains having fast inelastic relaxation decreases with decreasing $\mu Q$ and their distances to the protostar. Comparing to results in Figure \ref{fig:Barnett}, one can clearly see the importance of inelastic relaxation in driving the fast internal relaxation for suprathermal micron-sized and VLGs inside the outflow cavity wall and on the equatorial midplane. But similar to Figure \ref{fig:Barnett}, inelastic relaxation cannot help grains above $\sim 1\mum$ at low-\textit{J} to have fast internal relaxation due to their slow thermal rotation.} 
     \label{fig:Inelastic}
\end{figure*}

From the lower row of Figure \ref{fig:distribution_alignment}, one can see that RAT is the primary mechanism driving the magnetic alignment for PM grains (first panel), i.e., $a_{\rm max,JB}^{\rm DG,0.5} << a_{\rm align}$. In contrast, for SPM grains with $N_{\rm cl} = 5$ (second panel), micron-sized grains up to $\sim 10\mum$ inside the outflow cavity can experience MRAT mechanism owing to the enhanced magnetic relaxation mechanism by iron inclusions. The region where VLGs can be aligned with $\B$ by MRAT is extended outward with increasing iron inclusions locked inside dust grains. Particularly, for SPM grains with high $N_{\rm cl} = 10^{4}$ (fifth panel), $100\%$ of aligned dust grains will be aligned with $\B$ by MRAT mechanism, with $f_{\rm high-J} = 0.5$ regardless of their position to the protostar.

\subsection{Internal Alignment: Barnett Relaxation and Inelastic Relaxation}
We show in Figure \ref{fig:Barnett} the internal alignment state of dust grains driven by Barnett relaxation. The upper row presents the minimum size with fast Barnett relaxation at high-\textit{J} attractors, $a_{\rm min,aJ}^{\rm Bar, highJ}$, and the lower row presents the maximum size with fast Barnett relaxation at low-\textit{J} attractors, $a_{\rm max,aJ}^{\rm Bar, lowJ}$. Results for PM and SPM grains with $N_{\rm cl} = 5, 10^{2}, 10^{3}$, and $10^{4}$ are shown from left to right, respectively. One can see that for grains aligning with $\B$ at high-\textit{J} attractors (upper row), only aligned PM grains inside the outflow cavity can have fast Barnett relaxation owing to their higher rotational rate by efficient RATs, i.e., $a_{\rm min,aJ}^{\rm highJ} \sim 0.01\mu m$. By increasing the number of iron atoms per cluster ($N_{cl}$), more large grains beyond the outflow cavity can have fast internal relaxation due to enhanced Barnett relaxation strength. For example, all aligned dust grains inside the envelope and outflow cavity can have efficient internal alignment by Barnett relaxation if they are SPM with $N_{\rm cl} = 10^{4}$ (fifth panel). However, even for such highly magnetized ones, micron-sized grains beyond $\sim 1\mum$ inside the outflow cavity wall still have slow internal relaxation because of the strong gas randomization effect in this area. 

For grains aligned with $\B$ at low-\textit{J} attractors (lower row), the maximum size of fast Barnett relaxation increases with increasing the grain magnetic susceptibility. However, iron inclusions cannot help micron-sized grains and VLGs to have fast internal relaxation when they rotate thermally regardless of their positions to the protostar (\citealt{Hoang_2022}, \citealt{Hoang+2022}, \citealt{Giang_2023a}).
 
Then, we show in Figure \ref{fig:Inelastic} the critical size for grains having fast internal relaxation by inelastic relaxation. The upper row illustrates the minimum size for fast inelastic relaxation at high-\textit{J} attractors, $a_{\rm min,aJ}^{\rm iNER, highJ}$, and the lower row illustrates the maximum size for fast inelastic relaxation at low-\textit{J} attractors, $a_{\rm max,aJ}^{\rm iNER, lowJ}$. The first and second columns show results for inelastic material with low $\mu Q = 3\times 10^{9}$ and $3\times 10^{10}\erg\cm^{-3}$, and the third column shows results for elastic material with high $\mu Q = 3\times 10^{12}\erg\cm^{-3}$. Generally, more grains can have fast inelastic relaxation at both high and low-\textit{J} attractors with increasing the inelasticity of dust grains, i.e., decreasing $\mu Q$. The range for fast inelastic relaxation is narrower toward the protostar as increasing gaseous damping. In particular, grains beyond $a_{\rm min,aJ}^{\rm iNER, highJ} \sim 0.5\mum$ in the envelope or beyond $a_{\rm min,aJ}^{\rm iNER, highJ} \sim 2\mum$ inside the outflow cavity wall can have fast inelastic relaxation at high-\textit{J} if they have $\mu Q \sim 3\times 10^{9} \erg\cm^{-3}$. These values slightly increase to $a_{\rm min,aJ}^{\rm iNER,highJ} \sim 2\mum$ and $\sim 10\mum$ for higher $\mu Q \sim 3\times 10^{12}\erg\cm^{-3}$, respectively.

To understand in detail the joint effect of Barnett and inelastic relaxation in driving the grain internal alignment in the protostellar core, we compare results obtained from inelastic relaxation (Figure \ref{fig:Inelastic}) with Barnett relaxation (Figure \ref{fig:Barnett}). For grains at high-\textit{J} attractors (upper row), one can see that for PM and SPM grains with $N_{\rm cl}\leq 10^{2}$ (first and third panels), inelastic relaxation is the major mechanism driving the fast internal relaxation of micron-sized grains above $1\mum$ inside the outflow cavity wall, inner envelope, envelope, and on the equatorial midplane owing to the weak Barnett relaxation efficiency in these high density regions. By increasing $N_{\rm cl}$ to $\sim 10^{3}-10^{4}$ (third and fifth panels), the enhanced Barnett relaxation efficiency by iron inclusions leads them to become the major internal alignment mechanism for aligned dust grains inside the envelope. However, inside the outflow cavity wall, inner envelope, and on the equatorial plane, inelastic relaxation still be the major mechanism driving the efficient IA for micron-sized grains above $\geq 1-10 \mum$. The effective of inelastic relaxation on large grains in these dense areas is caused by the weak dependence of their timescale $\tau_{\rm iNER}$ on grain sizes but strong dependence on the grain rotational energy, i.e., $\tau_{\rm iNER} \sim a^{11/2} S_{\rm t}^{-3}$ with $S_{\rm t} = \Omega/\Omega_{\rm ther}$ (\citealt{Hoang+2022}). In contrast, the Barnett relaxation timescale $\tau_{\rm Bar}$ increases rapidly with grain sizes but decreases slower with $S_{\rm t}$ ($\tau_{\rm Bar} \sim a^{7} S_{\rm t}^{-2}$), resulting in their weaker effect on micron-sized grains and VLGs inside the dense outflow cavity wall, inner envelope, and the equatorial midplane.

\begin{figure*}
\centering
 \hspace*{-12cm}
    \includegraphics[width=0.25\textwidth,height=0.25\textheight,keepaspectratio]{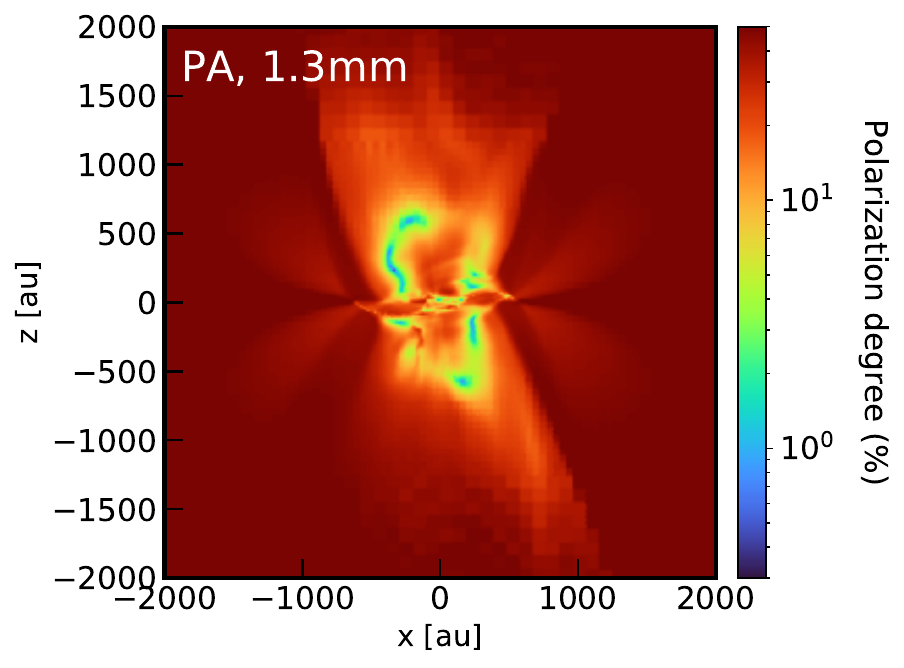}
\centering

    \includegraphics[width=\textwidth,height=\textheight,keepaspectratio]{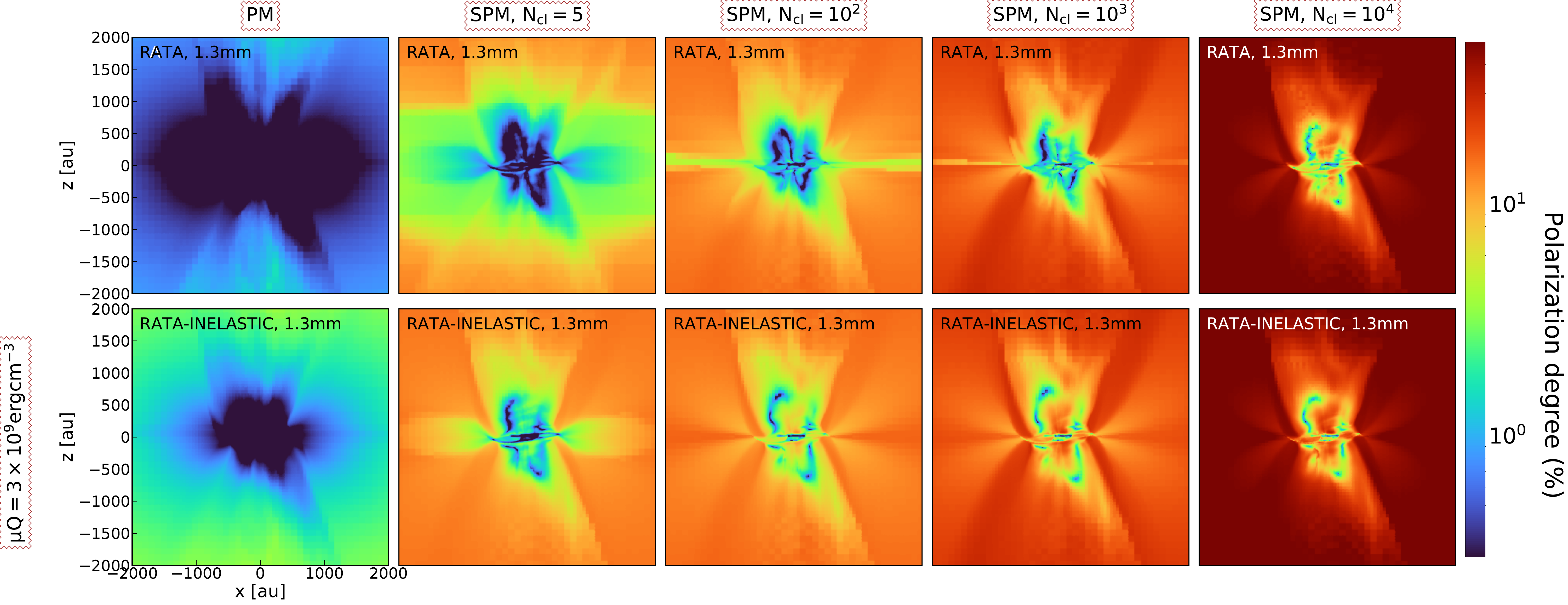}
    \caption{Upper row: Predicted polarization degree map obtained at 1.3mm within 2000 au around the protostar, assuming all dust grains above $a_{\rm align}$ are perfectly aligned with $\B$ (model PA). Middle row: effect of iron inclusions on the polarization degree map obtained from model RATA. Lower row: effect of inelastic relaxation on $p(\%)$ map obtained from model RATA$-$INELASTIC, assuming highly inelastic material with $\mu Q = 3\times 10^{9}\erg\cm^{-3}$. Results for PM and SPM grains with $N_{\rm cl} = 5, 10^{2}, 10^{3}$, and $10^{4}$ are shown from left to right, respectively. We fix the colorbar in all panels to visualize the important effect of iron inclusions and inelastic relaxation on strengthening the grain alignment, which enhances the observed degree of polarized dust emission.}
     \label{fig:p_map}
\end{figure*}

For grains at low-\textit{J} attractors (lower row), inelastic relaxation is the major mechanism driving the fast internal relaxation for sub-micron PM and SPM grains with low $N_{\rm cl} \leq 5$ and low $\mu Q \leq 3\times 10^{10}\erg\cm^{-3}$. In contrast, Barnett relaxation is the major internal mechanism for SPM grains with higher $N_{\rm cl}\geq 10^{2}$ in the entire $2000$ au around the protostar. However, thermal grains larger than $\geq 2\mum$ always have slow internal relaxation within $\sim 1000$ au of core scale regardless of the joint effect of Barnett and inelastic relaxation. It is because the thermal rotation can lengthen both $\tau_{\rm Bar} >> \tau_{\rm gas}$ and $\tau_{\rm iNER}>>\tau_{\rm gas}$.
\section{ Effect of Iron Inclusions and Inelastic relaxation on dust polarization }\label{sec:Iron_Inelastic_properties}
To understand the effect of iron inclusions and inelastic relaxation on polarized dust emission, we compare synthetic results obtained from the model perfect alignment (PA, in which the perfect alignment is applicable for all dust grains beyond $a_{\rm align}$ shown in the upper panel of Figure \ref{fig:distribution_alignment}) to two realistic models of grain alignment, RATA (RAdiative Torque Alignment) only and RATA$-$INELASTIC. We only consider the effect of Barnett relaxation in model RATA, and consider both Barnett and inelastic relaxation in model RATA$-$INELASTIC. The summary of three considered models is shown in Table \ref{tab:model}.

We show in the upper row of Figure \ref{fig:p_map} the polarization degree map obtained within 2000 au around the protostar from model PA, assuming $a_{\rm max} = 50\mum$. The middle row shows the effect of iron inclusions on $p(\%)$ at 1.3mm (model RATA), with the first panel corresponding to the results of PM grains, and the second to fifth panels corresponding to SPM grains with $N_{\rm cl} = 5, 10^{2}, 10^{3}$, and $10^{4}$, respectively.
 
In model PA, the maximum polarization degree beyond $\sim 1000$ au is very high, i.e.,  $p \sim 40\%$, because all grains from $a_{\rm align} \sim 0.5\mum$ to $a_{\rm max} = 50\mum$ are considered to have perfect alignment with magnetic fields (Figure \ref{fig:distribution_alignment}, first panel). The polarization degree slightly decreases toward the center, reaching $p \sim 10\%$ around the protostar. When considering the realistic alignment model of PM grains, the resulting polarization degree is significantly low of $p \sim 1\%$ in the envelope due to the misalignment of VLGs and the weak magnetic alignment of micron-sized grains there (Figure \ref{fig:distribution_alignment}, first column). By increasing the grain magnetic susceptibility, the mean value of $p(\%)$ increases systematically as a result of improving the grain alignment degree with the help of iron inclusions (Figures \ref{fig:distribution_alignment} and \ref{fig:Barnett}). One can see that both models of PM and SPM grains show the decreasing polarization fraction toward the center, but it seems to be more extreme than the results obtained from model PA. For example, SPM grains with $N_{\rm cl} = 10^{4}$ (fifth panel) can produce very high polarization degree of $p \sim 40\%$ in the envelope (similar to model PA) because of their perfect alignment with $\B$ beyond 500 au (Figures \ref{fig:distribution_alignment} and \ref{fig:Barnett}). However, their polarization degree decreases significantly toward the center, reaching only $p \leq 1\%$ around the protostar.

The effect of inelastic relaxation on the polarization degree map is shown in the lower row of Figure \ref{fig:p_map} (model RATA$-$INELASTIC), considering inelastic grains with $\mu Q =  3\times 10^{9} \erg\cm^{-3}$. The order of the grain magnetic susceptibility is the same as the middle row. By considering inelastic relaxation, the polarization degree obtained from all models of PM and SPM grains within $\sim 500$ au clearly increases as improving the IA of micron-sized and VLGs inside the outflow cavity wall, the inner envelope, and the equatorial midplane (Figure \ref{fig:Inelastic}). $p(\%)$ produced by SPM grains with low $N_{\rm cl} = 5$ (second column) inside the inner envelope increases from $p \sim 0.5\%$ in model RATA to $p \sim 2\%$ in model RATA$-$INELASTIC. Similarly, $p(\%)$ produced by SPM grains with higher $N_{\rm cl} = 10^{3}$ (fourth column) increases from $p \sim 3\%$ (model RATA) to $p \sim 8\%$ (model RATA$-$INELASTIC). However, inelastic relaxation cannot help to increase the low $1\%$ of dust polarization inside the disk scale found in model RATA because micron-sized and VLGs still have slow internal relaxation by high gaseous damping there (see Appendix \ref{sec:appen_disk}).

\begin{figure*}
\centering
    \includegraphics[width=\textwidth,height=\textheight,keepaspectratio]{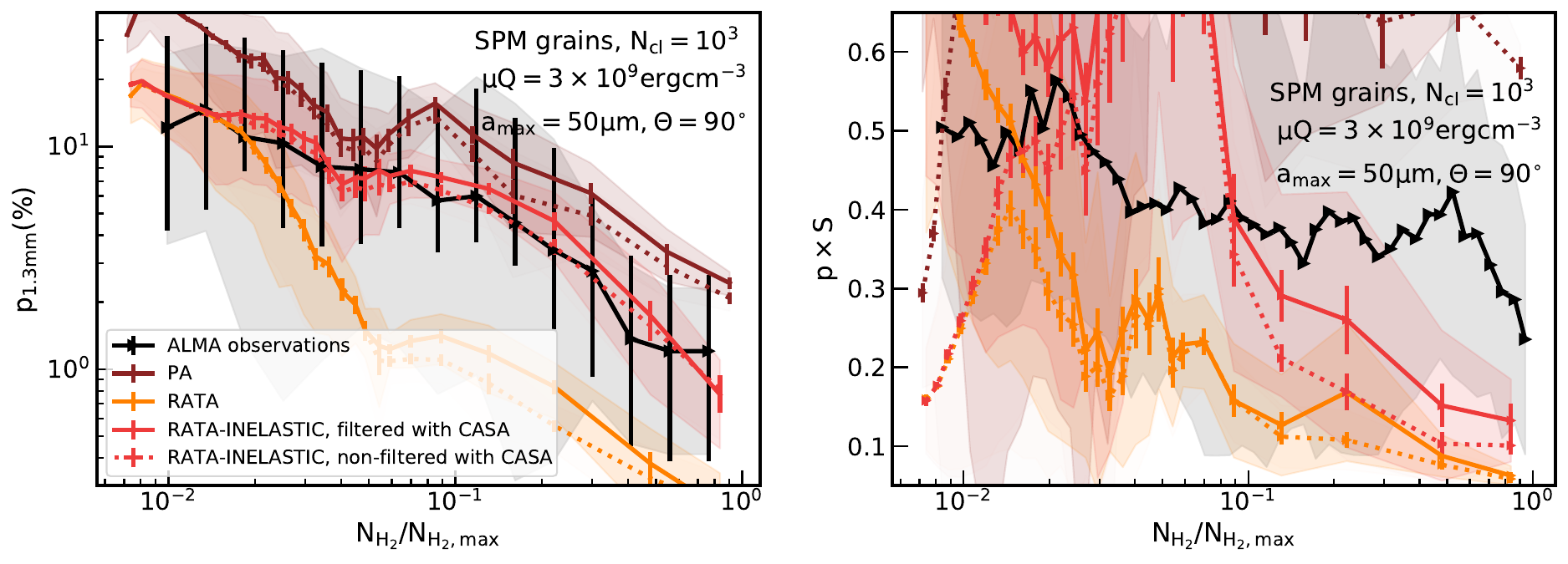}
    \caption{Left panel: comparison of the mean variation of $p(1.3 mm)$ with normalized gas column density $N_{\rm H_{2}}/N_{\rm H_{2},max}$ between ALMA observations of 13 YSOs from \cite{Valentin_2020} (black line) and synthetic modeling of polarized dust emission (colors lines). Synthetic results obtained directly from POLARIS are shown in the dashed lines, and results after filtering with CASA in the solid lines. The brown, orange, and red lines show model PA, RATA, and RATA$-$INELASTIC results, respectively. We consider the model of SPM grains with $N_{\rm cl} = 10^{3}$, $\mu Q = 3\times 10^{9}\erg\cm^{-3}$, $a_{\rm max} = 50\mum$, and $\Theta = 90^{\circ}$. Right panel: comparison of $p\times S$ and $N_{\rm H_{2}}/N_{\rm H_{2},max}$ obtained from synthetic modelling results with ALMA observations. The color code and dust parameters are the same as those on the left panel. The high polarization degree of $p \sim 10\%$ in the envelope and $p \sim 1-8\%$ observed in the inner envelope and Class 0 disk can be reproduced well by the efficient alignment of VLGs owing to the joint action of super-Barnett relaxation, inelastic relaxation, and MRAT mechanism (model RATA$-$INELASTIC). Our model can also reproduce the decrease level of $p(\%)$ from the envelope to the disk as revealed by ALMA. Different from the depolarization caused by the unresolved the gravo-magnetic field structure by low single disk resolution obtained from clouds to cores scale, this feature appears in our model because of the strong reduction of the IA and MRAT efficiency acting on VLGs with increasing gaseous damping toward the inner envelope and disk scale. }
    \label{fig:CASA_best_fit}
\end{figure*}

\section{Constraining Grain Alignment mechanism and Dust Properties with ALMA observations}\label{sec:ALMA_observations}

To constrain grain alignment mechanisms and dust properties with ALMA observations, we post-process synthetic polarization results obtained from model PA, RATA, and RATA$-$INELASTIC with CASA as described in Section \ref{sec:modelling_stokes} and compare them with ALMA observational data observed at 1.3mm in Figure \ref{fig:CASA_best_fit}. We will show the polarization degree map before and after filtering with CASA in Appendix \ref{sec:appen_filter} and the filtering effect on dust polarization in Section \ref{sec:discuss_filter}. Briefly, ALMA filter amplifies the original polarization degree obtained in the outermost, low-intensity region by $\sim 5-10\%$ (\citealt{Valentin_2020}, \citealt{Valentin_2023b}). However, it does not significantly change the mean value of dust polarization degree that can be detected.

To constrain grain alignment mechanisms, we show the comparison of the $p(\%)-N_{\rm H_{2}}/N_{\rm H_{2},max}$ relation obtained from the synthetic modeling with ALMA observation in the left panel of Figure \ref{fig:CASA_best_fit}. The black line shows the mean variation of $p(\%)$ from 13 Class 0/I YSOs observed by ALMA at 1.3mm (Figure B.4 in \citealt{Valentin_2020}). The errorbar of the black line shows the standard deviation of $p(\%)-N_{\rm H_{2}}/N_{\rm H_{2},max}$ relation from 13 objects to the mean trend of $p(\%)$, and the shaded area shows the standard deviation of polarization degree summarised from 13 objects. The dashed and solid color lines show the mean correlation obtained from the synthetic modeling before and after filtering with CASA, with the shaded areas presenting the standard deviation from the mean. We illustrate model PA in brown and models RATA and RATA$-$INELASTIC in orange and red, respectively. For two later models, we consider the scenario of SPM grains with $N_{\rm cl} = 10^{3}$, $\mu Q = 3\times 10^{9} \erg\cm^{-3}$,  $a_{\rm max} = 50\mu m$, and the observed angle of $\Theta = 90^{\circ}$ to the z$-$direction.

  \begin{figure*}
\centering
    \includegraphics[width=\textwidth,height=\textheight,keepaspectratio]{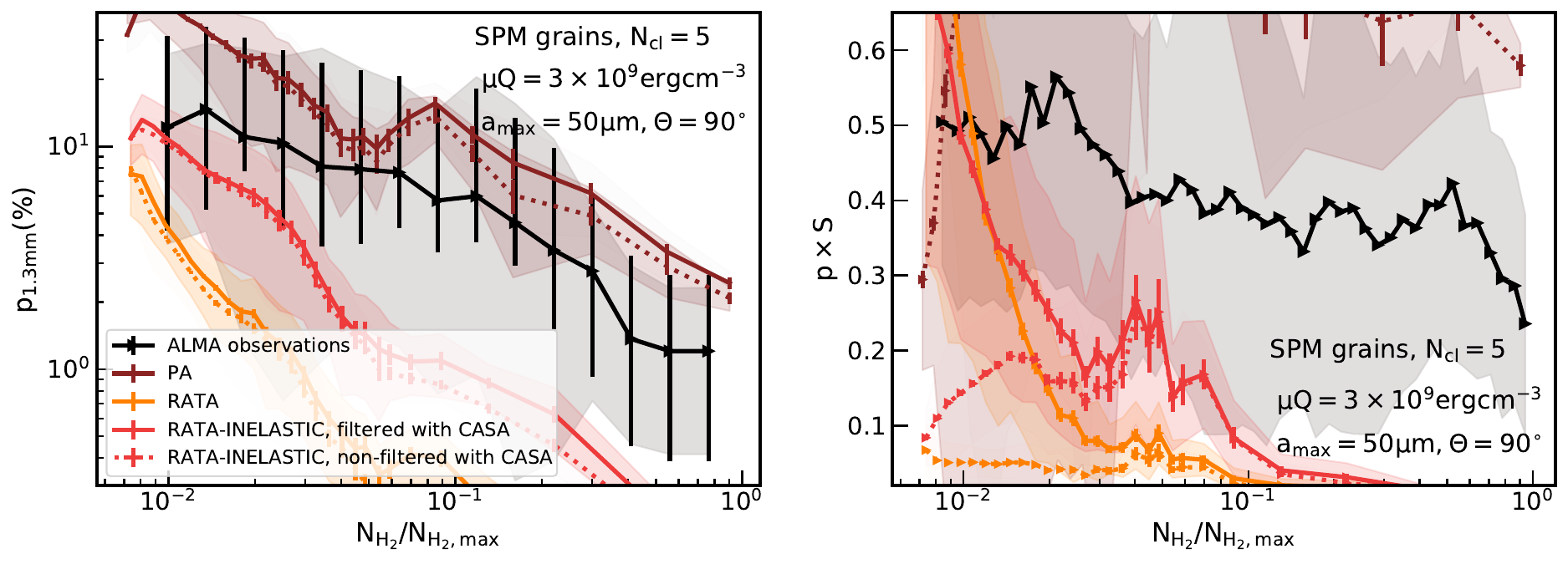}
 
    \caption{Similar results to Figure \ref{fig:CASA_best_fit} but for SPM grains with low $N_{\rm cl} = 5$. Inside the envelope, inelastic relaxation can help to maintain the efficient IA of VLGs and produce $p \sim 8-10\%$ observed by ALMA. However, moving toward the inner envelope and the disk, without the help of iron inclusions, the misalignment of VLGs and the alignment of micron-sized by only RATs lead the polarization signal to be undetectable by ALMA, i.e., $p << 1\%$.}
     \label{fig:CASA_Ncl}
 
\centering
    \includegraphics[width=\textwidth,height=\textheight,keepaspectratio]{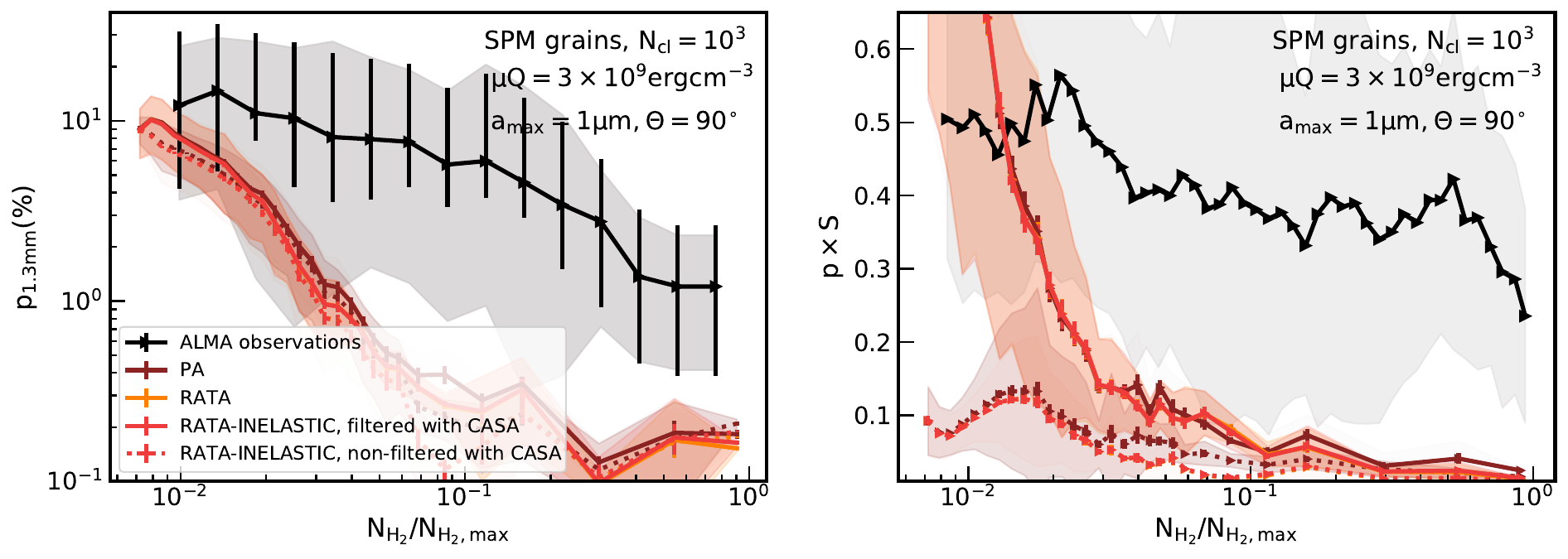}
    \caption{Similar results as Figure \ref{fig:CASA_best_fit} but for the maximum grain size $a_{\rm max} = 1\mum$. Without the presence of VLGs, micron-sized grains still can produce $p\sim 8\%$ in the envelope because of the perfect alignment of all aligned dust grains. But moving toward the inner envelope and the disk scale, both model RATA and RATA$-$INELASTIC with $a_{\rm max} = 1\mum$ fails to reproduce ALMA observations because of the significant reduction of the alignment range there, i.e., with $a_{\rm align} \sim 1\mum$ near the protostar. This result thus implies the importance of grain growth activities within a few hundred au around the protostar.}
    \label{fig:CASA_amax}
\end{figure*}

One can see that in the protostellar envelope (with $N_{\rm H_{2}}/N_{\rm H_{2},max} \sim 0.01$), both model RATA and RATA$-$INELASTIC of SPM grains with $N_{\rm cl} = 10^{3}$, $\mu Q = 3\times 10^{9}\erg\cm^{-3}$, and $a_{\rm max} = 50\mum$ can reproduce rather well the mean polarization degree of $p \sim 10 - 20\%$ observed by ALMA in this region. With this configuration, all sub-micron and micron-sized grains up to $10\mum$ can achieve efficient IA by fast Barnett relaxation (Figure \ref{fig:Barnett}) and efficient magnetic alignment by MRAT mechanism with $f_{\rm high-J} \geq 0.5$ (Figure \ref{fig:distribution_alignment}, lower row). Although VLGs within $\sim 10-50\mum$ only can be aligned with $\B$ by RATs with $f_{\rm high-J} = 0.25$, the net alignment degree of all aligned dust grains in the envelope is still very efficient to reproduce above $\geq 10\%$ of dust polarization there. However, moving toward the inner envelope and the disk scale ($N_{\rm H_{2}}/N_{\rm H_{2},max} > 3\times 10^{-2}$), model RATA (orange lines) totally fails to reproduce the observed mean polarization fraction of $p \sim 1-8\%$, even MRAT alignment is the major external alignment of all dust grains within the inner 500 au (Figure \ref{fig:distribution_alignment}, lower row). In contrast, model RATA$-$INELASTIC (red lines) can reproduce well the range of ALMA polarization fraction, given that micron-sized and VLGs inside the outflow cavity wall, inner envelope, and the equatorial plane can achieve efficient IA by fast inelastic relaxation (Figure \ref{fig:Inelastic}, upper left panel). The overlap between synthetic results from model RATA$-$INELASTIC and ALMA observations implies the important role of inelastic relaxation in maintaining the efficient internal alignment of micron-sized grains above $\geq 1\mum$ within hundreds au around the protostar. 
 
To constrain the magnetic properties of dust, we reduce the number of iron atoms per cluster from $N_{\rm cl} = 10^{3}$ to $N_{\rm cl} = 5$ and show the results in Figure \ref{fig:CASA_Ncl}, considering grains with $\mu Q = 3\times 10^{9}\erg\cm^{-3}$, $a_{\rm max} = 50\mum$, and $\Theta = 90^{\circ}$. Due to the reduction of magnetic susceptibility by decreasing $N_{\rm cl}$, inelastic relaxation becomes the major mechanism driving the efficient IA for almost aligned SPM grains with $N_{\rm cl} = 5$ (the orange line, see also Figures \ref{fig:Barnett} and \ref{fig:Inelastic}). Besides, SPM grains with low $N_{\rm cl} =5$ majorly experience RAT alignment with typical $f_{\rm high-J} = 0.25$ (Figure \ref{fig:distribution_alignment}, lower row). In the envelope where the tangling of $\B$ fields is still small, the weak external alignment by RATs but efficient IA of SPM grains with $N_{\rm cl} = 5$ still allow us to observe $\sim 5\%$ of polarization as detected by ALMA. However, moving toward the inner region where magnetic fields are strongly distorted by stellar feedback and the disk formation, the poor magnetic alignment of grains by RATs causes extra damping on the observed polarization signal, inducing the low $p < 1\%$ below the observed range by ALMA. Compared to results in Figure \ref{fig:CASA_best_fit}, MRAT alignment is more preferred than RATs in driving the magnetic grain alignment for the inner 500 au region. Thus, VLGs in this region need to have higher iron inclusions to activate the MRAT alignment there.
 
The last factor that affects the observed polarization fraction is the alignment range, which is partly controlled by the maximum grain size. To understand the importance of grain growth on polarized dust emission, we reduce the maximum grain size from $a_{\rm max} = 50\mum$ to $a_{\rm max} = 1\mum$ - the typical grain size found in the MCs, and show the comparison of our new results with ALMA observations in Figure \ref{fig:CASA_amax}. We consider similar setups as Figure \ref{fig:CASA_best_fit}, with SPM grains with $N_{\rm cl} = 10^{3}$, $\mu  Q = 3\times 10^{9}\erg\cm^{-3}$, and $\Theta = 90^{\circ}$. In the envelope scale, the model with $a_{\rm max} = 1\mum$ still can produce the observed ALMA range of $7 - 10\%$ in the envelope because sub-micron grains below $1\mum$ with $N_{\rm cl} = 10^{3}$ are perfectly aligned with $\B$ by super-Barnett relaxation and MRAT mechanism (Figures \ref{fig:distribution_alignment} and \ref{fig:Barnett}). But in the inner region, both models PA, RATA, and RATA$-$INELASTIC with $a_{\rm max} = 1\mum$ fail to reproduce ALMA polarization degree because the minimum alignment size $a_{\rm align}$ reaches the maximum grain size of $1\mum$ within 500 au around the protostar (Figure \ref{fig:distribution_alignment}). Grain growth must happen inside the inner envelope and disk in order to maintain the considerable range of efficiently aligned dust grains required to reproduce ALMA dust polarization fraction in these regions.

\section{Discussion}\label{sec:discussion}
As shown in Section \ref{sec:ALMA_observations}, the joint action of iron inclusions and inelastic relaxation can drive the efficient magnetic alignment of micron-sized and VLGs required to reproduce the general tendency of ALMA dust polarization observed in low/intermediate Class 0/I YSOs (\citealt{Valdivia_2019}, \citealt{Valentin_2020}, \citealt{Valentin_2023b}). This finding matches with previous predictions related to the physics of grain alignment in protostellar cores and disks found in series studies of \cite{Hoang_2022}, \cite{Hoang+2022}, \cite{Giang_2023a}, and \cite{Giang_2023b}. We note that our model of grain alignment not only works in our adopted snapshot of Class 0 YSO, but also be valid in all evolution stages of protostars because of the short gas damping timescale compared with the core dynamic timescale (with typical $\tau_{\rm gas} \sim 0.083$ years inside the envelope, and $\tau_{\rm gas}$ decreases quickly with increasing density toward the center). As model RATA$-$INELASTIC can reproduce the mean variation of the observed $p(\%)$ with normalized gas column density from ALMA, we continue to use this model to extend our parameter space to understand the configurations of dust grains supporting the ALMA dust polarization fraction. Alongside the discussion about grain physics constrained from ALMA observations, discussion about other potential alignment mechanisms in the YSOs environment is also carried out.

 \begin{figure*}
\centering
    \includegraphics[width=\textwidth,height=\textheight,keepaspectratio]{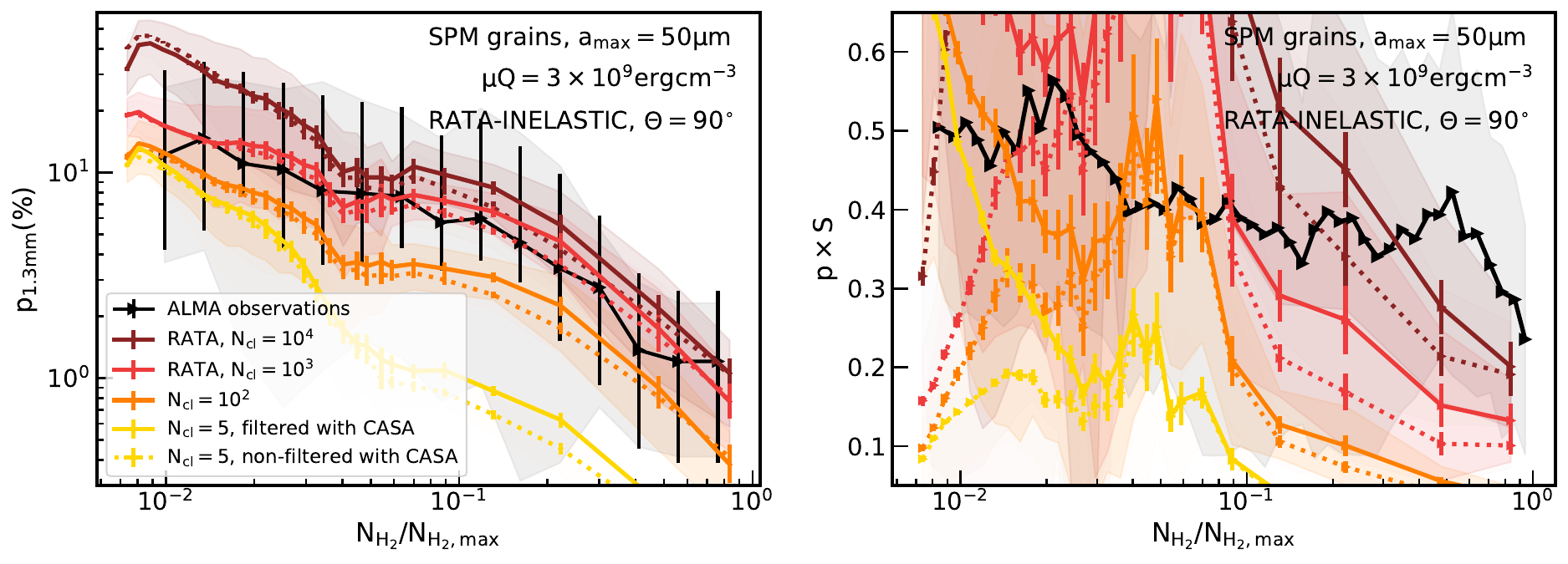}
    \includegraphics[width=\textwidth,height=\textheight,keepaspectratio]{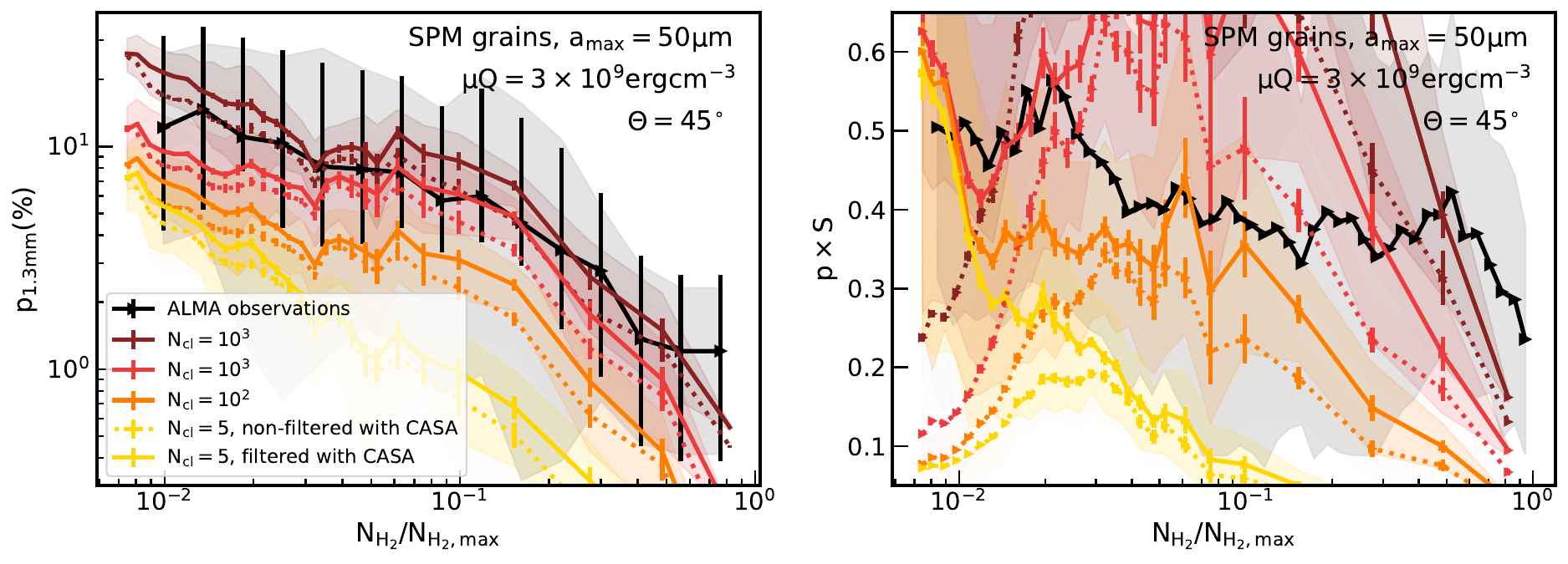}
    \caption{Effect of iron inclusions on $p(\%)$ and $p\times S$ with normalized gas density obtained at 1.3mm. We consider model RATA$-$INELASTIC for SPM grains with $N_{\rm cl}$ varying from $5$ to $10^{4}$, $\mu Q = 3\times 10^{9}\erg\cm^{-3}$, $a_{\rm max} = 50\mum$, and $\Theta = 90^{\circ}$ (first row), $\Theta = 45^{\circ}$ (lower row). The dashed lines show results obtained directly from POLARIS, and the solid lines show results after post-processing with CASA. Micron-sized and VLGs must be SPM, with $N_{\rm cl} \sim 5-10^{4}$ to reproduce the high polarization degree $\sim 10-20\%$ observed in the envelope, and higher $N_{\rm cl}\geq 10^{2}$ to produced the detected level of dust polarization of $\sim 1-8\%$ inside the inner envelope and the disk scale.
   }
    \label{fig:CASA_inelastic_Ncl}
\end{figure*}

\subsection{Envelope scale}\label{sec:discuss_envelope}
In the envelope scale with the typical density of $n_{\rm H_{2}}\sim 10^{6}-10^{7}\cm^{-3}$ and dust temperature of $T_{\rm d} \sim 20$ K, the wide dynamic range of the polarization fraction observed by ALMA from $p\sim 5\%$ to $\sim 30-40\%$ can be explained by different combinations of the maximum grain size (\citealt{Valdivia_2019}, \citealt{Valentin_2023b}, \citealt{Giang_2023b}), the number of iron atoms per cluster locked inside dust grains, $N_{cl}$ (Figures \ref{fig:p_map}, \citealt{Giang_2023b}), and the observed direction to the core (\citealt{Kataoka_2012}). By fixing $a_{\rm max} = 50\mu m$, we show in the left panel of Figure \ref{fig:CASA_inelastic_Ncl} the effect of iron inclusions on the relation of $p- N_{\rm H_{2}}/N_{\rm H_{2},max}$ obtained at 1.3mm, considering the low level of iron inclusions with $N_{\rm cl} = 5$ to high iron inclusions level with $N_{\rm cl} = 10^{4}$, assuming $\mu Q = 3\times 10^{9}\erg\cm^{-3}$. The upper row shows results obtained with $\Theta = 90^{\circ}$ (observing along the edge-on direction), and the lower row corresponds to results obtained with $\Theta = 30^{\circ}$ (observing near the pole-on direction). As illustrated in Figure \ref{fig:p_map}, $p(\%)$ increases with increasing $N_{\rm cl}$, and the wide observed polarization degree range of $p \sim 5-30\%$ can be explained by the presence of SPM grains with $[N_{\rm cl};\phi_{\rm sp}] = [5-10^{4};0.1]$. With this iron configuration, almost aligned dust grains in the envelope can achieve perfect alignment by super-Barnett relaxation/inelastic relaxation and MRAT mechanism (Figures \ref{fig:distribution_alignment}, \ref{fig:Barnett}, and \ref{fig:Inelastic}). The significant improvement of the magnetic alignment of VLGs by iron inclusions and inelastic relaxation can explain well the high grain alignment degree inferred from ALMA (via properties $p\times S$) found in \cite{Valentin_2020} and \cite{Valentin_2023b}, and matches with our predictions shown in \cite{Hoang+2022} and \cite{Giang_2023b}. 

However, there are some degeneracies in constraining the grain magnetic properties in the envelope due to the negative correlation between the observed polarization degree and the observed view (Figure \ref{fig:filtering_Ncl_theta}). For example, there are two configurations of $N_{\rm cl}$ and $\Theta$ that produce $10-30\%$ of polarization in the envelope, first is the combination of SPM grains with $N_{\rm cl} = 10^{2} - 10^{3}$ and $\Theta = 90^{\circ}$ (in which almost $\B$ fields are lie on the plane of sky POS), and second is for SPM grains with $N_{\rm cl} = 10^{4}$ and $\Theta = 30^{\circ}$ (in which almost $\B$ fields are on the line of sight LOS). The coupling between $N_{\rm cl} - \Theta - p(\%)$ can only be separated if the 3D magnetic fields in the protostellar envelope are constrained. \cite{Hull_Valentin_2020} estimated the mean inclination angle of $\B$ fields by comparing the observed Stokes Q and U maps of BHR71 IRS1 (with the clean hourglass-shape field) with the synthetic modeling of strong magnetized cores seen at different inclination angles from \cite{Frau_2011}. Their result is similar to the approach using the gas kinematic of IRS1 outflow by \cite{Yang_2017} and \cite{Tobin_2019}. \cite{Tahani_2019},  and \cite{Tahani_2022_Perseus} reconstructed the 3D magnetic fields inside the Orion' A and Perseus molecular cloud by combining measurement of $\B$ fields along the LOS from Faraday rotation (\citealt{Tahani_2018}) with $\B$ fields on the POS using dust polarization along with Galactic magnetic field models. None of them are tested for probing the 3D magnetic fields in protostellar environments, causing uncertainty when constraining the grain magnetic properties around the protostar. Besides the unclear 3D $\B$ field morphology, the tangling of projected $\B$ fields (on the POS) along the LOS may also cause extra damping on the observed polarization degree in the protostellar envelope (i.e., see \citealt{Chen_2016} for this effect in the clump scale), especially for objects formed inside strong turbulent cores. Yet, how much the disorganization of $\B$ fields contributes to suppressing the observed polarization degree in the protostellar envelope remains unclear.
 
  \begin{figure*}
\centering
    \includegraphics[width=\textwidth,height=\textheight,keepaspectratio]{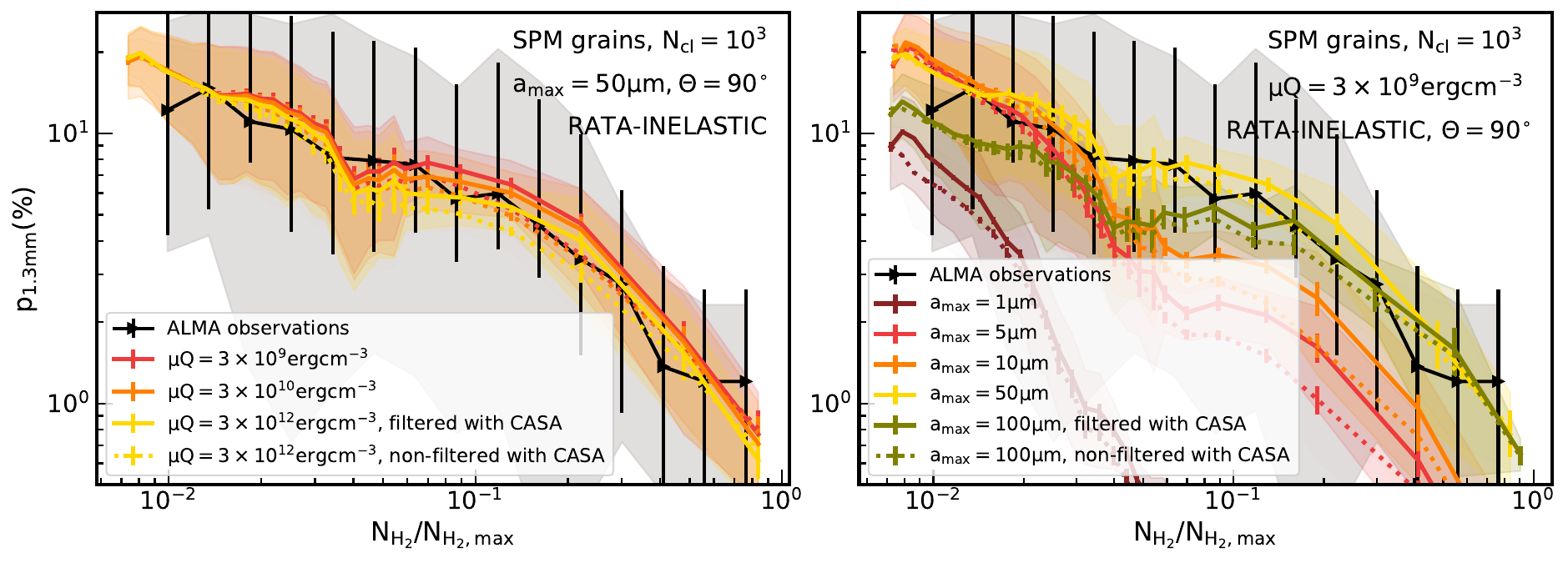}
    \caption{Left panel: Effect of inelastic relaxation on the polarization degree obtained at 1.3mm, considering model RATA$-$INELASTIC of SPM grains with $N_{\rm cl} = 10^{3}$, $\mu Q \sim 3\times 10^{9} - 3\times 10^{12}\erg\cm^{-3}$, $a_{\rm max} = 50\mum$, and $\Theta = 90^{\circ}$. The polarization degree observed in the inner envelope and disk decreases with decreasing the impact of inelastic relaxation acting on highly elastic grains, i.e., higher $\mu Q$. However, the difference is insignificant, implying the independence of the inelasticity of dust grains on polarized dust emission. Right panel: effect of grain growth on dust polarization, considering model RATA$-$INELASTIC of SPM grains with $N_{\rm cl} = 10^{3}$, $\mu Q = 3\times 10^{9}\erg\cm^{-3}$, $a_{\rm max}$ varying from $1\mum$ to $100\mum$, and $\Theta = 90^{\circ}$. The polarization degree increases with increasing micron-sized grains, but it reverts when grains grow beyond $50\mum$ due to the increased amount of VLGs with inefficient magnetic alignment. However, grains must grow beyond $\geq 10\mum$ in the inner envelope and beyond $\geq 50\mum$ in the disk to reproduce the observed range of $\sim 2-8\%$ and $\sim 1\%$ of polarization by ALMA there.  }
    \label{fig:CASA_inelastic_muQ_amax}
\end{figure*}

\subsection{Inner envelope of $\sim 500$ au}\label{sec:discuss_inner_envelope}
For the inner envelope with typical $n_{\rm H_{2}} \sim 10^{7}-10^{8}\cm^{-3}$ and $T_{\rm d} \sim 50-60$K, the observed polarization degree range from $p \sim 2\%$ to $\sim 10\%$ can be explained by the presence of SPM grains with $[N_{\rm cl};\phi_{\rm sp}] = [10^{2}-10^{4};0.1]$. With this configuration, micron-sized and VLGs up to $50\mum$ can maintain their fast Larmor precession around magnetic fields and achieve high external alignment with $f_{\rm high-J} \geq 0.5$ by MRAT mechanism (Figures \ref{fig:distribution_alignment}, \ref{fig:CASA_Ncl}, \ref{fig:CASA_inelastic_Ncl}). However, unlike SPM grains with $N_{\rm cl} \geq 10^{2}$ in the envelope which can achieve efficient IA by super-Barnett relaxation (Figure \ref{fig:Barnett}), the efficient IA of aligned large grains in the inner envelope is majorly led by inelastic relaxation regardless of the grain magnetic properties (Figure \ref{fig:CASA_best_fit}). We show in the left panel of Figure \ref{fig:CASA_inelastic_muQ_amax} the effect of the inelasticity of dust grains on the relation $p(\%)-N_{\rm H_{2}}/N_{\rm H_{2},max}$, considering SPM grains with $N_{\rm cl} = 10^{3}$, $\mu Q = 3\times10^{9} - 3\times10^{12}\erg\cm^{-3}$, $a_{\rm max} = 50\mu m$, and $\Theta = 90^{\circ}$. As illustrated in Figure \ref{fig:Inelastic}, lower inelasticity reduces the range of grains having efficient IA at high-\textit{J}, leading lower observed polarization fraction. However, increasing $\mu Q$ does not induce the huge decrease in $p(\%)$ as when we decrease $N_{\rm cl}$ (Figure \ref{fig:CASA_Ncl}) or remove the contribution of inelastic relaxation (Figure \ref{fig:CASA_best_fit}). It implies that large grains in the inner envelope can always achieve efficient IA by inelastic relaxation regardless of their inelasticity.
 
\subsection{Inside Class 0/I disk}\label{sec:discuss_disk}
 The most complicated region observed by ALMA is the disk scale of size $\sim 100$ au around the embedded protostar, with the typical density of $n_{\rm H} \sim 10^{9} - 10^{11}\cm^{-3}$. In contrast to the envelope and inner envelope where thermal emission of aligned dust grains is the major source of dust polarization, the complex convolving of grain growth, optical depth, and radiation fields can lead to either self-scattering of thermal dust emission (\citealt{Kataoka_2015}, \citealt{Kataoka_2016}), dichroic extinction (\citealt{Brauer_2016}, \citealt{Takahashi_2019}, \citealt{Liu_2021}), and dichroic emission from aligned dust grains to be the origin behind the observed dust polarization signal. Generally, the dust polarization pattern detected in half of Class 0/I disks by ALMA can be explained by self-scattering, while another may reveal features of polarized dust emission (\citealt{Cox_2018}, \citealt{Sadavoy_2019},  \citealt{Huang_2024}, \citealt{Zamponi_2024}). The typical polarization fraction obtained inside the disk is about $1-2\%$, but some disk even reveals very high $p \sim 3 - 10\%$ (i.e., the peusido disk OMC3 MMS6, \citealt{Takahashi_2019}, intermediate Class 0 disk IRAS4A, \citealt{Ko_2020}, \citealt{Liu_2021}), that challenges our understanding about the dust properties and disk conditions that produces such complicated behavior of ALMA dust polarization obtained in this scale. We first discuss the current limitation of our modeling in explaining dust polarization inside the disk in Section \ref{sec:discuss_scattering}. Then, we will discuss different scenarios that support the detection of dust emission polarization inside the disk scale in Sections \ref{sec:discuss_scenario} and \ref{sec:discuss_increase_Ncl}.

\subsubsection{Issue of polarized emission from magnetically aligned dust grains}\label{sec:discuss_scattering}
As shown in \cite{Giang_2023a}, \cite{Giang_2023b}, and Figures \ref{fig:p_map} and \ref{fig:CASA_inelastic_Ncl}, PM grains and SPM grains with $N_{\rm cl} \leq 10^{2}$ always produce $p << 1\%$ inside the protostellar disk owing to the alignment loss of VLGs and the weak MRAT alignment acting on micron-sized grains there (see Appendix \ref{sec:appen_disk} and \citealt{Giang_2023b} for details). Under this condition, self-scattering can dominate dust emission in producing the typical $p \sim 1\%$ observed by ALMA inside Class 0/I disks. However, it is unreasonable for large micron-sized grains having $N_{\rm cl} \geq 10^{2}$ beyond $\sim 500$ au (as constrained in Sections \ref{sec:discuss_envelope} and \ref{sec:discuss_inner_envelope}) to become PM or SPM with  $N_{\rm cl} \leq 10^{2}$ inside the protostellar disk. In case dust grains inside the disk have higher $N_{\rm cl} \sim 10^{3} - 10^{4}$, large micron-sized below $\sim 10\mum$ can experience MRAT mechanism, producing around $p \sim 1\%$ there (Figure \ref{fig:CASA_inelastic_Ncl}). In our MHD simulation, the polarization signal from magnetically aligned dust grains is not contaminated by self-scattering or dichroic extinction because the disk is optically thin at 1.3mm, and the maximum grain size of $50\mum$ is small compared to the observed wavelengths of $1.3$ mm. But in reality, since submillimeter grain sizes are observationally found (\citealt{Kwon_2009}, \citealt{Miotello_2014}, \citealt{Galametz_2019}, \citealt{Cacciapuoti_2023a}) and theoretically predicted (\citealt{Wong_2016}, \citealt{Yusuke_2022}) inside Class 0/I disks, self-scattering can feature polarized dust emission signal even at millimeter wavelengths if they share similar polarized intensity and $\sim 1\%$ of dust polarization. One typical case that shows the combined effect of self-scattering and (may) aligned dust grains is HL Tau (\citealt{Kataoka_2012}). While self-scattering can explain the uniform polarization pattern along the disk minor axis observed at Band 7 (0.87mm) by ALMA,  the origin behind the azimuthal pattern observed at Band 3 (3mm) and transition polarization pattern at Band 6 (1.3mm) remains unclear. Several scenarios are proposed, including the emission from radiative aligned dust grains (\citealt{Kataoka_2017}, \citealt{Stephens_2017}), emission from prolate grains aligned along the toroidal $\B$ fields inside HL Tau disk (\citealt{Mori_2021}, \citealt{Stephens_2024}, \citealt{Zhe-Yu_2024}), or emission from oblate grains aligned along the toroidal $\B$ fields due to the slow internal relaxation (\citealt{Thang_2024}). More detailed following-up papers of \cite{Yang_2016b}, which solves the Muller matrix of non-spherical grains accounting for the realistic alignment efficiency of large grains, must be performed to accurately diagnose the grain configurations and disk conditions that support the observed polarization pattern inside protostellar and protoplanetary disks.
 
Another issue inside the disk is that our model always points out the poor magnetic alignment for micron-sized and VLGs regardless of grain properties and alignment mechanisms (\citealt{Hoang+2021}, \citealt{Giang_2023b}). As shown in Appendix \ref{sec:appen_disk}, iron inclusions fail to let VLGs beyond $\geq 10\mum$ to have magnetic alignment, MRAT is deactivated for micron-sized grains beyond $\geq 6\mum$, and neither super-Barnett nor inelastic relaxation fail to drive the efficient IA for aligned dust grains inside the disk. Such inefficient alignment degree of dust grains is also indicated inside the $\beta$ Pic Debris Disk by \cite{Hull_2022}. However, this feature challenges our understanding in using RAT/MRAT alignment theory to explain the disk polarization flipping at ALMA Band 7 (in OMC3 MMS6 by \citealt{Takahashi_2019},  and IRAS4A in \citealt{Ko_2020}, \citealt{Liu_2021}) and the high $p \geq 2-10\%$ observed in the inner 100 au of the optically thick disks IRAS4A (\citealt{Ko_2020}). The authors indicate that these features could be a sign of dichroic extinction caused by the existence of VLGs. However, the current study by \cite{Giang_2023b} indicated that this mechanism is only activated if VLGs above $\geq 50\mum$ nearly have perfect magnetic alignment inside the optically thick disk. Given the weak magnetic alignment of grains as discussed above, unless the help of another mechanism that enhances the coupling of grains and magnetic fields inside the disk, it will deactivate the operator of dichroic extinction inside the optically thick disk even if VLGs can be formed inside there (\citealt{Giang_2023b}).  

\subsubsection{Potential scenarios driving efficient grain magnetic alignment inside the disk}\label{sec:discuss_scenario}
In this section, we will discuss several scenarios that can improve the alignment of micron-sized and VLGs inside the disk. The obvious scenario is to increase the magnetic field strength inside the disk or increase the grain magnetic susceptibility. However, higher magnetic field strength does not help to improve the IA degree - the major factor causing the poor alignment of large grains in this region. For the second scenario, we show in Figure \ref{fig:alignment_disk_phisp} the effect of $\phi_{\rm sp}$ the volume filling factor of iron clusters (from 0.1 to 0.3 - which corresponds to $\sim 30\%$ to $\sim 90\%$ of iron atoms depleted from the gas phase will appear inside dust grains under the cluster form, \citealt{Hoang_Lazarian_2016_mrat}) on the grain alignment degree distribution, assuming $N_{\rm cl} = 10^{4}$. By increasing $\phi_{\rm sp}$, more VLGs can be aligned with $\B$, and more micron-sized grains can experience MRAT alignment in the 100 au disk region. However, it does not improve super-Barnett relaxation efficiency, which maintains the low net alignment degree of grains there. 
 
Improving the inelastic relaxation efficiency is another way to enhance the IA of large grains inside the disk. In our study, we consider the lowest inelasticity of $\mu Q = 3\times 10^{9}\erg\cm^{-3}$, which is obtained from silicate rocks (\citealt{Efroimsky_Lazarian_2000}), vitreous silica (\citealt{Purcell_1979}), and cometary dust (\citealt{Seizinger_2013}, \citealt{Knapmeyer_2018}). The inelasticity of VLGs inside the protostellar disk remains unknown. However, given the flexibility of highly porous large grains created by grain coagulation and grain compression (\citealt{Seizinger_2013}, \citealt{Garcia_2020}, \citealt{Michoulier_2024}), VLGs inside the disk may have lower $\mu Q$ and experience stronger inelastic relaxation effect than our considered values.

Suppose neither super-Barnett nor inelastic relaxation efficiency is improved inside the disk. In that case, grains with slow internal relaxation still can produce higher polarization signal than we found if they obtain higher internal alignment degree $Q_{\rm X}^{\rm high-J}$ and $Q_{\rm X}^{\rm low-J}$ at high and low-\textit{J} attractors (Section \ref{sec:modelling_alignment}). In our study, we adopt low $Q_{\rm X}^{\rm high-J} = 0.15$ and  $Q_{\rm X}^{\rm low-J} = 0.05$, supposing grains with slow internal relaxation experience both the efficient gaseous damping and internal thermal wobbling (\citealt{Lazarian_1994}, \citealt{Hoang_Lazarian_2008}). However, \cite{Hoang_Lazarian_2009} indicated that when the grain symmetric axis is initially closely aligned to the grain angular momentum, grains without internal relaxation can still achieve perfect internal alignment. The high alignment degree of dust grains is also indicated in \cite{Hull_2022}, who argued that highly porous grains with low intrinsic polarization degree (\citealt{Kirchschlarger_2019})  inside $\beta$-Debris disk \footnote{\cite{Li_Greenberg_1999} indicated the grain porosity of $\sim 0.95$ inside $\beta$-Debris disk.} need to have $R \sim 0.3$ to reproduce the observed polarization degree of $p \sim 0.51\%$. Similarly, the model of wrong-aligned oblate grains by slow internal relaxation introduced by \cite{Thang_2024} can explain the polarization pattern observed at ALMA Band 7 for HL Tau. However, they must nearly achieve perfect wrong internal alignment ($Q_{\rm X,lowJ} \sim -0.5$) to reproduce $p\sim 1\%$ observed in ALMA Band 7. Indeed, \cite{Tazaki_2017} considered that grains with slow internal relaxation at high-\textit{J} attractors still can have perfect internal alignment (or $Q_{\rm X}^{\rm high-J} = 1$) owing to their suprathermal rotation. However, the validity of such an assumption has not been tested. Future studies from both the theoretical side and observational constraints are required to accurately understand the alignment dynamics of protostellar large grains.

Another parameter affecting the net polarization fraction is $f_{\rm high-J}$. In our study, we fix $f_{\rm high-J} = 0.25$ for grains aligning with $\B$ by RATs. However, this parameter can vary up to $f_{\rm high-J} \sim 0.7$ depending on the grain shapes, sizes, and their compositions (\citealt{Herranen_2021}). Similarly, we consider grains that experience moderate MRAT efficiency to have $f_{\rm high-J} = 0.5$, but this fraction can vary among $f_{\rm high-J} \sim 0.3 - 1$ (\citealt{Hoang_Lazarian_2016_mrat}). However, the above range of $f_{\rm high-J}$ is found from numerical studies of RATs/MRAT alignment for compact irregular grains. \cite{Jager_2024} presented a new work on ballistic aggregate grains. They found that the RAT efficiency $\Gamma_{\rm RAT}$ shares the same tendency with the ratio $\lambda/a$ as for compact irregular grains \footnote{RAT efficiency is a constant of $\sim 0.4$ for grains being comparable in size with interacting wavelengths $\lambda/a \sim 1$, and decreases with increasing ratio $\lambda/a$ (\citealt{Jager_2024})} (\citealt{Lazarian_Hoang_2007a}, \citealt{Herranen_2021}). However, its magnitude is $\sim 10-100$ times smaller in the range $\lambda/a \sim 1-100$, and thus aggregate-type grains are spun up by RATs weaker by twice times compared with compact grains. The potential range of $f_{\rm high-J}$ for aggregate grains has not yet been explored. Additionally, \cite{Jager_2024} stopped the study for grains of $0.25\mum$. Further studies of RATs/MRAT on aggregate grains beyond micron-sized to submillimeter grains are needed to accurately model the grain alignment dynamic inside protostellar cores.

Lastly, given the decoupling of large grains and gas inside the protostellar disk, grains may be coupled with magnetic fields better owing to the effect of MEchanical Torques (METs, \citealt{Lazarian_Hoang_2007a}, \citealt{Hoang_2018}) produced by the relation motion of gas and grains. As shown in \cite{Hoang+2022}, METs allow more sub-micron grains to be aligned with $\B$ by driving them to suprathermal alignment conditions. They also can improve the IA degree of protostellar grains by shortening both super-Barnett and inelastic relaxation timescale. Moreover, their alignment torque component also helps to drive $\J$ to align with $\B$ faster. Therefore, the MET may help the grain alignment process within the inner 100 au region. Difference from the RAT whose efficiency is rather well quantified based on grain size-wavelength ratio; the MET efficiency can change up to the orders of magnitude with varying grain shapes and sizes (\citealt{Reissl_2023}). However, it is worth studying to include both MET/RAT/MRAT together to fill in our understanding of the complicated picture of grain alignment in such a dynamic region. 

The following consequence of the MET is the alignment of dust grains with the gas flow direction (i.e., the mechanical alignment, or shortly v$-$MET, \citealt{Hoang_2018}). v$-$MET will happen when the mechanical precession of grains around the gas flow driven by METs is faster than the Larmor precession and gas randomization rate. Alongside v$-$MET, dust grains also can experience radiative alignment (or k$-$RAT, \citealt{Lazarian_Hoang_2007a}) if the radiative precession of grains around the radiation field driven by RATs is faster than Larmor precession and gas randomization. k$-$RAT is first modeled inside the protoplanetary disk by \cite{Tazaki_2017}, then it is applied to explain the transition of polarization pattern obtained between 3mm and 0.87mm in HL Tau (\citealt{Kataoka_2017}, \citealt{Stephens_2017}). However, this scenario is recently disfavored via the detailed modeling at 3mm by \cite{Yang_2019}. Recent synthetic modeling of k$-$RAT and B$-$RAT in $\beta$ Pic Debris disk by \cite{Hull_2022} flavored k$-$RAT in explaining the observed dust polarization at 0.87mm by ALMA. Similarly, \cite{Zhe-Yu_2020} suggested the combined effect of k$-$RAT and self-scattering to explain the azimuthal polarization pattern with stronger $p(\%)$ along the disk major axis in the gap of HD 163296.  However, one issue in k$-$RAT model of \cite{Tazaki_2017} is the condition for suprathermal grains to have radiative alignment, which is $\tau_{\rm rad} < \tau_{\rm Lar} < \tau_{\rm gas}$ with $\tau_{\rm rad}$ the radiative precession timescale determined for thermal grains. As indicated in \cite{Hoang+2022}, the radiative precession decreases with increasing the grain rotational energy. Using $\tau_{\rm rad}$ for thermal grains may overestimate the range of suprathermal grains having k$-$RAT alignment. As aligned grains at high-\textit{J} attractors contribute more to the net $p(\%)$, the above issue may cause inaccurate quantification of k$-$RAT inside the disk region. v$-$MET also shares properties similar to k$-$RAT for thermal and suprathermal grains. Given the complexity of grain dynamics inside the dense disk, accurately modeling with k$-$RAT, v$-$MET, and B$-$RAT/B$-$MET is required to understand the contribution from each scenario in producing the polarization signal observed inside the disk scale. 
 
\subsubsection{Whether increasing $N_{\rm cl}$ helps to improve the magnetic alignment?\label{sec:discuss_increase_Ncl}}
As discussed in the first paragraph of Section \ref{sec:discuss_scenario}, we can increase the volume filling factor of iron inclusions $\phi_{\rm sp}$ to strengthen the magnetic alignment of grains in the disk. However, it does not help to enhance the internal alignment degree of micron-sized and VLGs - the major reason causing the weak alignment degree inside the protostellar disk (Appendix \ref{sec:appen_disk}). Another case is to increase $N_{\rm cl}$ to beyond $10^{4}$ - the maximum value considered in our study. As shown in Appendix \ref{sec:appen_Ncr}, $N_{\rm cl} = 10^{4}$ is still inside the range of $N_{\rm cl}$ allowed for grains being SPM, where our grain alignment model is built in (except for magnetite $\rm Fe_{\rm 3}O_{\rm 4}$ in low dust temperature region $T_{\rm d} < 40$ K). However, given the unclear composition and structure of iron inclusions inside protostellar grains, it can be unrealistic to continue increasing $N_{\rm cl}$ to get better magnetic alignment. If $N_{\rm cl} > 10^{4}$ is really the upper limit of protostellar SPM grains in reality, further increasing $N_{\rm cl}$ will push grains to ferro/ferrimagnetic regism, in which iron inclusions can create their own magnetic domain and need lots of energy to reorient their net magnetic moment when applying to the external magnetic field (\citealt{Neel_1949}, \citealt{Bean_1959}, \citealt{Yang_2021}). Given the typical lifetime of Class 0/I YSOs of $10^{4}-10^{5}$ yr, ferro/ferrimagnetic material may not have enough time to achieve the magnetic moment by the Barnett effect as SPM grains. Their relaxation mechanism is also different from SPM (\citealt{Lazarian_Hoang_2019}); hence, the physics of grain alignment there does not apply to them. Moreover, it is unclear how such massive iron inclusions can form and be embedded inside grains in the disk scale. 

Indeed, iron cluster size may follow some distributions inside dust grains (\citealt{Yang_2021}) instead of sharing the same value $N_{\rm cl}$ as our model. This configuration may not significantly change the polarization degree obtained in the envelope (Section \ref{sec:ALMA_observations}). But for the inner envelope and protostellar disk, the cluster size contributing mostly to the magnetic susceptibility (we call $N_{\rm cl,peak}$) becomes very important to decide whether our model can reproduce ALMA polarization fraction. If grains contain some iron clusters with $N_{\rm cl} = 10^{4}$ but have $N_{\rm cl,peak} < 100$, they can produce $p \sim 5\%$ inside the inner envelope but cannot produce $p > 0.5\%$ in the disk region. If the peak slants more to bigger cluster sizes, the situation will approach what we obtain in model RATA$-$INELASTIC for $N_{\rm cl} \sim 10^{3}-10^{4}$. Further studies in the dust evolution inside YSOs may shed light on constraining the distribution and maximum available size of iron inclusions inside dust grains.

\subsection{In the outflow cavity and outflow cavity wall}\label{sec:discuss_outflow}
\subsubsection{Grain alignment state}\label{sec:grain_alignment_state}
The last remaining part of YSOs that we will discuss is the outflow cavity and outflow cavity wall. \cite{Valentin_2023b} found that by increasing the center luminosity from $L_{\odot}$ to $100L_{\odot}$, the minimum alignment size inside the dense outflow cavity wall of intermediate Class 0/I YSOs decreases from $5\mum$ to $1\mum$ owing to the enhanced RAT acting on small micron grains. However, as supposed in Section \ref{sec:Distribution}, VLGs above $10\mum$ inside the cavity wall must be SPM with at least $N_{\rm cl} \geq 5$ to stably maintain their magnetic alignment against the high gas randomization in this area (Figure \ref{fig:distribution_alignment}, second column). Aligned dust grains inside the outflow cavity will achieve fast internal relaxation at high-\textit{J} attractors by super-Barnett relaxation (Figure \ref{fig:Barnett}). While inside the dense outflow cavity wall, inelastic relaxation drives the efficient IA for aligned dust grains (Figure \ref{fig:Inelastic}). However, large grains above $\geq 1\mum$ in the entire protostellar outflow always have slow internal relaxation at low-\textit{J} attractors regardless of the joint action of Barnett and inelastic relaxation. 
 
\subsubsection{Alignment direction in outflow: B$-$RAT/MET, k$-$RAT, or v$-$MET?}
Similar to the complicated external alignment mechanism that can happen inside the disk (Section \ref{sec:discuss_disk}), grain inside the protostellar outflow also can experience either B$-$RAT, B$-$MET, v$-$MET, or k$-$RAT because of the intense radiation field strength and the strong gas-grain drifting inside outflows. \cite{Valentin_2023b} recently found that VLGs beyond $\geq 10\mum$ at low-\textit{J} attractors can experience k$-$RAT inside outflows. However, under the high center luminosity of $L_{\rm center} \geq 20L_{\odot}$ required to explain ALMA grain alignment efficiency, suprathermal VLGs driven by RATs can be totally fragmented to smaller sizes by their strong centrifugal force as described in theory of RAdiative Torque Disruption (RATD) mechanism (\citealt{Hoang_2019}). The removal of VLGs by RATD may reduce amount of polarized thermal dust emission (\citealt{Hoang_2019}, \citealt{Hoang_2021}), hence suppressing ALMA dust polarization observations at thousands au scale (\citealt{Valentin_2023b}). 
We will stress the effect of RATD on VLGs inside outflows in the forthcoming Paper II. 

In our study without RATD, we found that if the MRAT mechanism is the origin behind the grain alignment that would produce ALMA polarization degree of $p\geq 2-30\%$ beyond the disk, $\sim 50-100\%$ of dust grains inside outflows will have magnetic alignment at high-\textit{J} attractors (Section \ref{sec:Distribution}). k$-$RAT can become the alignment direction for the remaining grains rotating thermally inside outflows (\citealt{Hoang+2022}). However, as thermal grains always have inefficient IA by slow internal relaxation inside protostellar environments, their emission will be easily dominant by polarized emission from outflowing grains having magnetic alignment at high-\textit{J}. This issue reduces the possibility of detecting evidence of k$-$RAT inside the protostellar outflow. The search for k$-$RAT evidence also did toward the intense radiation source as Orion bar (\citealt{Chuss_2019}, \citealt{Valentin_2023c}) and OMC-1 (\citealt{Pattle_2021}), but none of them clearly show imprints from k$-$RAT. k$-$RAT may be detected around the C$-$rich AGB star, where carbonaceous grains without iron inclusions can be aligned with radiation fields and produce dust polarization (\citealt{Andersson_2022}, \citealt{Hoang_2023}, \citealt{Andersson_2024}). 
  
\subsection{On the origins of depolarization observed by ALMA from the envelope to disk scales} 
One of the most prominent features observed toward star-forming regions is the reduction of the polarization degree from a few tens percentages in the outer part to $p < 1\%$ in the center. This depolarization phenomenon is widely detected from the cloud and filament scale by JCMT (\citealt{Pattle_2019}, \citealt{Lyo_2021}, \citealt{Ngoc_2021}), down to the core scale by CARMA (\citealt{Hull_2014}), SMA (\citealt{Girart_2006}), JVLA (\citealt{Rao_2009}, \citealt{Liu_2018}), and then toward the envelope and disk scale by ALMA (\citealt{Cox_2015}, \citealt{Cox_2018},  \citealt{Sadavoy_2018}, \citealt{Valentin_2023a}). Regarding the depolarization effect in the large cloud and core scale, the unresolved structure of magnetic fields in the center may be important in suppressing the observed polarization degree there. It can answer why follow-up interferometric observations toward the core scale (by JVLA, SMA, CARMA observations, \citealt{Rao_2009} and envelope/disk scale (by ALMA observations, \citealt{Maury_2018}, \citealt{Galametz_2019}) with higher resolution come up again with few to few tens percent of dust polarization in the outer boundary. However, since ALMA can resolve the inner core up to a few tens to $\sim 100$ au, the beam size effect may contribute less to reducing $p(\%)$ in the central region. It thus lets either the geometrical effect of $\B$ fields or the properties of dust grains or grain alignment efficiency be more important in causing the reduction of $p(\%)$ with gas density. 

We show in the right panel from Figure \ref{fig:CASA_best_fit} to Figure \ref{fig:CASA_inelastic_Ncl} the comparison of the grain alignment efficiency described by product $p\times S$ with gas column density between our synthetic results and ALMA observation. \footnote{Our core is optically thin at 1.3mm, so we can neglect the effect of optical depth on the depolarization.} In model PA, the rise of $a_{\rm align}$ from $0.5\mum$ (in the envelope) to $\sim 2\mum$ (in the center) is negligible compared to the considered wide alignment range with $a_{\rm max} = 50\mum$. Therefore, the increasing $\B$ field tangling along the LOS and the change in magnetic field geometry by outflow activities and disk formation (Figure \ref{fig:CASA_inelastic_Ncl}, lower row with $\Theta = 30^{\circ}$) cause the reduction of $p(\%)$ with increasing gas density. However, model PA produces the polarization degree of $\sim 3\%$ and the product $p\times S > 0.7$ near the peak of gas column density, which is higher than values observed by ALMA. In contrast, results from model RATA$-$INELASTIC produce similar decreased levels of $p(\%)$ and $p\times S$ throughout YSOs as observational results. Therefore, we suppose the reduction of the IA degree and MRAT alignment efficiency by strong gaseous damping to be a possible scenario behind the reduction of $p(\%)$ observed in the ALMA scale. It converges with the conclusion from \cite{Giang_2023b}, who found the consistency of the depolarization index $p\sim I^{\alpha}$ observed by ALMA ($\alpha = -0.5$ to $\alpha = -1$, \citealt{Kwon_2019}, \citealt{Lai_2002}) with model taking the realistic alignment of dust grains into account. 

From the right panel of Figures \ref{fig:CASA_best_fit} and \ref{fig:CASA_inelastic_Ncl}, one can see that our synthetic results cannot exactly capture the tendency of $p\times S - N_{\rm H_{2}}/N_{\rm H_{2},max}$ revealed by ALMA as it does for the polarization fraction distribution. In particular, while ALMA shows the continuous decrease of $p\times S$ from $\sim 0.5$ in the envelope to $\sim 0.3$ in the center, our model reveals the increasing $p\times S$ when $N_{\rm H_{2}} < 0.1N_{\rm H_{2},max}$, following by the sharp decrease to $p\times S < 0.2$ at $N_{\rm H_{2},max}$. In our core, the magnetic field follows a clean hourglass-shaped morphology as turbulence is not injected when the core starts collapsing. It induces low $S$ and small $p\times S$ in the outermost region. Toward the inner region, the hourglass-shaped field is reshaped by the outflow, inducing the sudden rise of $S$ (see also the sudden drop of $p(\%)$ along the outflow cavity wall in Figure \ref{fig:p_map}) in region where $N_{\rm H_{2}} \sim 0.1N_{\rm H_{2},max}$. The field then becomes order again following the formation of the rotating disk on the equatorial midplane (see the inferred $\B$ field map from synthetic observations in \citealt{Valentin_2023b}). $S$ thus only slightly increases with increasing density. $p(\%)$ will control the tendency of $p\times S$, inducing the sharp decrease of $p\times S$ to $ \sim 0.1-0.2$ near the protostar. In reality, the magnetic field morphologies of Class 0/I objects are much more complicated (giving higher $S$) than our adopted model. It explains why we cannot well reproduce the observed trend of $p\times S$ revealed by ALMA. We note that $p(\%)$ also depends on the core magnetic field morphology. However, it is controlled also by the intrinsic polarization degree and the grain alignment efficiency. Grain alignment may share similar features among different low/intermediate-mass Class 0/I YSOs due to their similar gas density distribution (\citealt{Valdivia_2022}). It may be a reason why our model can capture well the observed tendency of $p(\%)$ and $N_{\rm H_{2}}/N_{\rm H_{2},max}$ regardless of the difference in $\B$ field morphologies.

Indeed, the decreasing tendency of $p\times S$ is unclear because of the large standard deviation and the small change of $p\times S$ from $\sim 0.5$ in the envelope to $\sim 0.3$ near the center. Therefore, it is hard to state whether the grain alignment efficiency decreases toward the center or is constant throughout the core. From the simulation side using the RAT/MRAT alignment framework, our studies always indicate that VLGs within a few hundred au around the protostar cannot achieve similar alignment degrees as grains at a thousand au scale regardless of being exposed to a higher radiation field. It thus leads to the prediction of decreasing $p\times S$ with density, and we suggest the grain dynamic to be more responsible in inducing the depolarization from the envelope to the center instead of magnetic field morphology. However, if the grain alignment efficiency is constant in YSOs as it suggested by ALMA data, it implies a lack of understanding when modeling the dynamic of grain alignment within the 500 au scale. Further studies with more accurate dust models (Section \ref{sec:discuss_dust}) and theoretical studies of grain alignment dynamics in extremely dense regions (Section \ref{sec:discuss_disk}) are required to fill in the gap between observations and simulations.

\begin{figure*}
\centering  \includegraphics[width=\textwidth,height=\textheight,keepaspectratio]{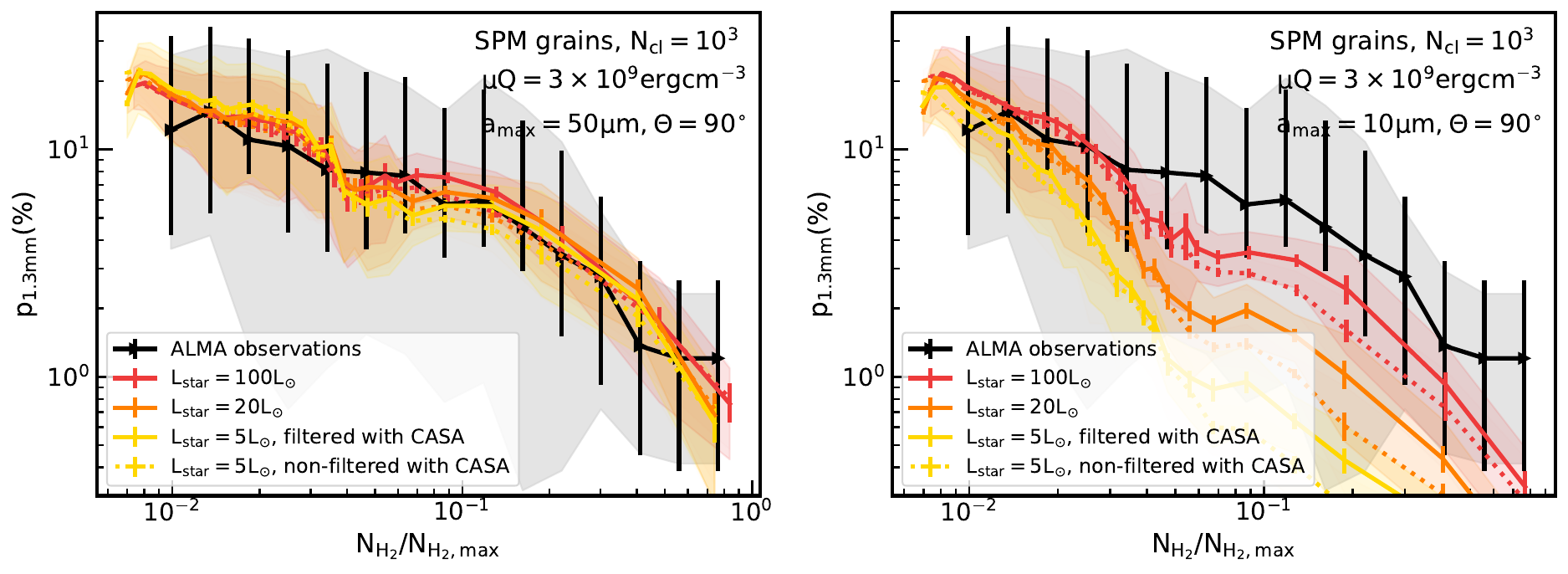}
    \caption{Similar results as Figure \ref{fig:CASA_inelastic_muQ_amax} but for the effect of stellar luminosity $L_{\rm star}$ on the variation of $p(\%)$ and $N_{\rm H_{2}}/N_{\rm H_{2},max}$, with $L_{\rm star}$ varying from $5L_{\odot}$ to $100L_{\odot}$. The left panel shows results for $a_{\rm max} = 50\mu m$, while the right panel corresponds to results with $a_{\rm max} = 10\mum$, assuming model RATA$-$INELASTIC of SPM grains with $N_{\rm cl} = 10^{3}$, $\mu Q = 3\times 10^{9}\erg\cm^{-3}$, and $\Theta = 90^{\circ}$. Stellar luminosity does not strongly affect the polarization fraction range around YSOs if protostellar grains can grow beyond $\sim 50\mum$ (left panel). As decreasing the maximum grain size, the reduction of the alignment range (i.e., larger $a_{\rm align}$ owing to lower $L_{\rm star}$, see Appendix \ref{sec:appen_Lstar}) is more prominent, producing lower $p < 1\%$ inside the inner envelope and disk scale. The result in the right panel is similar to finding in \cite{Valentin_2023b}.}
    \label{fig:CASA_Lstar}
\end{figure*}

Another issue related to the quantity $p\times S$ is whether it is still suitable for quantifying the grain alignment efficiency inside dense environments. The dispersion of $\B$ fields along the LOS is one of the reasons causing lower observed polarization fraction, but it is hard to directly measure from dust polarization (\citealt{Chen_2016}, \citealt{Hoang_Bao_2024}). In contrast, $S$ is determined by the dispersion of the polarization vectors on the POS. If the tangling of $\B$ fields on the POS is similar to their dispersion on the LOS, the quantity $p\times S$ can tell us about the grain alignment degree in the observed regions. But if not, $S$ may inaccurately tell us the impact of magnetic field geometry on dust polarization. Moreover, \cite{Giang_2023b} showed that if the alignment loss happens significantly inside the central region, $S$ cannot help to trace the disorganization of $\B$ fields close to the protostar. These issues may affect our understanding of the origin behind the depolarization feature in the ALMA scale.

\subsection{Effects of grain growth on dust polarization}\label{sec:grain_growth}
Besides the intrinsic polarization degree controlled by the physical properties of aligned dust grains (e.g., grain elongation) and the grain alignment efficiency, the maximum grain size is another important factor affecting the observed degree of polarized thermal dust emission. As shown in \cite{Valdivia_2019}, grains must grow beyond $\geq 10\mum$ inside the protostellar envelope to reproduce $p\geq 5\%$ observed by ALMA in this area. The recent study of \cite{Valentin_2023b} and results from Figure \ref{fig:CASA_amax} in Section \ref{sec:ALMA_observations} also clearly indicate the importance of grain growth in maintaining the considerable alignment range constrained by ALMA dust polarization. However, studies by \cite{Valdivia_2019} and \cite{Valentin_2023b} do not consider the inefficient alignment of dust grains under strong gaseous damping conditions in their synthetic modeling. It may overestimate the impact of grain growth on dust polarization, especially when grains grow beyond $\geq 10\mum$ (\citealt{Giang_2023b}). Taking this issue into account, and in order to constrain the lower limit of maximum grain size required to reproduce ALMA dust polarization, we show in the right panel of Figure \ref{fig:CASA_inelastic_muQ_amax} the effect of $a_{\rm max}$ on $p-N_{\rm H_{2}}/N_{\rm H_{2},max}$ observed at 1.3mm. We consider model RATA$-$INELASTIC of SPM grains with $N_{\rm cl} = 10^{3}$, $\mu Q = 3\times 10^{9}$, $\Theta = 90^{\circ}$, and $L_{\rm center} = 100L_{\odot}$. The polarization degree will increase systematically with increasing $a_{\rm max}$ from $1\mum$ to $50\mum$ as extending the alignment range toward larger sizes (similar to the finding in \citealt{Valdivia_2019} and \citealt{Valentin_2023b}). However, when grains grow beyond $a_{\rm max} \geq 50\mum$, $p(\%)$ turns to decrease with increasing $a_{\rm max}$ owing to the enhanced amount of VLGs with inefficient internal and external alignment (Section \ref{sec:Distribution}, see also \citealt{Giang_2023b}). But regardless of this issue, dust grains must grow beyond $\sim 10\mum$ in the inner envelope to reproduce $p \sim 2-8\%$ in this scale and beyond $\sim 50\mum$ inside the disk to reproduce $\sim 1\%$ of dust polarization there (Figures \ref{fig:CASA_best_fit} and \ref{fig:CASA_amax}). This finding confirms the previous conclusion of grain growth activities constrained via dust polarization by \cite{Valdivia_2019} and \cite{Valentin_2023b}. It also converges with the detection of VLGs based on the detection of low dust opacity index $\beta \sim 1$ inside the inner envelope and the disk scale by ALMA (i.e., \citealt{Kwon_2009}, \citealt{Miotello_2014}, \citealt{Bracco_2017}, \citealt{Galametz_2019}). However, given the worst alignment of VLGs inside the disk (Section \ref{sec:discuss_disk} and Appendix \ref{sec:appen_disk}), further detailed studies are required to accurately constrain the realistic impact of VLGs on disk polarization.

\begin{figure*}
\centering
\includegraphics[width=\textwidth,height=\textheight,keepaspectratio]{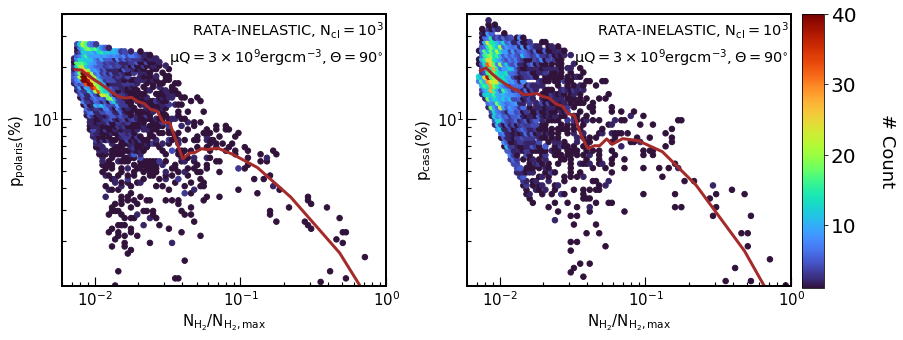}
    \caption{Scattered point map of the polarization degree obtained at 1.3mm with gas column density, before (left panel) and after filtering with CASA (right panel), considering model RATA$-$INELASTIC of SPM grains with $N_{\rm cl} = 10^{3}$, $\mu Q = 3 \times 10^{9}\erg\cm^{-3}$, $a_{\rm max} = 50\mum$, and $\Theta = 90^{\circ}$. The color bar shows the number of points inside each bin of $[p, N_{\rm H_{2}}/N_{\rm H_{2},max}$. The interferometric filtering tends to amplify the polarization fraction obtained in the outermost, low-intensity region of the envelope recovered by ALMA. However, the number of amplified points is just about $10-15$ cells, which does not significantly change the mean intrinsic value of $p(\%)$ obtained with the given gas column density value.}
     \label{fig:filtering_Ncl1e3}
\end{figure*}

\subsection{Impact of stellar radiation field strength on dust polarization}\label{sec:discuss_Lstar}

As argued in Section \ref{sec:modelling_setup}, we use the high bolometric luminosity of $100L_{\odot}$ to maximize the impact of stellar radiation on grain alignment and dust polarization. This value is beyond the average luminosity of the protostar samples used for comparison in Section \ref{sec:ALMA_observations}. However, some of these protostars are intermediate-mass Class 0/Is objects, and these type of objects can have very high bolometric luminosity of $\sim 100L_{\odot}$ (i.e., Serpens SMM1, \citealt{Valentin_2019}, HOPS 288 and 361, \citealt{Furlan_2016}, \citealt{Cheng_2022}). Moreover, the high mass-accretion rate $\dot{M}_{\rm acc} \sim 4\times 10^{-6}M_{\odot}/\rm yr$  corresponding to $L_{\rm center} = 100L_{\odot}$ still can happen during the accretion burst  \footnote{We note our MHD simulation is for intermediate-mass star formation, with the accretion rate onto the sink particle of $\dot{M}_{\rm acc} \sim 10^{-5}-10^{-4}M_{\odot}/\rm yr$, \cite{Valentin_2023b}}  (i.e., HH34 IRS, \citealt{Antoniucci_2008}, see also modeling of accretion activities by \citealt{Elbakyan_2016}). Therefore, it is acceptable to adopt $L_{\rm center} = 100L_{\odot}$ for understanding the grain alignment dynamic and dust polarization properties inside intermediate-mass cores. However, to better understand these activities inside protostars of less-intense irradiation, we consider two other cases of $L_{\rm center} = 20L_{\odot}$ ($\dot{M}_{\rm acc} \sim 8.38\times 10^{-7}M_{\odot}/\rm yr$) and $L_{\rm center} = 5L_{\odot}$ ($\dot{M}_{\rm acc}  \sim 2\times 10^{-7}M_{\odot}/\rm yr$). This simulation was already done in \cite{Valentin_2023b} for the ideal RAT alignment model. However, given the strong correlation of the grain rotational velocity and the IA shown in Figures \ref{fig:Barnett} and \ref{fig:Inelastic}, the effect of $L_{\rm center}$ on the net alignment degree need to be re-examined. The grain alignment state inside fainter protostars is shown in Appendix \ref{sec:appen_Lstar}, considering the case of SPM grains with $N_{\rm cl} = 10^{3}$ and $\mu Q = 3\times 10^{9}\erg\cm^{-3}$. We find that by decreasing $L_{\rm center}$, apart from the reduction of the alignment size range as increasing the minimum alignment size range (as found in \citealt{Valentin_2023b}), aligned dust grains still have efficient IA through inelastic relaxation, and further experience stronger MRAT owing to the enhanced magnetic relaxation resulting from the decreased dust temperature. 

We show in Figure \ref{fig:CASA_Lstar} the effect of $L_{\rm center}$ on the relation $p - N_{\rm H_{2}}/N_{\rm H_{2},max}$, considering $L_{\rm center} = 5L_{\odot}, 20L_{\odot}, 100L_{\odot}$. We consider model RATA$-$INELASTIC for SPM grains with $N_{\rm cl} = 10^{3}, \mu Q = 3\times 10^{9}\erg\cm^{-3}$, the observed angle of $\Theta = 90^{\circ}$, with the left showing results with $a_{\rm max} =50\mum$ and the right corresponding to results with $a_{\rm max}. =10\mum$. One can see that if VLGs up to $50\mum$ can be distributed over thousands au core scale (left panel), the central luminosity does not affect the observed polarization fraction due to the insignificant change in the alignment size range when varying $L_{\rm center}$ (Figures \ref{fig:alignment_Lstar50}). In contrast, if $a_{\rm max}$ is reduced to below $10\mum$ (right panel), the increase in the minimum alignment size around low-luminous protostar significantly narrowers the alignment size range (Figure \ref{fig:alignment_Lstar10}), which clearly induces the lower observed polarization fraction (as expected in \citealt{Valentin_2023b}). Our MRAT alignment model for SPM grains with $N_{\rm cl} = 10^{3}$ and $\mu Q = 3\times 10^{9}\erg\cm^{-3}$ fails to explain ALMA observations inside the inner envelope when the central luminosity reduces to below  $\leq 5L_{\odot}$(Figure \ref{fig:alignment_Lstar10}, first row). If grains have higher inelasticity, the situation worsens as models with $L_{\rm center} = 20L_{\odot}$ cannot reproduce the polarization degree range observed in low/intermediate Class 0 YSOs. 
 
Another issue that could affect the alignment efficiency of protostellar dust grains is the spectral energy distribution (SED) of the radiation source. In the POLARIS post-processing, we consider the source (including the stellar and accretion luminosity) as the black body and vary the stellar temperature from $T_{\rm star} = 16457$ K to $7800$ K to obtain $L_{\rm center} = 100L_{\odot}$ to $5L_{\odot}$. The corresponding SED will peak at far-ultraviolet of $0.176\mum$ or near-ultraviolet of $0.37\mum$. In reality, the accretion luminosity is produced by the interaction of supersonic accreting material and the protostar photosphere via the accretion shock (\citealt{GULLBRING_1998}, \citealt{Elbakyan_2016}) \footnote{The accretion luminosity is dominant the bolometric luminosity for our MHD run so that we can neglect the contribution of SED from the sink particle.}. Given the high shock temperature, the SED from accretion is much wider in X-ray, UV, and optical range. The exceeded UV and optical photons from the accretion shock can heat and produce stronger RATs on sub-micron grains close to the protostars than the radiation field from the black body object. However, as grains inside the disk face very efficient gaseous damping \footnote{We note the extra damping caused by the viscous heating inside the accretion disk is also considered when modeling the gas damping near the protostar (Section \ref{sec:modelling_mcrt}, we consider $T_{\rm g} = T_{\rm d,rt} + T_{\rm g,mhd}$}), such enhanced RATs acting on grains will not strongly improve the alignment of grains there. If dust grains close to the protostar are heated to higher temperatures when exposed to the SED of the accretion shock, micron-sized grains and VLGs far from the radiation source can reach higher alignment degrees and produce higher dust polarization fraction by exposing to stronger IR dust emission.

\subsection{Influence of dust model on dust polarization}\label{sec:discuss_dust}
In Figure \ref{fig:CASA_inelastic_muQ_amax}, we found a similar positive correlation between the observed polarization degree and the maximum grain size as found in \cite{Valdivia_2019} and \cite{Valentin_2023b}. These studies also show that when grains grow beyond $30\mum$, the relation reverts due to the inappropriate extrapolations of the emissivity of VLGs beyond $30\mum$. The revert correlation is also seen in our model (Figure \ref{fig:CASA_inelastic_muQ_amax}, right panel). However, it happens when grains grow beyond $50\mum$ and is caused by the dominance of inefficient aligned VLGs. On the other hand, while \cite{Valdivia_2019} indicated that grains must grow beyond $10\mum$ to reproduce $p \geq 5\%$ revealed by ALMA in the envelope, our model showed that $1\mum$ grains also can reproduce this polarization level by MRAT mechanism (Figure \ref{fig:CASA_amax} and the right panel of Figure \ref{fig:CASA_inelastic_muQ_amax}). The reason for the difference between two results is the alignment dust component, which is assigned to only silicate grains in studies of \cite{Valdivia_2019} and \cite{Valentin_2023b} (named as model A), while we consider both silicate and graphite components following the nature of composite dust model (named as model B). This assumption thus makes our obtained polarization fraction always twice times higher than that of model A. Given such a correlation, dust grains in model A must contain higher numbers of iron inclusions than model B to reproduce $1 - 8\%$ of dust polarization observed by ALMA toward the inner 500 au (Figure \ref{fig:CASA_inelastic_Ncl}). However, dust grains may not hold more than $\sim 10^{4}$ iron atoms per cluster to still be SPM (Section \ref{sec:discuss_increase_Ncl}). Iron clusters also cannot hold more than $30\%$ of the grain volume (which corresponds to the maximum $90\%$ of iron abundance depleted from the gas phase, \citealt{Jenkins.2009}), which limits the existence of model A inside the inner envelope and the disk. Furthermore, it is hard to maintain the separation of silicate and graphite grains inside such dense, turbulent disk environments.

 \begin{figure*}
\centering
    \includegraphics[width=\textwidth,height=\textheight,keepaspectratio]{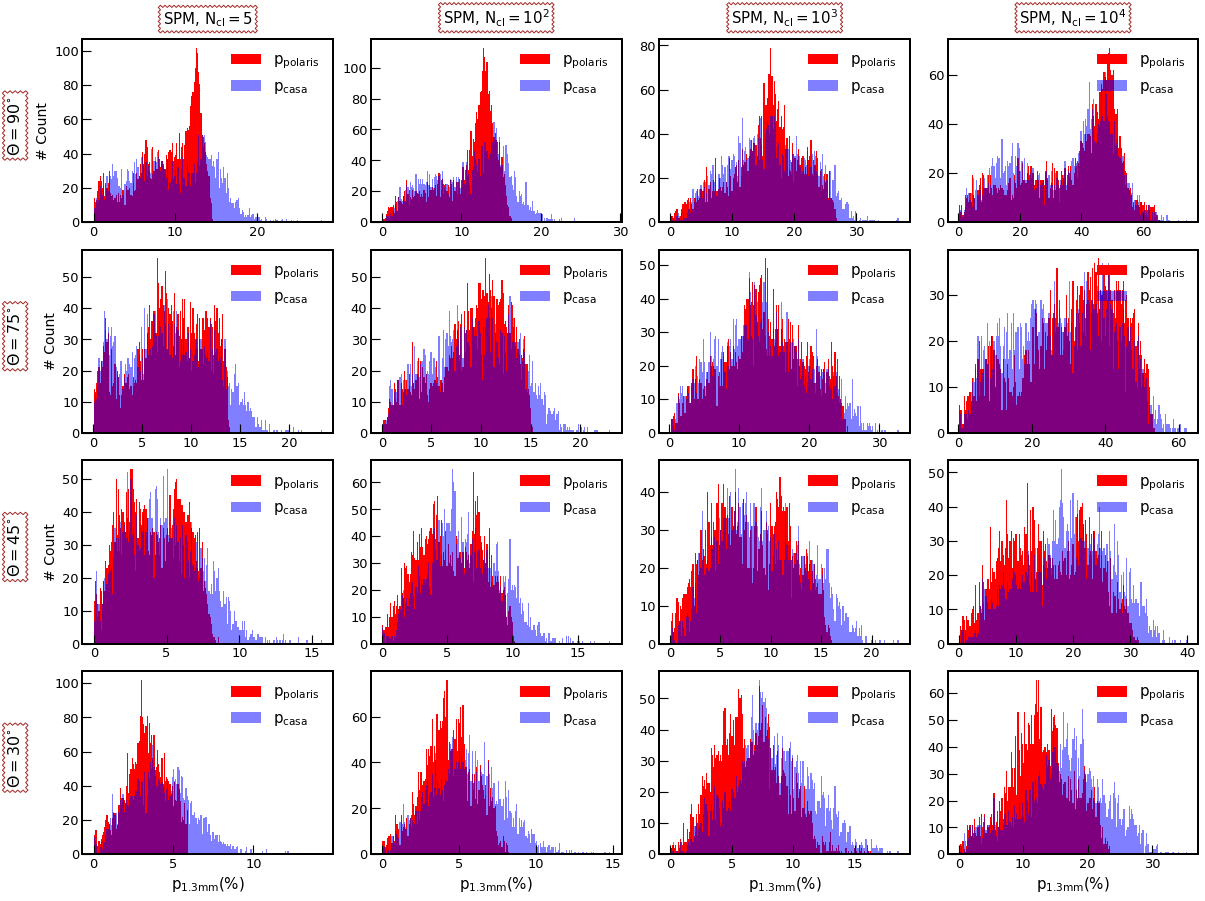}
    \caption{Histogram of the polarization degree obtained in the entire map recovered by CASA (see the map in Appendix \ref{sec:appen_filter}). We show in red results obtained from POLARIS and in blue results after post-processing with CASA for different combinations of $N_{\rm cl}$ and $\Theta$. Generally, the interferometric filtering amplifies the maximum intrinsic polarization degree found in the envelope by $\sim 5-10\%$. The combination between the emission from SPM grains with $N_{\rm cl} \sim 10^{2} - 10^{4}$ (which produces the intrinsic $p \sim 10-20\%$) and the amplification by ALMA filtering explains well why we obtain very high $p \sim 20-40\%$ in the envelope of low/intermediate Class 0/I YSOs.}
     \label{fig:filtering_Ncl_theta}
\end{figure*}

Regardless of the difference in the choice of dust component producing polarized dust emission, models A and B are built based on the \cite{Draine_Lee_1984} dust model, which is a mixture of astronomical silicate and graphite grains. However, given the turbulent nature of the ISM and the complex evolution of dust grains from the dust formation zones around Asymmetric Giant Branch/Red Giant Branch/Supernova Remnants to the dense protostellar cores and disks, it is unreasonable for grains having simple compact, separate/composite structure. Therefore, our adopted dust model may lead to some inaccurate predictions of polarized dust emission properties.

Astrodust \citep{Draine_Hensley_2021c,Draine_Hensley_2021b,Draine_Hensley_2021a} and THEMIS1 \citep{Jones_2013,Jones_2017}/THEMIS2(\citealt{Ysard_2024}) are two replacements that potentially better describes the realistic physical properties of dust grains in ISM, and maybe denser star-forming regions. Difference from \cite{Draine_Lee_1984} model, the Astrodust model has information on the porosity of a mixture of silicate, graphite, and iron inclusions. In the meanwhile, THEMIS1/THEMIS2 dust has the core/mantle structure motivated by observation \citep{Ishii_2018, David_2015} and all accretion and photo-processing processes from UV radiation acting on dust grains during their life-cycle in the ISM \citep{Jones_2013,  Ysard_2015}. However, the maximum grain size of both Astrodust and THEMIS/THEMIS2 grains currently is about $5\mum$ and $2\mum$, which is not enough to model both thermal and polarized dust emission in protostellar cores, where grains can grow beyond $10\mum$. Further development of THEMIS dust model toward large micron-sized and VLGs in dense star-forming regions is required to seek the light in accurately constraining the physical properties of dust grains around the protostar.

The grain aspect ratio $s$ contributes to a minor effect on the grain alignment dynamic (\citealt{Hoang+2022}), but clearly induces smaller observed polarization fraction if grains become more spherical (\citealt{Draine_Hensley_2021a}, \citealt{Draine_Hensley_2021c}) toward the center (\citealt{Juarez_2017}). However, \cite{Hoang_2022} recently found that if dust grains can have the magnetic alignment, the hit-and-stick of grain monomers drifting perpendicular to $\B$ can increase the elongation of large grains. The match between our synthetic results with oblate grains having $s = 0.5$ and ALMA observation in 500 au scales may support this scenario. The grain aspect ratio may affect more $p(\%)$ in the envelope where the grain growth is limited to micron-sized grains (\citealt{Wong_2016}, \citealt{Lebreuilly_2023}, \citealt{Marchand_2023}), given the typical aspect ratio (for sub-micron grains) found inside the ISM is $s = 0.7143$ (or s = 1.4 in
\citealt{Draine_Hensley_2021b} if defining s to be the ratio of the major to minor axis). Besides the change in the grain elongation, the increasing porosity of large grains by grain coagulation process inside protostellar environments (\citealt{Garcia_2020}, \citealt{Michoulier_2024}) also induces lower polarization degrees than expected for compact grains in our study (\citealt{Kirchschlarger_2019}). Given the size- and location-dependence porosity around the protostar, further detailed modeling of the alignment dynamics for realistic protostellar grains is required to accurately understand the influence of dust properties on dust polarization.

\subsection{Effect of ALMA interferometric filtering on dust polarization}\label{sec:discuss_filter}
The last factor affecting the observed polarization degree is the interferometric filtering of ALMA observation mentioned in \cite{Valentin_2020} and \cite{Valentin_2023b}. We show in Figure \ref{fig:filtering_Ncl1e3} the scattered map of the polarization degree at 1.3mm as a function of normalized gas column density, for results obtained from POLARIS $p_{\rm polaris}$ (left panel) and results after post-processing with CASA $p_{\rm casa}$ (right panel). The colorbar counts the number of points inside each bin of $[p-N_{\rm H_{2}}/N_{\rm H_{2},max}]$. We consider results obtained from model RATA$-$INELASTIC of SPM grains with $N_{\rm cl} = 10^{3}$, $\mu Q = 3\times 10^{9}\erg\cm^{-3}$, $a_{\rm max} = 50\mu m$, and $\Theta = 90^{\circ}$ (Figure \ref{fig:CASA_best_fit}). As expected in \cite{Valentin_2020} and \cite{Valentin_2023b}, ALMA filtering amplifies the intrinsic polarization degree in the entire protostellar core, especially in the outermost, low-intensity region owing to the strong leakage of the extended Stokes I emission (see the polarization degree map after filtering with CASA in Appendix \ref{sec:appen_filter}). For grains with $N_{\rm cl} = 10^{3}$, the maximum polarization fraction increases from $\sim 20\%$ to $p \sim 40\%$. However, the number of amplified points is $< 10$, which explains why the mean variation of $p(\%)$ with gas density is near unchanged under the filtering effect (see Figures \ref{fig:CASA_best_fit}, \ref{fig:CASA_amax}, \ref{fig:CASA_Ncl}).

Regarding very high polarization fraction with $p \sim 20-30\%$ observed in the envelope of Class 0/I YSOs by ALMA (\citealt{Hull_2017b}, \citealt{Maury_2018}, \citealt{Kwon_2019}), \cite{Valentin_2020} suggested that it may be an artifact caused by the interferometric filtering. However, numerical modeling in \cite{Giang_2023a} and \cite{Giang_2023b} showed that such a high polarization can be explained by SPM grains with $N_{\rm cl} = 10^{4}$ and $a_{\rm max} \geq 50\mum$ which are aligned via the MRAT mechanism. To understand how much the filtering amplifies $p(\%)$ obtained in the envelope, we show in Figure \ref{fig:filtering_Ncl_theta} the histogram of $p_{\rm polaris}$ and $p_{\rm casa}$ for different combinations of $N_{\rm cl}$ and $\Theta$. Generally, ALMA increases the intrinsic polarization degree to about $5-10\%$. But to be more detailed, it cannot amplify $p_{\rm polaris,max} \sim 15\%$ produced by SPM grains with low $N_{\rm cl} = 5$ to $30\%$. On the other hand, with $\Theta = 90^{\circ}$, the polarization fraction produced by SPM grains with $N_{\rm cl} = 10^{4}$ (upper right panel) can be amplified from $\sim 50\%$ to $\sim 60\%$ by ALMA filtering, which is far beyond the maximum observed $p \sim 30\%$. Given lower $\Theta <30^{\circ}$ or lower $N_{\rm cl} \leq 10^{3}$, the combination between the emission of SPM grains (which produce the maximum intrinsic $p_{\rm polaris} \sim 10-20\%$) and filtering effect can explain well the maximum observed $\sim 20-30\%$ of polarization. Indeed, the number of cells that show very high $p > 15\%$ only occupy a small part of the envelope recovered by ALMA (see the statistic distribution of $p(\%)$ in 13 low mass Class 0/I YSOs at 1.3 mm in \citealt{Valentin_2020}, Figure 1). However, it is still meaningful because it indirectly tells us about the presence of SPM grains with large iron inclusions in protostellar environments, similar to what concluded in Sections \ref{sec:ALMA_observations}, \ref{sec:discuss_envelope}, and \ref{sec:discuss_inner_envelope}.

\section{Summary}\label{sec:summary}
In paper I, we focus on understanding the grain alignment mechanisms and constraining the grain physical properties inside intermediate Class 0/I YSOs. We first perform the synthetic modeling of polarized dust emission with the updated POLARIS code from \cite{Giang_2023a}, considering all well-known alignment mechanisms such as Barnett relaxation, inelastic relaxation, RATs, and MRAT mechanism. The synthetic results are then post-processed with CASA and confronted with ALMA observations of low/intermediate Class 0/I YSOs. We summarize our findings as follows:

 \begin{enumerate}
\item  By considering that $30\%$ of iron abundance is locked inside dust grains under the cluster form, we found that the high polarization degree $p \sim 5 - 30\%$ observed in the envelope by ALMA can be explained by the emission of superparamagnetic grains (SPM grains) of size above $1\mum$ with $N_{\rm cl} \sim  5 - 10^{4}$ iron atoms/cluster. With these configurations, almost micron-sized and VLGs in the envelope can achieve efficient internal alignment (IA) by super-Barnett/inelastic relaxation and be aligned with $\B$ by MRAT mechanism with $f_{\rm high-J} \sim 0.5-1$. 

\item  Inside the inner envelope, the outflow cavity wall, and the equatorial midplane, the joint effect of inelastic relaxation, large iron inclusions with $N_{\rm cl} = 10^{2}-10^{4}$, and grain grow beyond $10\mum$ are required to maintain the efficient IA and MRAT alignment of dust grains in the region with high density of $n_{\rm H_{2}}\geq 10^{7}\cm^{-3}$. Such a wide alignment range from micron-sized up to VLGs with efficient magnetic alignment can explain the observed range of $p \sim 1-8\%$ by ALMA toward this area.  

\item However, our study shows that, because of increasing gaseous damping at higher densities, both the IA and MRAT mechanisms remain inefficient to align a large fraction of the micron-sized and VLGs grains, even when increasing the amount of iron inclusions. This could explain the low polarization fractions observed by ALMA at disk scale.

\item We show that levels $p >1\%$ of dust polarization in optically thin wavelengths at disk scales require to include SPM grains with high $N_{\rm cl} = 10^{4}, \phi_{\rm sp}\geq 0.1$ and grains with large sizes (beyond $\geq 50 \mu m$). But such a low polarization fraction is very easy to be feather or even dominant by self-scattering, depending on the comparison of grain size with the observed wavelengths. Besides, our current RAT/MRAT alignment model fails to explain the dichroic extinction feature with high $p \geq 2-10\%$ observed in the inner 100 au region of OMC3 MMS6 and IRAS4A disks. Further studies that include Mechanical Torques (METs), RATs on accurate dust models (potentially Astrodust, THEMIS, THEMIS2 dust model), and grain dynamics (for grains with slow internal relaxation) must be performed to accurately interpret the complicated polarization features observed toward Class 0/I disks.

\item We examine the effect of bolometric luminosity $L_{\rm center}$ on the grain alignment and found that excepting cause the increase of the minimum alignment size, $L_{\rm center}$ does not reduce the internal and external alignment driven by super-Barnett relaxation/inelastic relaxation/MRAT alignment acting on micron-sized and VLGs. We found that if VLGs up to $a_{\rm max} = 50\mum$ can distribute over the envelope to the inner envelope, MRAT alignment always can reproduce ALMA dust polarization fraction regardless of $L_{\rm center}$. But if $a_{\rm max}$ is reduced to below $10\mum$ beyond 100 au, the center luminosity must exceed $\sim 20L_{\odot}$ to explain the high $p \sim 1-10\%$ observed in low/intermediate Class 0 YSOs as found in \cite{Valentin_2023b}.

\item  Lastly, we found that the detection of $20-30\%$ of dust polarization by ALMA in the protostellar envelope is not entirely an artifact usually attributed to the interferometric filtering effect. Such very high polarization signal could be reproduced by thermal emission of aligned SPM grains with $N_{\rm cl} \sim 10^{2} - 10^{3}$, which produces the intrinsic polarization degree of $p \sim 10-15\%$, and the ALMA filtering effect, which further amplifies the intrinsic $p \sim 10-15\%$ to $p \sim 20-30\%$. Besides, we show that the inclination angle can change the observed polarization fraction inside the envelope twice (when changing the observed view from the edge$-$on to face$-$on direction). Thus, it is very important to constrain the 3D magnetic fields in the envelope scale to probe the magnetic properties of dust grains there accurately.

\end{enumerate}
\section*{Acknowledgements}
T.H. is supported by the National Research Foundation of Korea (NRF) grant funded by the Korean government (MSIT (No. 2019R1A2C1087045). This work was partly supported by a grant from the Simons Foundation to IFIRSE, ICISE (916424, N.H.). We appreciate the useful discussion from Professors Alex Lazarian and Michael Efroimsky about the inelastic relaxation on dust grain alignment.  

\bibliography{ms}

\begin{thebibliography}{}
\expandafter\ifx\csname natexlab\endcsname\relax\def\natexlab#1{#1}\fi
\providecommand{\url}[1]{\href{#1}{#1}}
\providecommand{\dodoi}[1]{doi:~\href{http://doi.org/#1}{\nolinkurl{#1}}}
\providecommand{\doeprint}[1]{\href{http://ascl.net/#1}{\nolinkurl{http://ascl.net/#1}}}
\providecommand{\doarXiv}[1]{\href{https://arxiv.org/abs/#1}{\nolinkurl{https://arxiv.org/abs/#1}}}

\bibitem[{Andersson {et~al.}(2024)Andersson, Karoly, Bastien, Soam, Coudé, Tahani, Gordon, \& Fox-Middleton}]{Andersson_2024}
Andersson, B.-G., Karoly, J., Bastien, P., {et~al.} 2024, The Astrophysical Journal, 963, 76, \dodoi{10.3847/1538-4357/ad1835}

\bibitem[{{Andersson} {et~al.}(2015){Andersson}, {Lazarian}, \& {Vaillancourt}}]{Anderson_2015}
{Andersson}, B.~G., {Lazarian}, A., \& {Vaillancourt}, J.~E. 2015, \araa, 53, 501, \dodoi{10.1146/annurev-astro-082214-122414}

\bibitem[{Andersson {et~al.}(2022)Andersson, Lopez-Rodriguez, Medan, Soam, Hoang, Vaillancourt, Lazarian, Sandin, Mattsson, \& Tahani}]{Andersson_2022}
Andersson, B.-G., Lopez-Rodriguez, E., Medan, I., {et~al.} 2022, \apj, 931, 80, \dodoi{10.3847/1538-4357/ac64a4}

\bibitem[{{Antoniucci} {et~al.}(2008){Antoniucci}, {Nisini}, {Giannini}, \& {Lorenzetti}}]{Antoniucci_2008}
{Antoniucci}, S., {Nisini}, B., {Giannini}, T., \& {Lorenzetti}, D. 2008, \aap, 479, 503, \dodoi{10.1051/0004-6361:20077468}

\bibitem[{Barnett(1915)}]{Barnett_1915}
Barnett, S.~J. 1915, Phys. Rev., 6, 239, \dodoi{10.1103/PhysRev.6.239}

\bibitem[{Bean \& Livingston(1959)}]{Bean_1959}
Bean, C.~P., \& Livingston, J.~D. 1959, Journal of Applied Physics, 30, S120, \dodoi{10.1063/1.2185850}

\bibitem[{{Bleuler} \& {Teyssier}(2014)}]{Bleuler_2014}
{Bleuler}, A., \& {Teyssier}, R. 2014, \mnras, 445, 4015, \dodoi{10.1093/mnras/stu2005}

\bibitem[{{Bracco} {et~al.}(2017){Bracco}, {Palmeirim}, {Andr{\'e}}, {Adam}, {Ade}, {Bacmann}, {Beelen}, {Beno{\^\i}t}, {Bideaud}, {Billot}, {Bourrion}, {Calvo}, {Catalano}, {Coiffard}, {Comis}, {D'Addabbo}, {D{\'e}sert}, {Didelon}, {Doyle}, {Goupy}, {K{\"o}nyves}, {Kramer}, {Lagache}, {Leclercq}, {Mac{\'\i}as-P{\'e}rez}, {Maury}, {Mauskopf}, {Mayet}, {Monfardini}, {Motte}, {Pajot}, {Pascale}, {Peretto}, {Perotto}, {Pisano}, {Ponthieu}, {Rev{\'e}ret}, {Rigby}, {Ritacco}, {Rodriguez}, {Romero}, {Roy}, {Ruppin}, {Schuster}, {Sievers}, {Triqueneaux}, {Tucker}, \& {Zylka}}]{Bracco_2017}
{Bracco}, A., {Palmeirim}, P., {Andr{\'e}}, P., {et~al.} 2017, \aap, 604, A52, \dodoi{10.1051/0004-6361/201731117}

\bibitem[{{Brauer} {et~al.}(2016){Brauer}, {Wolf}, \& {Reissl}}]{Brauer_2016}
{Brauer}, R., {Wolf}, S., \& {Reissl}, S. 2016, \aap, 588, A129, \dodoi{10.1051/0004-6361/201527546}

\bibitem[{{Cacciapuoti} {et~al.}(2023){Cacciapuoti}, {Macias}, {Maury}, {Chandler}, {Sakai}, {Tychoniec}, {Viti}, {Natta}, {De Simone}, {Miotello}, {Codella}, {Ceccarelli}, {Podio}, {Fedele}, {Johnstone}, {Shirley}, {Liu}, {Bianchi}, {Zhang}, {Pineda}, {Loinard}, {M{\'e}nard}, {Lebreuilly}, {Klessen}, {Hennebelle}, {Molinari}, {Testi}, \& {Yamamoto}}]{Cacciapuoti_2023a}
{Cacciapuoti}, L., {Macias}, E., {Maury}, A.~J., {et~al.} 2023, \aap, 676, A4, \dodoi{10.1051/0004-6361/202346204}

\bibitem[{Chen {et~al.}(2016)Chen, King, \& Li}]{Chen_2016}
Chen, C.-Y., King, P.~K., \& Li, Z.-Y. 2016, The Astrophysical Journal, 829, 84, \dodoi{10.3847/0004-637X/829/2/84}

\bibitem[{{Cheng} {et~al.}(2022){Cheng}, {Gutermuth}, {Offner}, {Heyer}, {Zinnecker}, {Megeath}, \& {Pokhrel}}]{Cheng_2022}
{Cheng}, Y., {Gutermuth}, R.~A., {Offner}, S., {et~al.} 2022, \mnras, 512, 960, \dodoi{10.1093/mnras/stac436}

\bibitem[{Chuss {et~al.}(2019)Chuss, Andersson, Bally, Dotson, Dowell, Guerra, Harper, Houde, Jones, Lazarian, Rodriguez, Michail, Morris, Novak, Siah, Staguhn, Vaillancourt, Volpert, Werner, Wollack, Benford, Berthoud, Cox, Crutcher, Dale, Fissel, Goldsmith, Hamilton, Hanany, Henning, Looney, Moseley, Santos, Stephens, Tassis, Trinh, Camp, Ward-Thompson, \& Team)}]{Chuss_2019}
Chuss, D.~T., Andersson, B.-G., Bally, J., {et~al.} 2019, The Astrophysical Journal, 872, 187, \dodoi{10.3847/1538-4357/aafd37}

\bibitem[{Coey(2010)}]{Coey_2010}
Coey, J. M.~D. 2010, Magnetism and Magnetic Materials (Cambridge University Press)

\bibitem[{{Costantini} {et~al.}(2005){Costantini}, {Freyberg}, \& {Predehl}}]{Costantini_2005}
{Costantini}, E., {Freyberg}, M.~J., \& {Predehl}, P. 2005, \aap, 444, 187, \dodoi{10.1051/0004-6361:20042562}

\bibitem[{Cox {et~al.}(2018)Cox, Harris, Looney, Li, Yang, Tobin, \& Stephens}]{Cox_2018}
Cox, E.~G., Harris, R.~J., Looney, L.~W., {et~al.} 2018, The Astrophysical Journal, 855, 92, \dodoi{10.3847/1538-4357/aaacd2}

\bibitem[{{Cox} {et~al.}(2015){Cox}, {Harris}, {Looney}, {Segura-Cox}, {Tobin}, {Li}, {Tychoniec}, {Chandler}, {Dunham}, {Kratter}, {Melis}, {Perez}, \& {Sadavoy}}]{Cox_2015}
{Cox}, E.~G., {Harris}, R.~J., {Looney}, L.~W., {et~al.} 2015, \apjl, 814, L28, \dodoi{10.1088/2041-8205/814/2/L28}

\bibitem[{{Davidson} {et~al.}(2015){Davidson}, {Nittler}, {Stroud}, {Takigawa}, {De Gregorio}, {Alexander}, {Kilcoyne}, \& {Cody}}]{David_2015}
{Davidson}, J., {Nittler}, L.~R., {Stroud}, R.~M., {et~al.} 2015, in 46th Annual Lunar and Planetary Science Conference, Lunar and Planetary Science Conference, 1609

\bibitem[{Davis~Jr \& Greenstein(1951)}]{David_1951}
Davis~Jr, L., \& Greenstein, J.~L. 1951, \apj, 114, 206

\bibitem[{{Davoisne} {et~al.}(2006){Davoisne}, {Djouadi}, {Leroux}, {D'Hendecourt}, {Jones}, \& {Deboffle}}]{Davoisne_2006}
{Davoisne}, C., {Djouadi}, Z., {Leroux}, H., {et~al.} 2006, \aap, 448, L1, \dodoi{10.1051/0004-6361:200600002}

\bibitem[{{Dolginov} \& {Mitrofanov}(1976)}]{Dolginov_1976}
{Dolginov}, A.~Z., \& {Mitrofanov}, I.~G. 1976, \apss, 43, 291, \dodoi{10.1007/BF00640010}

\bibitem[{{Draine}(1996)}]{Draine_1996}
{Draine}, B.~T. 1996, in Astronomical Society of the Pacific Conference Series, Vol.~97, Polarimetry of the Interstellar Medium, ed. W.~G. {Roberge} \& D.~C.~B. {Whittet}, 16, \dodoi{10.48550/arXiv.astro-ph/9603053}

\bibitem[{{Draine} \& {Hensley}(2021{\natexlab{a}})}]{Draine_Hensley_2021a}
{Draine}, B.~T., \& {Hensley}, B.~S. 2021{\natexlab{a}}, \apj, 919, 65, \dodoi{10.3847/1538-4357/ac0050}

\bibitem[{{Draine} \& {Hensley}(2021{\natexlab{b}})}]{Draine_Hensley_2021c}
---. 2021{\natexlab{b}}, \apj, 909, 94, \dodoi{10.3847/1538-4357/abd6c6}

\bibitem[{{Draine} \& {Hensley}(2021{\natexlab{c}})}]{Draine_Hensley_2021b}
---. 2021{\natexlab{c}}, \apj, 910, 47, \dodoi{10.3847/1538-4357/abddb7}

\bibitem[{Draine \& Lee(1984)}]{Draine_Lee_1984}
Draine, B.~T., \& Lee, H.~M. 1984, \apj, 285, 89

\bibitem[{{Efroimsky}(2000)}]{Efroimky_2000}
{Efroimsky}, M. 2000, Journal of Mathematical Physics, 41, 1854, \dodoi{10.1063/1.533216}

\bibitem[{{Efroimsky} \& {Lazarian}(2000)}]{Efroimsky_Lazarian_2000}
{Efroimsky}, M., \& {Lazarian}, A. 2000, \mnras, 311, 269, \dodoi{10.1046/j.1365-8711.2000.03036.x}

\bibitem[{{Elbakyan} {et~al.}(2016){Elbakyan}, {Vorobyov}, \& {Glebova}}]{Elbakyan_2016}
{Elbakyan}, V.~G., {Vorobyov}, E.~I., \& {Glebova}, G.~M. 2016, Astronomy Reports, 60, 879, \dodoi{10.1134/S1063772916100012}

\bibitem[{{Frank} {et~al.}(2014){Frank}, {Ray}, {Cabrit}, {Hartigan}, {Arce}, {Bacciotti}, {Bally}, {Benisty}, {Eisl{\"o}ffel}, {G{\"u}del}, {Lebedev}, {Nisini}, \& {Raga}}]{Frank_2014}
{Frank}, A., {Ray}, T.~P., {Cabrit}, S., {et~al.} 2014, in Protostars and Planets VI, ed. H.~{Beuther}, R.~S. {Klessen}, C.~P. {Dullemond}, \& T.~{Henning}, 451--474, \dodoi{10.2458/azu_uapress_9780816531240-ch020}

\bibitem[{{Frau} {et~al.}(2011){Frau}, {Galli}, \& {Girart}}]{Frau_2011}
{Frau}, P., {Galli}, D., \& {Girart}, J.~M. 2011, \aap, 535, A44, \dodoi{10.1051/0004-6361/201117813}

\bibitem[{{Furlan} {et~al.}(2016){Furlan}, {Fischer}, {Ali}, {Stutz}, {Stanke}, {Tobin}, {Megeath}, {Osorio}, {Hartmann}, {Calvet}, {Poteet}, {Booker}, {Manoj}, {Watson}, \& {Allen}}]{Furlan_2016}
{Furlan}, E., {Fischer}, W.~J., {Ali}, B., {et~al.} 2016, \apjs, 224, 5, \dodoi{10.3847/0067-0049/224/1/5}

\bibitem[{{Galametz} {et~al.}(2019){Galametz}, {Maury}, {Valdivia}, {Testi}, {Belloche}, \& {Andr{\'e}}}]{Galametz_2019}
{Galametz}, M., {Maury}, A.~J., {Valdivia}, V., {et~al.} 2019, \aap, 632, A5, \dodoi{10.1051/0004-6361/201936342}

\bibitem[{{Galli} {et~al.}(2006){Galli}, {Lizano}, {Shu}, \& {Allen}}]{Galli_2006}
{Galli}, D., {Lizano}, S., {Shu}, F.~H., \& {Allen}, A. 2006, \apj, 647, 374, \dodoi{10.1086/505257}

\bibitem[{{Garcia} \& {Gonzalez}(2020)}]{Garcia_2020}
{Garcia}, A. J.~L., \& {Gonzalez}, J.-F. 2020, \mnras, 493, 1788, \dodoi{10.1093/mnras/staa382}

\bibitem[{Giang(2024)}]{POLARIS_link}
Giang, N.~C. 2024, {Updated POLARIS code with detailed grain alignment physics and radiative torques disruption - a tool for synthetic dust polarization and modeling grain dynamic in star-forming regions.}, v1,  Zenodo, \dodoi{10.5281/zenodo.14199014}

\bibitem[{{Giang} \& {Hoang}(2023)}]{Giang_2023b}
{Giang}, N.~C., \& {Hoang}, T. 2023, arXiv e-prints, arXiv:2307.16829, \dodoi{10.48550/arXiv.2307.16829}

\bibitem[{Giang {et~al.}(2023)Giang, Hoang, Kim, \& Tram}]{Giang_2023a}
Giang, N.~C., Hoang, T., Kim, J.-G., \& Tram, L.~N. 2023, \mnras, 520, 3788, \dodoi{10.1093/mnras/stad020}

\bibitem[{{Girart} {et~al.}(2006){Girart}, {Rao}, \& {Marrone}}]{Girart_2006}
{Girart}, J.~M., {Rao}, R., \& {Marrone}, D.~P. 2006, Science, 313, 812, \dodoi{10.1126/science.1129093}

\bibitem[{{Greenberg}(1968)}]{Greenberg_1968}
{Greenberg}, J.~M. 1968, in Nebulae and Interstellar Matter, ed. B.~M. {Middlehurst} \& L.~H. {Aller}, 221

\bibitem[{{Gullbring} {et~al.}(1998){Gullbring}, {Hartmann}, {Brice{\~n}o}, \& {Calvet}}]{GULLBRING_1998}
{Gullbring}, E., {Hartmann}, L., {Brice{\~n}o}, C., \& {Calvet}, N. 1998, \apj, 492, 323, \dodoi{10.1086/305032}

\bibitem[{{Herranen} {et~al.}(2021){Herranen}, {Lazarian}, \& {Hoang}}]{Herranen_2021}
{Herranen}, J., {Lazarian}, A., \& {Hoang}, T. 2021, \apj, 913, 63, \dodoi{10.3847/1538-4357/abf096}

\bibitem[{{Hoang}(2021)}]{Hoang_2021}
{Hoang}, T. 2021, \apj, 921, 21, \dodoi{10.3847/1538-4357/ac185d}

\bibitem[{Hoang(2022)}]{Hoang_2022}
Hoang, T. 2022, \apj, 928, 102, \dodoi{10.3847/1538-4357/ac5408}

\bibitem[{Hoang {et~al.}(2018)Hoang, Cho, \& Lazarian}]{Hoang_2018}
Hoang, T., Cho, J., \& Lazarian, A. 2018, The Astrophysical Journal, 852, 129, \dodoi{10.3847/1538-4357/aa9edc}

\bibitem[{Hoang \& Lazarian(2008)}]{Hoang_Lazarian_2008}
Hoang, T., \& Lazarian, A. 2008, \mnras, 388, 117, \dodoi{10.1111/j.1365-2966.2008.13249.x}

\bibitem[{Hoang \& Lazarian(2009)}]{Hoang_Lazarian_2009}
---. 2009, \apj, 697, 1316, \dodoi{10.1088/0004-637x/697/2/1316}

\bibitem[{{Hoang} \& {Lazarian}(2014)}]{Hoang_Lazarian_2014}
{Hoang}, T., \& {Lazarian}, A. 2014, \mnras, 438, 680, \dodoi{10.1093/mnras/stt2240}

\bibitem[{Hoang \& Lazarian(2016)}]{Hoang_Lazarian_2016}
Hoang, T., \& Lazarian, A. 2016, The Astrophysical Journal, 821, 91, \dodoi{10.3847/0004-637x/821/2/91}

\bibitem[{{Hoang} \& {Lazarian}(2016)}]{Hoang_Lazarian_2016_mrat}
{Hoang}, T., \& {Lazarian}, A. 2016, \apj, 831, 159, \dodoi{10.3847/0004-637X/831/2/159}

\bibitem[{Hoang {et~al.}(2023)Hoang, Phan, \& Tram}]{Hoang_2023}
Hoang, T., Phan, V. H.~M., \& Tram, L.~N. 2023, The Astrophysical Journal, 954, 216, \dodoi{10.3847/1538-4357/ace788}

\bibitem[{{Hoang} {et~al.}(2019){Hoang}, {Tram}, {Lee}, \& {Ahn}}]{Hoang_2019}
{Hoang}, T., {Tram}, L.~N., {Lee}, H., \& {Ahn}, S.-H. 2019, Nature Astronomy, 3, 766, \dodoi{10.1038/s41550-019-0763-6}

\bibitem[{{Hoang} {et~al.}(2021){Hoang}, {Tram}, {Lee}, {Diep}, \& {Ngoc}}]{Hoang+2021}
{Hoang}, T., {Tram}, L.~N., {Lee}, H., {Diep}, P.~N., \& {Ngoc}, N.~B. 2021, \apj, 908, 218, \dodoi{10.3847/1538-4357/abd54f}

\bibitem[{{Hoang} {et~al.}(2022){Hoang}, {Tram}, {Minh Phan}, {Giang}, {Phuong}, \& {Dieu}}]{Hoang+2022}
{Hoang}, T., {Tram}, L.~N., {Minh Phan}, V.~H., {et~al.} 2022, \aj, 164, 248, \dodoi{10.3847/1538-3881/ac9af5}

\bibitem[{Hoang \& Truong(2024)}]{Hoang_Bao_2024}
Hoang, T., \& Truong, B. 2024, The Astrophysical Journal, 965, 183, \dodoi{10.3847/1538-4357/ad2a56}

\bibitem[{{Huang} {et~al.}(2024){Huang}, {Girart}, {Stephens}, {Fern{\'a}ndez L{\'o}pez}, {Arce}, {Carpenter}, {Cortes}, {Cox}, {Friesen}, {Le Gouellec}, {Hull}, {Karnath}, {Kwon}, {Li}, {Looney}, {Megeath}, {Myers}, {Murillo}, {Pineda}, {Sadavoy}, {S{\'a}nchez-Monge}, {Sanhueza}, {Tobin}, {Zhang}, {Jackson}, \& {Segura-Cox}}]{Huang_2024}
{Huang}, B., {Girart}, J.~M., {Stephens}, I.~W., {et~al.} 2024, \apjl, 963, L31, \dodoi{10.3847/2041-8213/ad27d4}

\bibitem[{{Hull} {et~al.}(2020){Hull}, {Le Gouellec}, {Girart}, {Tobin}, \& {Bourke}}]{Hull_Valentin_2020}
{Hull}, C. L.~H., {Le Gouellec}, V. J.~M., {Girart}, J.~M., {Tobin}, J.~J., \& {Bourke}, T.~L. 2020, \apj, 892, 152, \dodoi{10.3847/1538-4357/ab5809}

\bibitem[{{Hull} \& {Zhang}(2019)}]{Hull_2019}
{Hull}, C. L.~H., \& {Zhang}, Q. 2019, Frontiers in Astronomy and Space Sciences, 6, 3, \dodoi{10.3389/fspas.2019.00003}

\bibitem[{Hull {et~al.}(2014)Hull, Plambeck, Kwon, Bower, Carpenter, Crutcher, Fiege, Franzmann, Hakobian, Heiles, Houde, Hughes, Lamb, Looney, Marrone, Matthews, Pillai, Pound, Rahman, Sandell, Stephens, Tobin, Vaillancourt, Volgenau, \& Wright}]{Hull_2014}
Hull, C. L.~H., Plambeck, R.~L., Kwon, W., {et~al.} 2014, The Astrophysical Journal Supplement Series, 213, 13, \dodoi{10.1088/0067-0049/213/1/13}

\bibitem[{{Hull} {et~al.}(2017){Hull}, {Girart}, {Tychoniec}, {Rao}, {Cort{\'e}s}, {Pokhrel}, {Zhang}, {Houde}, {Dunham}, {Kristensen}, {Lai}, {Li}, \& {Plambeck}}]{Hull_2017b}
{Hull}, C. L.~H., {Girart}, J.~M., {Tychoniec}, {\L}., {et~al.} 2017, \apj, 847, 92, \dodoi{10.3847/1538-4357/aa7fe9}

\bibitem[{{Hull} {et~al.}(2022){Hull}, {Yang}, {Cort{\'e}s}, {Dent}, {Kral}, {Li}, {Le Gouellec}, {Hughes}, {Milli}, {Teague}, \& {Wyatt}}]{Hull_2022}
{Hull}, C. L.~H., {Yang}, H., {Cort{\'e}s}, P.~C., {et~al.} 2022, \apj, 930, 49, \dodoi{10.3847/1538-4357/ac6023}

\bibitem[{{Ishii} {et~al.}(2018){Ishii}, {Bradley}, {Bechtel}, {Brownlee}, {Bustillo}, {Ciston}, {Cuzzi}, {Floss}, \& {Joswiak}}]{Ishii_2018}
{Ishii}, H.~A., {Bradley}, J.~P., {Bechtel}, H.~A., {et~al.} 2018, Proceedings of the National Academy of Science, 115, 6608, \dodoi{10.1073/pnas.1720167115}

\bibitem[{{J{\"a}ger} {et~al.}(2024){J{\"a}ger}, {Reissl}, \& {Klessen}}]{Jager_2024}
{J{\"a}ger}, J.~A., {Reissl}, S., \& {Klessen}, R.~S. 2024, arXiv e-prints, arXiv:2407.09968, \dodoi{10.48550/arXiv.2407.09968}

\bibitem[{Jenkins(2009)}]{Jenkins.2009}
Jenkins, E.~B. 2009, \apj, 700, 1299

\bibitem[{{Jones} {et~al.}(2013){Jones}, {Fanciullo}, {K{\"o}hler}, {Verstraete}, {Guillet}, {Bocchio}, \& {Ysard}}]{Jones_2013}
{Jones}, A.~P., {Fanciullo}, L., {K{\"o}hler}, M., {et~al.} 2013, \aap, 558, A62, \dodoi{10.1051/0004-6361/201321686}

\bibitem[{{Jones} {et~al.}(2017){Jones}, {K{\"o}hler}, {Ysard}, {Bocchio}, \& {Verstraete}}]{Jones_2017}
{Jones}, A.~P., {K{\"o}hler}, M., {Ysard}, N., {Bocchio}, M., \& {Verstraete}, L. 2017, \aap, 602, A46, \dodoi{10.1051/0004-6361/201630225}

\bibitem[{{Ju{\'a}rez} {et~al.}(2017){Ju{\'a}rez}, {Girart}, {Frau}, {Palau}, {Estalella}, {Morata}, {Alves}, {Beltr{\'a}n}, \& {Padovani}}]{Juarez_2017}
{Ju{\'a}rez}, C., {Girart}, J.~M., {Frau}, P., {et~al.} 2017, \aap, 597, A74, \dodoi{10.1051/0004-6361/201628608}

\bibitem[{{Kataoka} {et~al.}(2012){Kataoka}, {Machida}, \& {Tomisaka}}]{Kataoka_2012}
{Kataoka}, A., {Machida}, M.~N., \& {Tomisaka}, K. 2012, \apj, 761, 40, \dodoi{10.1088/0004-637X/761/1/40}

\bibitem[{{Kataoka} {et~al.}(2017){Kataoka}, {Tsukagoshi}, {Pohl}, {Muto}, {Nagai}, {Stephens}, {Tomisaka}, \& {Momose}}]{Kataoka_2017}
{Kataoka}, A., {Tsukagoshi}, T., {Pohl}, A., {et~al.} 2017, \apjl, 844, L5, \dodoi{10.3847/2041-8213/aa7e33}

\bibitem[{{Kataoka} {et~al.}(2015){Kataoka}, {Muto}, {Momose}, {Tsukagoshi}, {Fukagawa}, {Shibai}, {Hanawa}, {Murakawa}, \& {Dullemond}}]{Kataoka_2015}
{Kataoka}, A., {Muto}, T., {Momose}, M., {et~al.} 2015, \apj, 809, 78, \dodoi{10.1088/0004-637X/809/1/78}

\bibitem[{{Kataoka} {et~al.}(2016){Kataoka}, {Tsukagoshi}, {Momose}, {Nagai}, {Muto}, {Dullemond}, {Pohl}, {Fukagawa}, {Shibai}, {Hanawa}, \& {Murakawa}}]{Kataoka_2016}
{Kataoka}, A., {Tsukagoshi}, T., {Momose}, M., {et~al.} 2016, \apjl, 831, L12, \dodoi{10.3847/2041-8205/831/2/L12}

\bibitem[{{Kirchschlager} {et~al.}(2019){Kirchschlager}, {Bertrang}, \& {Flock}}]{Kirchschlarger_2019}
{Kirchschlager}, F., {Bertrang}, G. H.~M., \& {Flock}, M. 2019, \mnras, 488, 1211, \dodoi{10.1093/mnras/stz1763}

\bibitem[{{Knapmeyer} {et~al.}(2018){Knapmeyer}, {Fischer}, {Knollenberg}, {Seidensticker}, {Thiel}, {Arnold}, {Faber}, \& {M{\"o}hlmann}}]{Knapmeyer_2018}
{Knapmeyer}, M., {Fischer}, H.~H., {Knollenberg}, J., {et~al.} 2018, \icarus, 310, 165, \dodoi{10.1016/j.icarus.2017.12.002}

\bibitem[{{Ko} {et~al.}(2020){Ko}, {Liu}, {Lai}, {Ching}, {Rao}, \& {Girart}}]{Ko_2020}
{Ko}, C.-L., {Liu}, H.~B., {Lai}, S.-P., {et~al.} 2020, \apj, 889, 172, \dodoi{10.3847/1538-4357/ab5e79}

\bibitem[{{K{\"o}hler} {et~al.}(2014){K{\"o}hler}, {Jones}, \& {Ysard}}]{Kohler_2014}
{K{\"o}hler}, M., {Jones}, A., \& {Ysard}, N. 2014, \aap, 565, L9, \dodoi{10.1051/0004-6361/201423985}

\bibitem[{{Konigl} \& {Pudritz}(2000)}]{Konigl_2000}
{Konigl}, A., \& {Pudritz}, R.~E. 2000, in Protostars and Planets IV, ed. V.~{Mannings}, A.~P. {Boss}, \& S.~S. {Russell}, 759, \dodoi{10.48550/arXiv.astro-ph/9903168}

\bibitem[{{Krumholz} {et~al.}(2004){Krumholz}, {McKee}, \& {Klein}}]{Krumholz_2004}
{Krumholz}, M.~R., {McKee}, C.~F., \& {Klein}, R.~I. 2004, \apj, 611, 399, \dodoi{10.1086/421935}

\bibitem[{{Kwon} {et~al.}(2009){Kwon}, {Looney}, {Mundy}, {Chiang}, \& {Kemball}}]{Kwon_2009}
{Kwon}, W., {Looney}, L.~W., {Mundy}, L.~G., {Chiang}, H.-F., \& {Kemball}, A.~J. 2009, \apj, 696, 841, \dodoi{10.1088/0004-637X/696/1/841}

\bibitem[{Kwon {et~al.}(2019)Kwon, Stephens, Tobin, Looney, Li, van~der Tak, \& Crutcher}]{Kwon_2019}
Kwon, W., Stephens, I.~W., Tobin, J.~J., {et~al.} 2019, \apj, 879, 25, \dodoi{10.3847/1538-4357/ab24c8}

\bibitem[{{Lai} {et~al.}(2002){Lai}, {Crutcher}, {Girart}, \& {Rao}}]{Lai_2002}
{Lai}, S.-P., {Crutcher}, R.~M., {Girart}, J.~M., \& {Rao}, R. 2002, \apj, 566, 925, \dodoi{10.1086/338336}

\bibitem[{{Lam} {et~al.}(2021){Lam}, {Chen}, {Li}, {Yang}, {Cox}, {Looney}, \& {Stephens}}]{Lam_2021}
{Lam}, K.~H., {Chen}, C.-Y., {Li}, Z.-Y., {et~al.} 2021, \mnras, 507, 608, \dodoi{10.1093/mnras/stab2105}

\bibitem[{{Lazarian}(1994)}]{Lazarian_1994}
{Lazarian}, A. 1994, \mnras, 268, 713, \dodoi{10.1093/mnras/268.3.713}

\bibitem[{{Lazarian}(2007)}]{Lazarian_2007}
---. 2007, \jqsrt, 106, 225, \dodoi{10.1016/j.jqsrt.2007.01.038}

\bibitem[{{Lazarian} {et~al.}(2015){Lazarian}, {Andersson}, \& {Hoang}}]{Lazarian_2015}
{Lazarian}, A., {Andersson}, B.~G., \& {Hoang}, T. 2015, in Polarimetry of Stars and Planetary Systems, 81

\bibitem[{{Lazarian} \& {Draine}(1999)}]{Lazarian_Draine_1999}
{Lazarian}, A., \& {Draine}, B.~T. 1999, \apjl, 520, L67, \dodoi{10.1086/312137}

\bibitem[{{Lazarian} \& {Efroimsky}(1999)}]{Lazarian_Efroisky_1999}
{Lazarian}, A., \& {Efroimsky}, M. 1999, \mnras, 303, 673, \dodoi{10.1046/j.1365-8711.1999.02235.x}

\bibitem[{Lazarian \& Hoang(2007)}]{Lazarian_Hoang_2007a}
Lazarian, A., \& Hoang, T. 2007, Monthly Notices of the Royal Astronomical Society, 378, 910, \dodoi{10.1111/j.1365-2966.2007.11817.x}

\bibitem[{{Lazarian} \& {Hoang}(2008)}]{Lazarian_Hoang_2008}
{Lazarian}, A., \& {Hoang}, T. 2008, \apjl, 676, L25, \dodoi{10.1086/586706}

\bibitem[{{Lazarian} \& {Hoang}(2019)}]{Lazarian_Hoang_2019}
---. 2019, \apj, 883, 122, \dodoi{10.3847/1538-4357/ab3d39}

\bibitem[{Lazarian \& Roberge(1997)}]{Lazarian_Roberge_1997}
Lazarian, A., \& Roberge, W. 1997, \apj, 484, 230

\bibitem[{{Le Gouellec} {et~al.}(2023{\natexlab{a}}){Le Gouellec}, {Maury}, \& {Hull}}]{Valentin_2023a}
{Le Gouellec}, V.~J.~M., {Maury}, A.~J., \& {Hull}, C.~L.~H. 2023{\natexlab{a}}, \aap, 671, A167, \dodoi{10.1051/0004-6361/202244865}

\bibitem[{{Le Gouellec} {et~al.}(2023{\natexlab{b}}){Le Gouellec}, {Maury}, {Hull}, {Verliat}, {Hennebelle}, \& {Valdivia}}]{Valentin_2023b}
{Le Gouellec}, V.~J.~M., {Maury}, A.~J., {Hull}, C.~L.~H., {et~al.} 2023{\natexlab{b}}, \aap, 675, A133, \dodoi{10.1051/0004-6361/202245346}

\bibitem[{{Le Gouellec} {et~al.}(2019){Le Gouellec}, {Hull}, {Maury}, {Girart}, {Tychoniec}, {Kristensen}, {Li}, {Louvet}, {Cortes}, \& {Rao}}]{Valentin_2019}
{Le Gouellec}, V. J.~M., {Hull}, C. L.~H., {Maury}, A.~J., {et~al.} 2019, \apj, 885, 106, \dodoi{10.3847/1538-4357/ab43c2}

\bibitem[{{Le Gouellec} {et~al.}(2020){Le Gouellec}, {Maury}, {Guillet}, {Hull}, {Girart}, {Verliat}, {Mignon-Risse}, {Valdivia}, {Hennebelle}, {Gonz{\'a}lez}, \& {Louvet}}]{Valentin_2020}
{Le Gouellec}, V.~J.~M., {Maury}, A.~J., {Guillet}, V., {et~al.} 2020, \aap, 644, A11, \dodoi{10.1051/0004-6361/202038404}

\bibitem[{{Le Gouellec} {et~al.}(2023{\natexlab{c}}){Le Gouellec}, {Andersson}, {Soam}, {Schirmer}, {Michail}, {Lopez-Rodriguez}, {Flores}, {Chuss}, {Vaillancourt}, {Hoang}, \& {Lazarian}}]{Valentin_2023c}
{Le Gouellec}, V. J.~M., {Andersson}, B.~G., {Soam}, A., {et~al.} 2023{\natexlab{c}}, \apj, 951, 97, \dodoi{10.3847/1538-4357/accff7}

\bibitem[{{Lebreuilly} {et~al.}(2023){Lebreuilly}, {Vallucci-Goy}, {Guillet}, {Lombart}, \& {Marchand}}]{Lebreuilly_2023}
{Lebreuilly}, U., {Vallucci-Goy}, V., {Guillet}, V., {Lombart}, M., \& {Marchand}, P. 2023, \mnras, 518, 3326, \dodoi{10.1093/mnras/stac3220}

\bibitem[{{Li} \& {Greenberg}(1998)}]{Li_Greenberg_1999}
{Li}, A., \& {Greenberg}, J.~M. 1998, \aap, 331, 291

\bibitem[{{Li} {et~al.}(2014){Li}, {Goodman}, {Sridharan}, {Houde}, {Li}, {Novak}, \& {Tang}}]{Li_2014}
{Li}, H.~B., {Goodman}, A., {Sridharan}, T.~K., {et~al.} 2014, in Protostars and Planets VI, ed. H.~{Beuther}, R.~S. {Klessen}, C.~P. {Dullemond}, \& T.~{Henning}, 101--123, \dodoi{10.2458/azu_uapress_9780816531240-ch005}

\bibitem[{{Lin} {et~al.}(2020){Lin}, {Li}, {Yang}, {Looney}, {Stephens}, \& {Hull}}]{Zhe-Yu_2020}
{Lin}, Z.-Y.~D., {Li}, Z.-Y., {Yang}, H., {et~al.} 2020, \mnras, 496, 169, \dodoi{10.1093/mnras/staa1499}

\bibitem[{{Lin} {et~al.}(2024){Lin}, {Li}, {Stephens}, {Fern{\'a}ndez-L{\'o}pez}, {Carrasco-Gonz{\'a}lez}, {Chandler}, {Pasetto}, {Looney}, {Yang}, {Harrison}, {Sadavoy}, {Henning}, {Hughes}, {Kataoka}, {Kwon}, {Muto}, \& {Segura-Cox}}]{Zhe-Yu_2024}
{Lin}, Z.-Y.~D., {Li}, Z.-Y., {Stephens}, I.~W., {et~al.} 2024, \mnras, 528, 843, \dodoi{10.1093/mnras/stae040}

\bibitem[{{Liu}(2021)}]{Liu_2021L}
{Liu}, H.~B. 2021, \apj, 914, 25, \dodoi{10.3847/1538-4357/abf8b6}

\bibitem[{Liu(2021)}]{Liu_2021}
Liu, H.~B. 2021, The Astrophysical Journal, 914, 25, \dodoi{10.3847/1538-4357/abf8b6}

\bibitem[{{Liu} {et~al.}(2018){Liu}, {Hasegawa}, {Ching}, {Lai}, {Hirano}, \& {Rao}}]{Liu_2018}
{Liu}, H.~B., {Hasegawa}, Y., {Ching}, T.-C., {et~al.} 2018, \aap, 617, A3, \dodoi{10.1051/0004-6361/201832699}

\bibitem[{{Lucy}(1999)}]{Lucy_1999}
{Lucy}, L.~B. 1999, \aap, 344, 282

\bibitem[{{Lyo} {et~al.}(2021){Lyo}, {Kim}, {Sadavoy}, {Johnstone}, {Berry}, {Pattle}, {Kwon}, {Bastien}, {Onaka}, {Di Francesco}, {Kang}, {Furuya}, {Hull}, {Tamura}, {Koch}, {Ward-Thompson}, {Hasegawa}, {Hoang}, {Arzoumanian}, {Won Lee}, {Lee}, {Byun}, {Kirchschlager}, {Doi}, {Kim}, {Hwang}, {Diep}, {Fanciullo}, {Lee}, {Park}, {Yoo}, {Chung}, {Whitworth}, {Mairs}, {Soam}, {Liu}, {Tang}, {Coud{\'e}}, {Andr{\'e}}, {Bourke}, {Vivien Chen}, {Chen}, {Ping Chen}, {Chen}, {Ching}, {Cho}, {Choi}, {Choi}, {Chrysostomou}, {Dai}, {Dowell}, {Duan}, {Duan}, {Eden}, {Eswaraiah}, {Eyres}, {Fiege}, {Fissel}, {Franzmann}, {Friberg}, {Friesen}, {Fuller}, {Gledhill}, {Graves}, {Greaves}, {Griffin}, {Gu}, {Han}, {Hatchell}, {Hayashi}, {Houde}, {Inoue}, {Inutsuka}, {Iwasaki}, {Jeong}, {Kang}, {Kataoka}, {Kawabata}, {Kemper}, {Kim}, {Kim}, {Kim}, {Kim}, {Kirk}, {Kobayashi}, {K{\"o}nyves}, {Kusune}, {Kwon}, {Lacaille}, {Lai}, {Law}, {Lee}, {Lee}, {Lee}, {Li}, {Li}, {Li}, {Liu}, {Liu}, {Liu}, {Lu}, {Matsumura}, {Matthews},
  {Moriarty-Schieven}, {Nagata}, {Nakamura}, {Nakanishi}, {Bich Ngoc}, {Ohashi}, {Parsons}, {Peretto}, {Priestley}, {Pyo}, {Qian}, {Qiu}, {Rao}, {Rawlings}, {Rawlings}, {Retter}, {Richer}, {Rigby}, {Saito}, {Savini}, {Scaife}, {Seta}, {Shimajiri}, {Shinnaga}, {Tahani}, {Tang}, {Tomisaka}, {Tram}, {Tsukamoto}, {Viti}, {Wang}, {Wang}, {Xie}, {Yen}, {Yuan}, {Yun}, {Zenko}, {Zhang}, {Zhang}, {Zhang}, {Zhou}, {Zhu}, {de Looze}, {Dowell}, {Falle}, {Robitaille}, \& {van Loo}}]{Lyo_2021}
{Lyo}, A.~R., {Kim}, J., {Sadavoy}, S., {et~al.} 2021, \apj, 918, 85, \dodoi{10.3847/1538-4357/ac0ce9}

\bibitem[{{Machida} {et~al.}(2011){Machida}, {Inutsuka}, \& {Matsumoto}}]{Machida_2010}
{Machida}, M.~N., {Inutsuka}, S.-I., \& {Matsumoto}, T. 2011, \pasj, 63, 555, \dodoi{10.1093/pasj/63.3.555}

\bibitem[{{Marchand} {et~al.}(2023){Marchand}, {Lebreuilly}, {Mac Low}, \& {Guillet}}]{Marchand_2023}
{Marchand}, P., {Lebreuilly}, U., {Mac Low}, M.~M., \& {Guillet}, V. 2023, \aap, 670, A61, \dodoi{10.1051/0004-6361/202244291}

\bibitem[{{Masson} {et~al.}(2012){Masson}, {Teyssier}, {Mulet-Marquis}, {Hennebelle}, \& {Chabrier}}]{Masson_2012}
{Masson}, J., {Teyssier}, R., {Mulet-Marquis}, C., {Hennebelle}, P., \& {Chabrier}, G. 2012, \apjs, 201, 24, \dodoi{10.1088/0067-0049/201/2/24}

\bibitem[{{Mathis} {et~al.}(1983){Mathis}, {Mezger}, \& {Panagia}}]{Mathis_1983}
{Mathis}, J.~S., {Mezger}, P.~G., \& {Panagia}, N. 1983, \aap, 128, 212

\bibitem[{{Mathis} {et~al.}(1977){Mathis}, {Rumpl}, \& {Nordsieck}}]{Mathis_1977}
{Mathis}, J.~S., {Rumpl}, W., \& {Nordsieck}, K.~H. 1977, \apj, 217, 425, \dodoi{10.1086/155591}

\bibitem[{Maury {et~al.}(2018)Maury, Girart, Zhang, Hennebelle, Keto, Rao, Lai, Ohashi, \& Galametz}]{Maury_2018}
Maury, A.~J., Girart, J.~M., Zhang, Q., {et~al.} 2018, \mnras, 477, 2760

\bibitem[{{Mellon} \& {Li}(2008)}]{Melon_2008}
{Mellon}, R.~R., \& {Li}, Z.-Y. 2008, \apj, 681, 1356, \dodoi{10.1086/587542}

\bibitem[{Mestel \& Spitzer~Jr(1956)}]{Mestel_Spitzer_1956}
Mestel, L., \& Spitzer~Jr, L. 1956, \mnras, 116, 503

\bibitem[{{Michoulier} {et~al.}(2024){Michoulier}, {Gonzalez}, \& {Price}}]{Michoulier_2024}
{Michoulier}, S., {Gonzalez}, J.-F., \& {Price}, D.~J. 2024, \aap, 688, A31, \dodoi{10.1051/0004-6361/202449719}

\bibitem[{{Miotello} {et~al.}(2014){Miotello}, {Testi}, {Lodato}, {Ricci}, {Rosotti}, {Brooks}, {Maury}, \& {Natta}}]{Miotello_2014}
{Miotello}, A., {Testi}, L., {Lodato}, G., {et~al.} 2014, \aap, 567, A32, \dodoi{10.1051/0004-6361/201322945}

\bibitem[{Mori \& Kataoka(2021)}]{Mori_2021}
Mori, T., \& Kataoka, A. 2021, The Astrophysical Journal, 908, 153, \dodoi{10.3847/1538-4357/abd08a}

\bibitem[{N{\'e}el(1949)}]{Neel_1949}
N{\'e}el, L. 1949, Comptes Rendus Hebdomadaires Des Seances De L Academie Des Sciences, 228, 664

\bibitem[{{Ngoc} {et~al.}(2021){Ngoc}, {Diep}, {Parsons}, {Pattle}, {Hoang}, {Ward-Thompson}, {Tram}, {Hull}, {Tahani}, {Furuya}, {Bastien}, {Qiu}, {Hasegawa}, {Kwon}, {Doi}, {Lai}, {Coud{\'e}}, {Berry}, {Ching}, {Hwang}, {Soam}, {Wang}, {Arzoumanian}, {Bourke}, {Byun}, {Chen}, {Chen}, {Chen}, {Chen}, {Cho}, {Choi}, {Choi}, {Chrysostomou}, {Chung}, {Dai}, {Di Francesco}, {Duan}, {Duan}, {Eden}, {Eswaraiah}, {Fanciullo}, {Fiege}, {Fissel}, {Franzmann}, {Friberg}, {Friesen}, {Fuller}, {Gledhill}, {Graves}, {Greaves}, {Griffin}, {Gu}, {Han}, {Hatchell}, {Hayashi}, {Houde}, {Inoue}, {Inutsuka}, {Iwasaki}, {Jeong}, {Johnstone}, {Kang}, {Kang}, {Kang}, {Kataoka}, {Kawabata}, {Kemper}, {Kim}, {Kim}, {Pyo}, {Qian}, {Rao}, {Rawlings}, {Rawlings}, {Retter}, {Richer}, {Rigby}, {Sadavoy}, {Saito}, {Savini}, {Scaife}, {Seta}, {Kim}, {Kim}, {Kim}, {Kim}, {Kirchschlager}, {Kirk}, {Kobayashi}, {Koch}, {Konyves}, {Kusune}, {Kwon}, {Lacaille}, {Law}, {Lee}, {Lee}, {Lee}, {Lee}, {Lee}, {Lee}, {Li}, {Li}, {Li}, {Liu}, {Liu},
  {Liu}, {Liu}, {Lu}, {Lyo}, {Mairs}, {Matsumura}, {Matthews}, {Moriarty-Schieven}, {Nagata}, {Nakamura}, {Nakanishi}, {Ohashi}, {Onaka}, {Park}, {Peretto}, {Shimajiri}, {Shinnaga}, {Tamura}, {Tang}, {Tang}, {Tomisaka}, {Tsukamoto}, {Viti}, {Wang}, {Whitworth}, {Xie}, {Yen}, {Yoo}, {Yuan}, {Yun}, {Zenko}, {Zhang}, {Zhang}, {Zhang}, {Zhou}, {Zhu}, {de Looze}, {Andr{\'e}}, {Dowell}, {Eyres}, {Falle}, {Robitaille}, \& {van Loo}}]{Ngoc_2021}
{Ngoc}, N.~B., {Diep}, P.~N., {Parsons}, H., {et~al.} 2021, \apj, 908, 10, \dodoi{10.3847/1538-4357/abd0fc}

\bibitem[{{Nguyen Tat} {et~al.}(2023){Nguyen Tat}, {Diep}, {Hoang}, {Tram}, {Bich Ngoc}, {Phuong}, \& {Truong}}]{Thang_2024}
{Nguyen Tat}, T., {Diep}, P.~N., {Hoang}, T., {et~al.} 2023, arXiv e-prints, arXiv:2401.00220, \dodoi{10.48550/arXiv.2401.00220}

\bibitem[{Offner \& Chaban(2017)}]{Offner_2017}
Offner, S. S.~R., \& Chaban, J. 2017, The Astrophysical Journal, 847, 104, \dodoi{10.3847/1538-4357/aa8996}

\bibitem[{Pattle {et~al.}(2019)Pattle, Lai, Hasegawa, Wang, Furuya, Ward-Thompson, Bastien, Coudé, Eswaraiah, Fanciullo, di~Francesco, Hoang, Kim, Kwon, Lee, Liu, Liu, Matsumura, Onaka, Sadavoy, \& Soam}]{Pattle_2019}
Pattle, K., Lai, S.-P., Hasegawa, T., {et~al.} 2019, The Astrophysical Journal, 880, 27, \dodoi{10.3847/1538-4357/ab286f}

\bibitem[{{Pattle} {et~al.}(2021){Pattle}, {Lai}, {Wright}, {Coud{\'e}}, {Plambeck}, {Hoang}, {Tang}, {Bastien}, {Eswaraiah}, {Furuya}, {Hwang}, {Inutsuka}, {Kim}, {Kirchschlager}, {Kwon}, {Lee}, {Liu}, {Lyo}, {Ohashi}, {Rawlings}, {Tahani}, {Tamura}, {Soam}, {Wang}, \& {Ward-Thompson}}]{Pattle_2021}
{Pattle}, K., {Lai}, S.-P., {Wright}, M., {et~al.} 2021, \mnras, 503, 3414, \dodoi{10.1093/mnras/stab608}

\bibitem[{{Purcell}(1979)}]{Purcell_1979}
{Purcell}, E.~M. 1979, \apj, 231, 404, \dodoi{10.1086/157204}

\bibitem[{{Rao} {et~al.}(2009){Rao}, {Girart}, {Marrone}, {Lai}, \& {Schnee}}]{Rao_2009}
{Rao}, R., {Girart}, J.~M., {Marrone}, D.~P., {Lai}, S.-P., \& {Schnee}, S. 2009, \apj, 707, 921, \dodoi{10.1088/0004-637X/707/2/921}

\bibitem[{{Reissl} {et~al.}(2023){Reissl}, {Meehan}, \& {Klessen}}]{Reissl_2023}
{Reissl}, S., {Meehan}, P., \& {Klessen}, R.~S. 2023, \aap, 674, A47, \dodoi{10.1051/0004-6361/202142528}

\bibitem[{{Reissl} {et~al.}(2016){Reissl}, {Wolf}, \& {Brauer}}]{Reissl_2016}
{Reissl}, S., {Wolf}, S., \& {Brauer}, R. 2016, \aap, 593, A87, \dodoi{10.1051/0004-6361/201424930}

\bibitem[{{Reissl} {et~al.}(2014){Reissl}, {Wolf}, \& {Seifried}}]{Reissl_2014}
{Reissl}, S., {Wolf}, S., \& {Seifried}, D. 2014, \aap, 566, A65, \dodoi{10.1051/0004-6361/201323116}

\bibitem[{{Sadavoy} {et~al.}(2018){Sadavoy}, {Myers}, {Stephens}, {Tobin}, {Kwon}, {Segura-Cox}, {Henning}, {Commer{\c{c}}on}, \& {Looney}}]{Sadavoy_2018}
{Sadavoy}, S.~I., {Myers}, P.~C., {Stephens}, I.~W., {et~al.} 2018, \apj, 869, 115, \dodoi{10.3847/1538-4357/aaef81}

\bibitem[{Sadavoy {et~al.}(2019)Sadavoy, Stephens, Myers, Looney, Tobin, Kwon, Commerçon, Segura-Cox, Henning, \& Hennebelle}]{Sadavoy_2019}
Sadavoy, S.~I., Stephens, I.~W., Myers, P.~C., {et~al.} 2019, The Astrophysical Journal Supplement Series, 245, 2, \dodoi{10.3847/1538-4365/ab4257}

\bibitem[{{Seifried} {et~al.}(2015){Seifried}, {Banerjee}, {Pudritz}, \& {Klessen}}]{Seifried_2015}
{Seifried}, D., {Banerjee}, R., {Pudritz}, R.~E., \& {Klessen}, R.~S. 2015, \mnras, 446, 2776, \dodoi{10.1093/mnras/stu2282}

\bibitem[{{Seizinger} {et~al.}(2013){Seizinger}, {Speith}, \& {Kley}}]{Seizinger_2013}
{Seizinger}, A., {Speith}, R., \& {Kley}, W. 2013, \aap, 559, A19, \dodoi{10.1051/0004-6361/201322046}

\bibitem[{{Shu} {et~al.}(1987){Shu}, {Adams}, \& {Lizano}}]{Shu_1987}
{Shu}, F.~H., {Adams}, F.~C., \& {Lizano}, S. 1987, \araa, 25, 23, \dodoi{10.1146/annurev.aa.25.090187.000323}

\bibitem[{Stephens {et~al.}(2013)Stephens, Looney, Kwon, Hull, Plambeck, Crutcher, Chapman, Novak, Davidson, Vaillancourt, Shinnaga, \& Matthews}]{Stephens_2013}
Stephens, I.~W., Looney, L.~W., Kwon, W., {et~al.} 2013, \apjl, 769, L15, \dodoi{10.1088/2041-8205/769/1/L15}

\bibitem[{{Stephens} {et~al.}(2017){Stephens}, {Yang}, {Li}, {Looney}, {Kataoka}, {Kwon}, {Fern{\'a}ndez-L{\'o}pez}, {Hull}, {Hughes}, {Segura-Cox}, {Mundy}, {Crutcher}, \& {Rao}}]{Stephens_2017}
{Stephens}, I.~W., {Yang}, H., {Li}, Z.-Y., {et~al.} 2017, \apj, 851, 55, \dodoi{10.3847/1538-4357/aa998b}

\bibitem[{{Stephens} {et~al.}(2023){Stephens}, {Lin}, {Fern{\'a}ndez-L{\'o}pez}, {Li}, {Looney}, {Yang}, {Harrison}, {Kataoka}, {Carrasco-Gonzalez}, {Okuzumi}, \& {Tazaki}}]{Stephens_2024}
{Stephens}, I.~W., {Lin}, Z.-Y.~D., {Fern{\'a}ndez-L{\'o}pez}, M., {et~al.} 2023, \nat, 623, 705, \dodoi{10.1038/s41586-023-06648-7}

\bibitem[{{Tahani} {et~al.}(2018){Tahani}, {Plume}, {Brown}, \& {Kainulainen}}]{Tahani_2018}
{Tahani}, M., {Plume}, R., {Brown}, J.~C., \& {Kainulainen}, J. 2018, \aap, 614, A100, \dodoi{10.1051/0004-6361/201732219}

\bibitem[{{Tahani} {et~al.}(2019){Tahani}, {Plume}, {Brown}, {Soler}, \& {Kainulainen}}]{Tahani_2019}
{Tahani}, M., {Plume}, R., {Brown}, J.~C., {Soler}, J.~D., \& {Kainulainen}, J. 2019, \aap, 632, A68, \dodoi{10.1051/0004-6361/201936280}

\bibitem[{{Tahani} {et~al.}(2022){Tahani}, {Lupypciw}, {Glover}, {Plume}, {West}, {Kothes}, {Inutsuka}, {Lee}, {Robishaw}, {Knee}, {Brown}, {Doi}, {Grenier}, \& {Haverkorn}}]{Tahani_2022_Perseus}
{Tahani}, M., {Lupypciw}, W., {Glover}, J., {et~al.} 2022, \aap, 660, A97, \dodoi{10.1051/0004-6361/202141170}

\bibitem[{Takahashi {et~al.}(2019)Takahashi, Machida, Tomisaka, Ho, Fomalont, Nakanishi, \& Girart}]{Takahashi_2019}
Takahashi, S., Machida, M.~N., Tomisaka, K., {et~al.} 2019, \apj, 872, 70, \dodoi{10.3847/1538-4357/aaf6ed}

\bibitem[{{Tazaki} {et~al.}(2017){Tazaki}, {Lazarian}, \& {Nomura}}]{Tazaki_2017}
{Tazaki}, R., {Lazarian}, A., \& {Nomura}, H. 2017, \apj, 839, 56, \dodoi{10.3847/1538-4357/839/1/56}

\bibitem[{{Teyssier}(2002)}]{Teyssier_2002}
{Teyssier}, R. 2002, \aap, 385, 337, \dodoi{10.1051/0004-6361:20011817}

\bibitem[{{Tobin} {et~al.}(2019){Tobin}, {Bourke}, {Mader}, {Kristensen}, {Arce}, {Gueth}, {Gusdorf}, {Codella}, {Leurini}, \& {Chen}}]{Tobin_2019}
{Tobin}, J.~J., {Bourke}, T.~L., {Mader}, S., {et~al.} 2019, \apj, 870, 81, \dodoi{10.3847/1538-4357/aaef87}

\bibitem[{{Tram} \& {Hoang}(2022)}]{Tram_Hoang_2022}
{Tram}, L.~N., \& {Hoang}, T. 2022, Frontiers in Astronomy and Space Sciences, 9, 923927, \dodoi{10.3389/fspas.2022.923927}

\bibitem[{{Tsukamoto} {et~al.}(2022){Tsukamoto}, {Maury}, {Commer{\c{c}}on}, {Alves}, {Cox}, {Sakai}, {Ray}, {Zhao}, \& {Machida}}]{Yusuke_2022}
{Tsukamoto}, Y., {Maury}, A., {Commer{\c{c}}on}, B., {et~al.} 2022, arXiv e-prints, arXiv:2209.13765, \dodoi{10.48550/arXiv.2209.13765}

\bibitem[{{Valdivia} {et~al.}(2019){Valdivia}, {Maury}, {Brauer}, {Hennebelle}, {Galametz}, {Guillet}, \& {Reissl}}]{Valdivia_2019}
{Valdivia}, V., {Maury}, A., {Brauer}, R., {et~al.} 2019, \mnras, 488, 4897, \dodoi{10.1093/mnras/stz2056}

\bibitem[{{Valdivia} {et~al.}(2022){Valdivia}, {Maury}, \& {Hennebelle}}]{Valdivia_2022}
{Valdivia}, V., {Maury}, A., \& {Hennebelle}, P. 2022, \aap, 668, A83, \dodoi{10.1051/0004-6361/202243633}

\bibitem[{{Verliat} {et~al.}(2022){Verliat}, {Hennebelle}, {Gonz{\'a}lez}, {Lee}, \& {Geen}}]{Verliat_2022}
{Verliat}, A., {Hennebelle}, P., {Gonz{\'a}lez}, M., {Lee}, Y.-N., \& {Geen}, S. 2022, \aap, 663, A6, \dodoi{10.1051/0004-6361/202141765}

\bibitem[{{Westphal} {et~al.}(2014){Westphal}, {Stroud}, {Bechtel}, {Brenker}, {Butterworth}, {Flynn}, {Frank}, {Gainsforth}, {Hillier}, {Postberg}, {Simionovici}, {Sterken}, {Nittler}, {Allen}, {Anderson}, {Ansari}, {Bajt}, {Bastien}, {Bassim}, {Bridges}, {Brownlee}, {Burchell}, {Burghammer}, {Changela}, {Cloetens}, {Davis}, {Doll}, {Floss}, {Gr{\"u}n}, {Heck}, {Hoppe}, {Hudson}, {Huth}, {Kearsley}, {King}, {Lai}, {Leitner}, {Lemelle}, {Leonard}, {Leroux}, {Lettieri}, {Marchant}, {Ogliore}, {Ong}, {Price}, {Sandford}, {Tresseras}, {Schmitz}, {Schoonjans}, {Schreiber}, {Silversmit}, {Sol{\'e}}, {Srama}, {Stadermann}, {Stephan}, {Stodolna}, {Sutton}, {Trieloff}, {Tsou}, {Tyliszczak}, {Vekemans}, {Vincze}, {Von Korff}, {Wordsworth}, {Zevin}, {Zolensky}, \& {aff14}}]{Westphal_2014}
{Westphal}, A.~J., {Stroud}, R.~M., {Bechtel}, H.~A., {et~al.} 2014, Science, 345, 786, \dodoi{10.1126/science.1252496}

\bibitem[{{Wong} {et~al.}(2016){Wong}, {Hirashita}, \& {Li}}]{Wong_2016}
{Wong}, Y. H.~V., {Hirashita}, H., \& {Li}, Z.-Y. 2016, \pasj, 68, 67, \dodoi{10.1093/pasj/psw066}

\bibitem[{{Yang}(2021)}]{Yang_2021}
{Yang}, H. 2021, \apj, 911, 125, \dodoi{10.3847/1538-4357/abebde}

\bibitem[{{Yang} {et~al.}(2016){Yang}, {Li}, {Looney}, {Cox}, {Tobin}, {Stephens}, {Segura-Cox}, \& {Harris}}]{Yang_2016b}
{Yang}, H., {Li}, Z.-Y., {Looney}, L.~W., {et~al.} 2016, \mnras, 460, 4109, \dodoi{10.1093/mnras/stw1253}

\bibitem[{{Yang} {et~al.}(2019){Yang}, {Li}, {Stephens}, {Kataoka}, \& {Looney}}]{Yang_2019}
{Yang}, H., {Li}, Z.-Y., {Stephens}, I.~W., {Kataoka}, A., \& {Looney}, L. 2019, \mnras, 483, 2371, \dodoi{10.1093/mnras/sty3263}

\bibitem[{{Yang} {et~al.}(2017){Yang}, {Evans}, {Green}, {Dunham}, \& {J{\o}rgensen}}]{Yang_2017}
{Yang}, Y.-L., {Evans}, Neal~J., I., {Green}, J.~D., {Dunham}, M.~M., \& {J{\o}rgensen}, J.~K. 2017, \apj, 835, 259, \dodoi{10.3847/1538-4357/835/2/259}

\bibitem[{{Ysard} {et~al.}(2015){Ysard}, {K{\"o}hler}, {Jones}, {Miville-Desch{\^e}nes}, {Abergel}, \& {Fanciullo}}]{Ysard_2015}
{Ysard}, N., {K{\"o}hler}, M., {Jones}, A., {et~al.} 2015, \aap, 577, A110, \dodoi{10.1051/0004-6361/201425523}

\bibitem[{{Ysard} {et~al.}(2024){Ysard}, {Jones}, {Guillet}, {Demyk}, {Decleir}, {Verstraete}, {Choubani}, {Miville-Desch{\^e}nes}, \& {Fanciullo}}]{Ysard_2024}
{Ysard}, N., {Jones}, A.~P., {Guillet}, V., {et~al.} 2024, \aap, 684, A34, \dodoi{10.1051/0004-6361/202348391}

\bibitem[{{Zamponi} {et~al.}(2024){Zamponi}, {Maureira}, {Liu}, {Zhao}, {Segura-Cox}, {Ko}, \& {Caselli}}]{Zamponi_2024}
{Zamponi}, J., {Maureira}, M.~J., {Liu}, H.~B., {et~al.} 2024, \aap, 682, A56, \dodoi{10.1051/0004-6361/202244628}

\end{thebibliography}

\appendix
\section{Implementation of Inelastic relaxation on POLARIS}\label{sec:inelastic_theory}

Difference from Barnett relaxation which dissipates the grain rotational energy via the precession of the magnetic moment around the grain symmetric axis, inelastic relaxation reduces rotational energy via the deformation of spinning dust grains (\citealt{Purcell_1979},\citealt{Lazarian_Efroisky_1999},\citealt{Efroimky_2000}, \citealt{Efroimsky_Lazarian_2000}). Following \cite{Hoang+2022}, the inelastic relaxation timescale of grains of size $a$ rotating with the angular velocity $\Omega$ is given by:
\bea 
\tau_{\rm iNER} = \frac{\mu Q}{\rho a^{2} \Omega^{3}} g(s),\label{eq:t_iNER}
\ena
where $g(s)$ is the geometrical factor that depends on grain axial ratio, $\mu$ is the shear modulus which characterizes the deformation of materials, and $Q$ is the Q factor which characterizes the energy dissipation rate when grains are strained. The value of $\mu$ can vary from $\mu \sim 3.6\times 10^{7} - 3.46\times 10^{9}\erg\cm^{-3}$ (\citealt{Knapmeyer_2018}), and $Q$ can range from $Q = 100$ for silicate rocks (\citealt{Efroimsky_Lazarian_2000}) to $Q \sim 400-2000$ for vitreous silicate (\citealt{Purcell_1979}). Large aggregate$-$type grains formed inside the protostellar disks may have lower values of $\mu$ and $Q$ than the above samples of compact grains owing to the large portion of voids between dust monomers.  

Grains will have fast internal relaxation by inelastic relaxation if $\tau_{\rm iNER} \leq \tau_{\rm gas}$. For grains rotating thermally, given $\Omega_{\rm ther} = \sqrt{k_{\rm B} T_{\rm gas}/I}$ with $\k_{\rm B}$ the Boltzmann constant and $I$ the inertia moment of grains, grains with small sizes will have faster thermal rotation and thus be easier to experience fast inelastic relaxation. We call the maximum size for grains at low-\textit{J} having efficient inelastic relaxation is $a_{\rm max,aJ}^{\rm iNER, lowJ}$. By plugging $\Omega_{\rm ther}$ into Equation (\ref{eq:t_iNER}), the value of $a_{\rm max,aJ}^{\rm iNER, lowJ}$ will be determined by comparing $\tau_{\rm iNER}$ with $\tau_{\rm gas}$ over the grain size distribution. All dust grains beyond $a_{\rm max,aJ}^{\rm iNER, lowJ}$ will have slow internal relaxation by inelastic relaxation.

For grains rotating suprathermal, sub-micron grains and VLGs tend to have slower maximum rotation rate by RATs owing to low RAT efficiency acting on small sub-micron grains and high internal moment of VLGs (\citealt{Hoang_2019}), respectively. Consequently, we will have the lower and upper threshold for grains having efficient inelastic relaxation at high-\textit{J} attractors, denoted as $a_{\rm min,aJ}^{\rm iNER,highJ}$ and $a_{\rm max,aJ}^{\rm iNER, highJ}$. Considering the constant stellar luminosity and grains are at rest, the RAT can continuously spin up grains to the maximum rotation rate $\Omega_{\rm RAT}$ given by:
\bea 
\Omega_{\rm RAT} = \frac{\Gamma_{\rm RAT} \tau_{\rm damp}}{I}
\ena 
where $\Gamma_{\rm RAT}$ is the radiative torque acting on grains and the gas damping timescale and $\tau_{\rm damp}$ is the gas damping timescale (\citealt{Hoang_2019}). Then, by plugging $\Omega_{\rm RAT}$ into Equation (\ref{eq:t_iNER}) and comparing them with $\tau_{\rm gas}$ over the size distribution, we can determine grains with $\tau_{\rm iNER} \leq \tau_{\rm gas}$ to be the range for grains having efficient inelastic relaxation at high-\textit{J} attractors.

\section{Upper limit of iron inclusions inside SPM grains}\label{sec:appen_Ncr}
Here we follow the calculation in \cite{Yang_2021} to get the upper critical $N_{\rm cl}$ value of SPM grains in our YSOs. The net magnetic moment of SPM grains is the sum of all iron cluster magnetic moments, which is determined by the magnetic moment of $N_{\rm cl}$ iron atoms embedded inside each cluster. Theoretically, non-spinning SPM grains in non-magnetized environments are not magnetized owing to the random orientation of cluster magnetic moments. The rotation of SPM grains can lead to the formation of the Barnett-equivalent magnetic field, that forces cluster magnetic moments to align along the grain angular momentum direction (Barnett effect, \citealt{Barnett_1915}). However, the magnetic energy produced by the Barnett-equivalent $\B$ field is usually not strong enough to overcome the anisotropy barrier energy required to reorient the cluster magnetic moment. \cite{Neel_1949} found that the Brownian-like motion of iron clusters can help to break such barrier energy, allowing SPM grains to have the net magnetic moment. Considering $K$ the anisotropy constant of material instituting iron inclusions \footnote{The anisotropic constant depends on the shape (shape anisotropy) and the crystalline structure (magnetocrystalline anisotropy) of material building iron cluster. The values of $K$ adopted from \cite{Yang_2021} is the magnetocrystalline anisotropy. If iron clusters are non-spherical, the extra contribution from the shape anisotropy will increase the barrier energy required to reorient the cluster magnetic moment, which limits $N_{\rm cl}$ allowed inside each iron cluster.}, the timescale required to break its barrier energy under the dust temperature $T_{\rm d}$ is (\citealt{Neel_1949}, \citealt{Bean_1959}):
\bea 
\tau = \tau_{\rm 0} e^{KV_{\rm cl}/k_{\rm B}T_{\rm d}},\label{eq:tau_barrier}
\ena
with $\tau_{\rm 0} = 10^{-9}$s and $V_{\rm cl}$ the volume of iron cluster including $N_{\rm cl}$ iron atoms, which is given by:
\bea 
V_{\rm cl} = \frac{m_{\rm Fe}}{\rho_{\rm iron}} = \frac{56 m_{\rm p} N_{\rm cl}}{\rho f_{\rm Fe}},\label{eq:V_cluster}
\ena

\begin{table}
\centering
\caption{Critical size $N_{\rm cr}$ of metallic iron nano-particle (upper row), $\rm Fe_{\rm 3}\rm O_{\rm 4}$ (middle row), and $\gamma-\rm Fe_{\rm 2}\rm O_{\rm 3}$ (lower row). The first, second, and third columns show the anisotropic constant $K$, mass density $\rho$, and fraction of iron atoms inside magnetized material $f_{\rm FE}$. The fourth, fifth, and sixth columns show the value of $N_{\rm cr}$ found inside the envelope, inner envelope, and the disk scale of 100 au for protostar with $L_{\rm center} = 5 - 100L_{\odot}$. We note that the value of $N_{\rm cr}$ in the disk may not be accurate because the anisotropy barrier energy of three considered materials is adopted when $T_{\rm d} < 150$ K.}
\begin{tabular}{llllllll}
\hline
Region ($T_{\rm d}$) & & & &  Envelope (15-30 K) & Inner envelope (40-90 K) &  Disk (200-500 K) \\
\hline
Material &  $K (\rm erg cm^{-3})$ & $\rho (\rm g cm^{-3})$ & $f_{\rm Fe}$ &  $N_{\rm cr}$ &  $N_{\rm cr}$ &   $N_{\rm cr}$    \\  
\hline
Metallic iron & $1.35\times 10^{5}$ & 7.87 & 1 &   $6.25\times 10^{4}-1.25\times 10^{5}$ & $1.67\times 10^{5}-3.75\times 10^{5}$ & $8.34\times 10^{5}-2.08\times10^{6}$ \\

$\rm Fe_{\rm 3}\rm O_{\rm 4}$  &  $6.1\times 10^{5}$ & 5.17 & 0.72 & $6.55\times 10^{3}-1.31\times 10^{4}$ & $1.74\times 10^{4}-3.93\times 10^{4}$ & $8.73\times 10^{4}-2.18\times 10^{5}$ \\ 

$\gamma-\rm Fe_{\rm 2}\rm O_{\rm 3}$ &  $ 1.9\times 10^{5}$ & 4.9 & 0.69 & $1.91\times 10^{4}-3.82\times 10^{4}$ & $5.09\times10^{4}-1.14\times 10^{5}$ & $2.54\times10^{5}-6.36\times 10^{5}$ \\
  \hline
     \label{tab:Ncr}
    \end{tabular}   
\end{table} 
 
where $56m_{\rm p}$ is the mass of the single iron atom with $m_{\rm p}$ the proton mass, $\rho$ is the mass density of ferromagnetic material building iron cluster, and $f_{\rm Fe}$ is the mass fraction of iron atoms inside iron cluster.

  \begin{figure*}
 \centering
    \includegraphics[width=0.45 \textwidth,height=0.45\textheight,keepaspectratio]{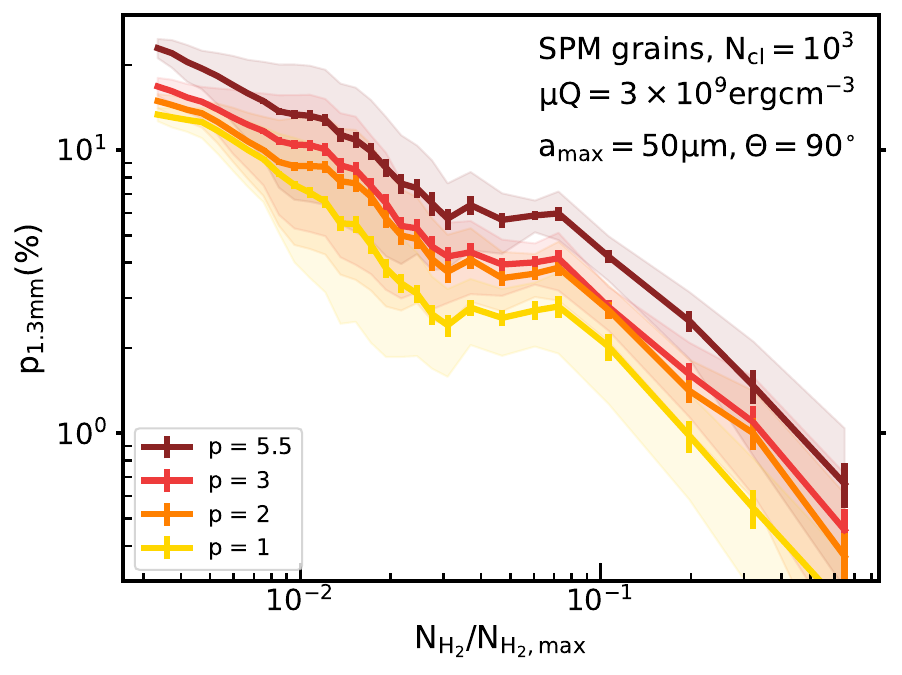}

    \caption{Effect of $p$ the magnetic moment per iron atom on the relation of $p-N_{\rm H_{2}}/N_{\rm H_{2},max}$, considering model RATA$-$INELASTIC of SPM grains with $N_{\rm cl} = 10^{3}, \mu Q = 3\times 10^{9}\erg\cm^{-3}, a_{\rm max} = 50\mu m$. Results are not filtered with CASA. By reducing $p$ from 5.5 (our adopted value) to $p = 1$ (for $\rm Fe_{\rm 3} \rm O_{\rm 4}$, \citealt{Yang_2021}), the polarization degree reduces $\sim 2$ times, giving $p \sim 3\%$ in the inner envelope and $p < 0.3\%$ in the disk scale. SPM grains must have higher $N_{\rm cl}$ than our estimation in Section \ref{sec:discuss_inner_envelope} to reproduce ALMA dust polarization of $> 5\%$ in the inner envelope and $\sim 1\%$ inside the disk scale.}
     \label{fig:magnetic_moment}
\end{figure*}

Considering the age of our YSO of $\tau_{\rm YSOs} = 38.52$ kyr (Section \ref{sec:MHD})\footnote{Note that the typical timescale for Class 0 YSOs is $\sim 10^{4}$ yrs and for Class I YSOs if $\sim 10^{5}$ yrs. Given longer considered timescales, bigger iron clusters can have more time to reorient their magnetic moment to the applied magnetic field direction, which slightly increases the boundary for defining SPM.}, the maximum cluster volume that Brownian-like motions of clusters can overcome the anisotropy barrier energy to produce the net grain magnetic alignment is given by:
\bea 
V_{\rm cr} = \frac{k_{\rm B}T_{\rm d}}{K}\ln \Bigg(\frac{\tau_{\rm YSOs}}{\tau_{\rm 0}}\Bigg),
\ena
which gives the critical number of iron atoms per cluster $N_{\rm cr}$ (Equation \ref{eq:V_cluster}) of:
\bea 
N_{\rm cr} = \frac{\rho f_{\rm Fe}k_{\rm B}T_{\rm d}}{56 m_{\rm p} K} \ln \Bigg(\frac{\tau_{\rm YSOs}}{\tau_{\rm 0}}\Bigg).\label{eq:Ncr}
\ena

 \begin{figure*}
 \centering
    \includegraphics[width= \textwidth,height=\textheight,keepaspectratio]{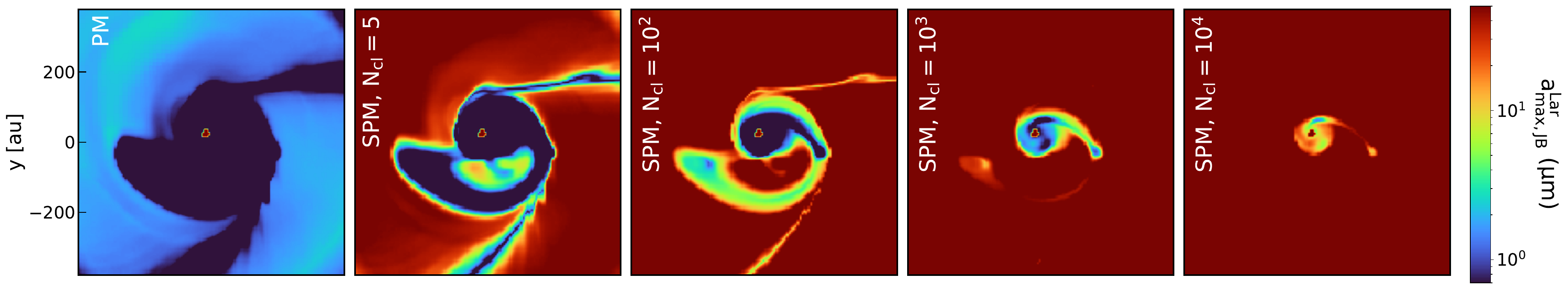}
    \includegraphics[width=\textwidth,height=\textheight,keepaspectratio]{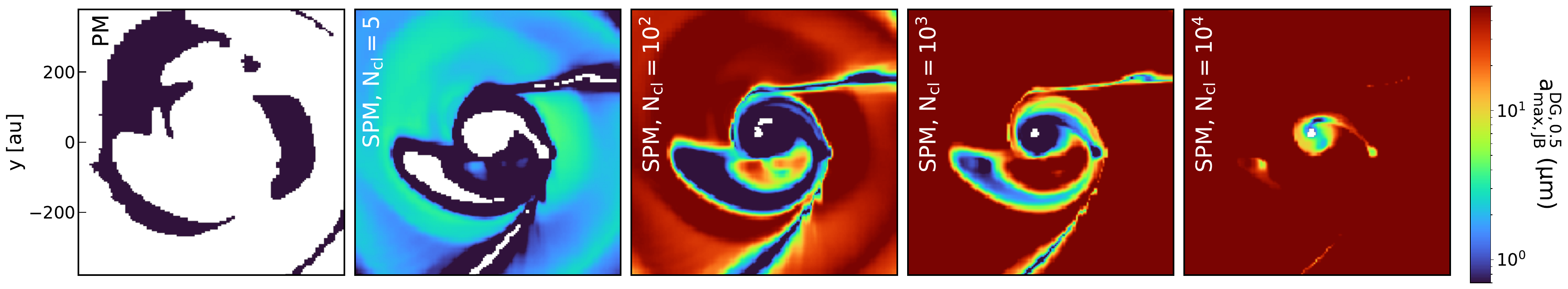}
    \includegraphics[width=\textwidth,height=\textheight,keepaspectratio]{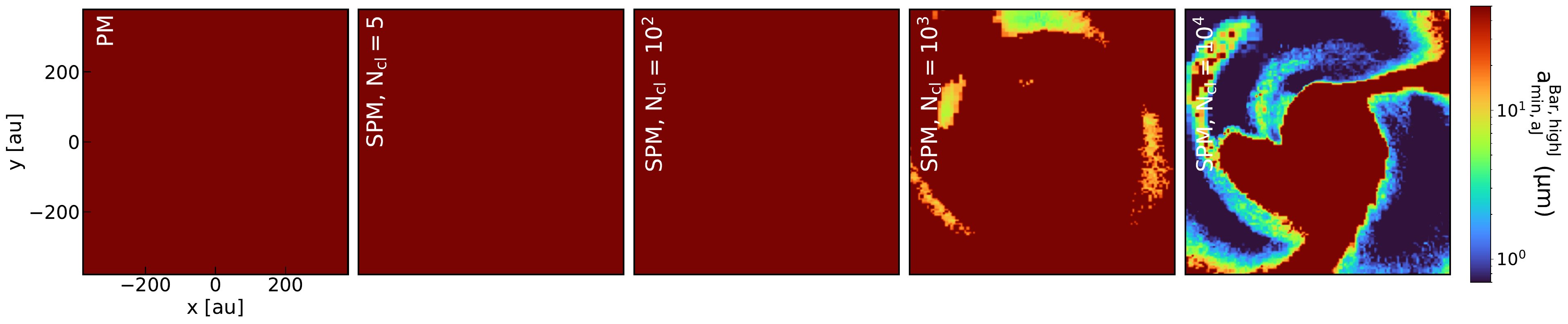}
    \centering
    \includegraphics[width=\textwidth,height=\textheight,keepaspectratio]{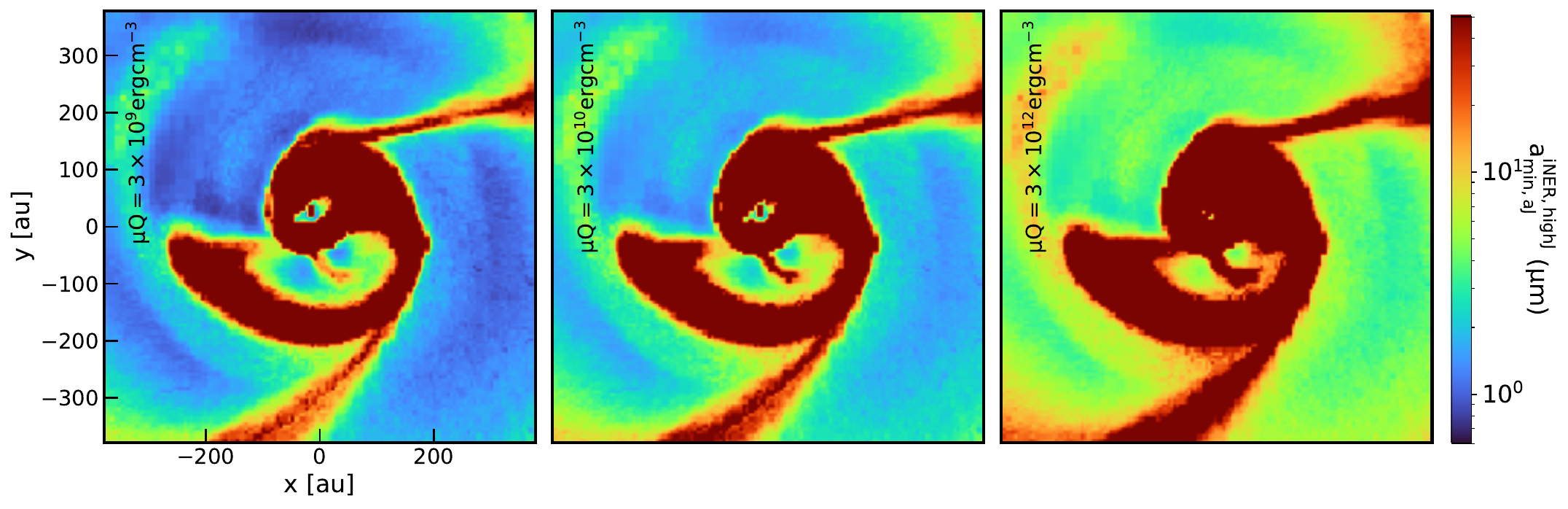}
    \caption{First, second, third rows: Distribution of $a_{\rm max,JB}^{\rm Lar}$, $a_{\rm max,JB}^{\rm DG,0.5}$, and $a_{\rm min,aJ}^{\rm Bar, highJ}$ obtained within 400 au on the disk midplane for PM grains (first column), SPM grains with $N_{\rm cl} = 5, 10^{2}, 10^{3}$ and $10^{4}$ (second to fifth columns), respectively. Fourth row: distribution of $a_{\rm min,aJ}^{\rm iNER, highJ}$ on the disk midplane for highly inelastic grains with $\mu Q = 3\times 10^{9}, 3\times 10^{10}\erg\cm^{-3}$ and highly elastic grains with $\mu Q = 3\times 10^{12}\erg\cm^{-3}$. Iron inclusions and inelastic relaxation can help to maintain the efficient internal alignment and MRAT mechanism of micron-sized and VLGs beyond 100 au to the protostar. However, inside the disk scale,  iron inclusions cannot help maintain the magnetic alignment of VLGs and MRAT alignment for micron-sized grains due to the efficient gaseous damping there. Furthermore, the joint action of super-Barnett and inelastic relaxation cannot help to drive the fast internal relaxation for aligned dust grains at both high and low-\textit{J} attractors in this dense region.} 
     \label{fig:alignment_disk}
\end{figure*}

There are several candidate forms of iron clusters inside dust grains, including metallic iron (Fe), iron surphide nanoparticle (FeS), magnetite ($\rm Fe_{\rm 3}\rm O_{\rm 4}$), maghemite ($\gamma- \rm Fe_{\rm 2}\rm O_{\rm 3}$). The form of iron inclusions inside protostellar dust grains remains unclear. Observations in ISM and dust collections from Stardust mission indicate the presence of metallic iron and iron sulphide nanoparticles inside amorphous Mg-rich silicate (\citealt{Davoisne_2006}, \citealt{Costantini_2005}, \citealt{Westphal_2014}). The Fe/FeS nanoparticles included in THEMIS/THEMIS2 dust model (i.e., \citealt{Kohler_2014}, \citealt{Jones_2017}, \citealt{Ysard_2015}, \citealt{Ysard_2024}) can reproduce well the SiO-emission feature and the observed SED in ISM. However, how the grain composition and iron inclusions evolve from ISM to protostellar environments remains unclear. Thus, we consider three different cases of metallic iron, $\rm Fe_{\rm 3}\rm O_{\rm 4}$, and $\gamma-\rm Fe_{\rm 2}\rm O_{\rm 3}$ similar to \cite{Yang_2021} to estimate the critical $N_{\rm cr}$ allowed for SPM grains inside our YSOs. Results of $N_{\rm cr}$ are shown in Table \ref{tab:Ncr} for each material in the upper, middle, and lower rows, respectively. The second, third, and fourth columns show the anisotropy barrier constant $K$, the mass density $\rho$, and the fraction of iron atoms inside magnetized material $f_{\rm FE}$ for each material. To understand how $N_{\rm cr}$ changes inside our considered YSO, we list in the last three columns values of $N_{\rm cr}$ inside the envelope, inner envelope, and the disk scale of 100 au. We consider YSOs with $L_{\rm center} = 5L_{\odot}-100L_{\odot}$, giving the typical dust temperature inside three considered regions of $T_{\rm d} = 15-30$ K, $40-90$K, and $200-500$K. Given higher central luminosity, $N_{\rm cr}$ increases owing to the stronger Browian-like motion of iron inclusions that shortens the timescale for overcoming the anisotropy barrier energy. 

  \begin{figure*}
 \centering
    \includegraphics[width= \textwidth,height=\textheight,keepaspectratio]{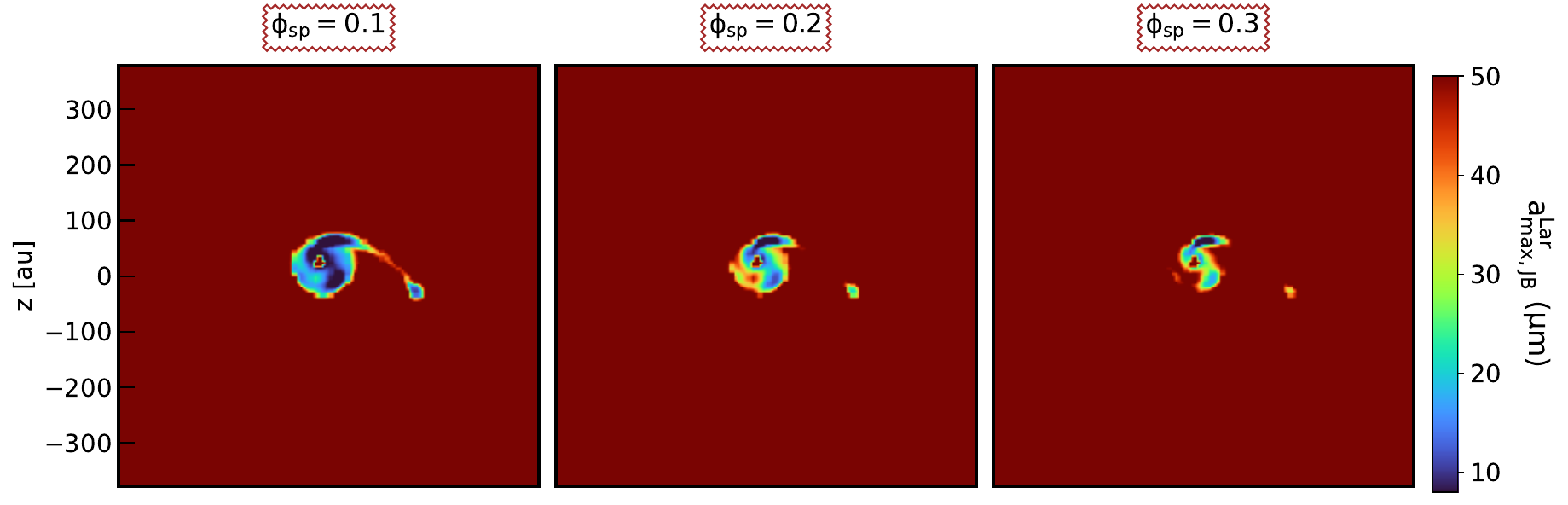}
    \includegraphics[width= \textwidth,height=\textheight,keepaspectratio]{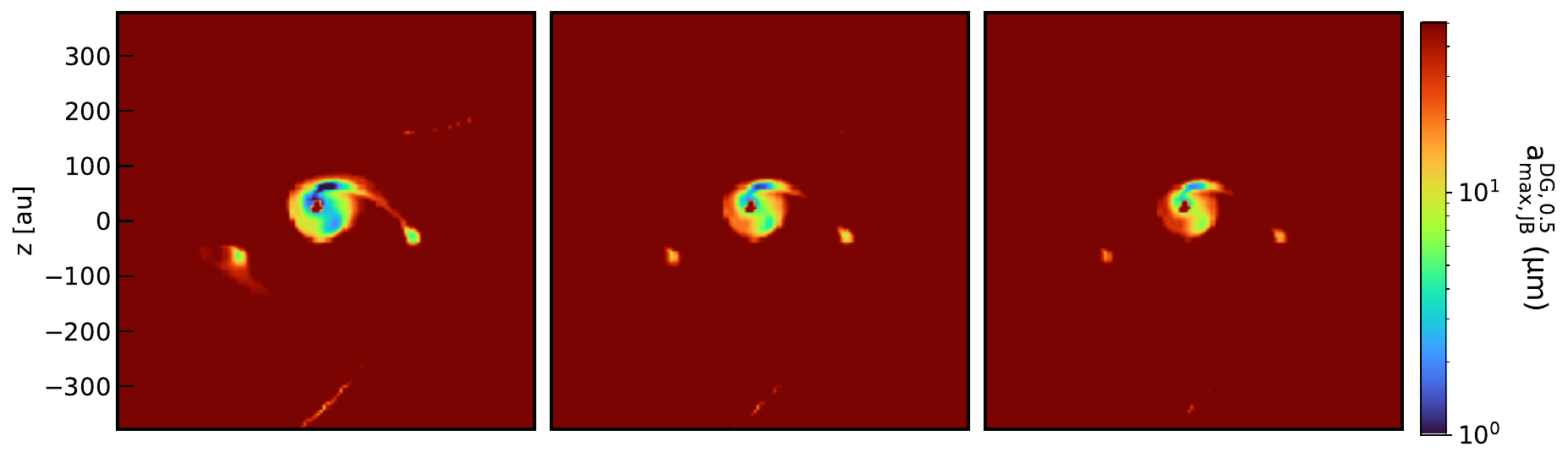}
    \includegraphics[width= \textwidth,height=\textheight,keepaspectratio]{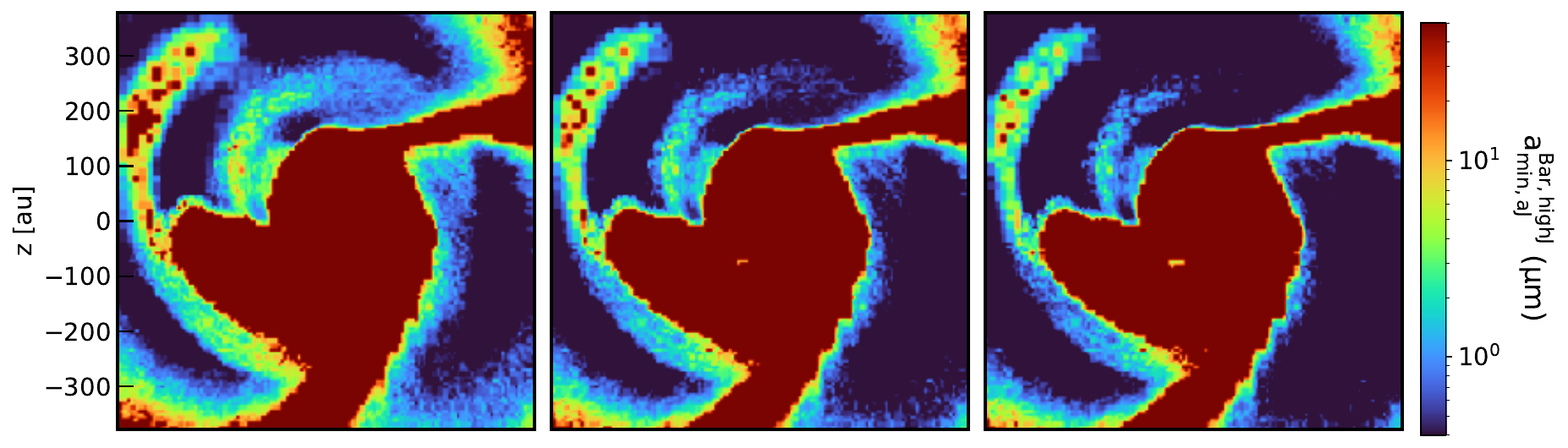}
    \caption{Effect of $\phi_{\rm sp}$ on the distribution of $a_{\rm max,JB}^{\rm Lar}$ (first row), $a_{\rm max,JB}^{\rm DG,0.5}$ (lower row), and $a_{\rm min,aJ}^{\rm Bar,highJ}$ (lower row) inside the equatorial midplane of 400 au scale. The volume filling factor of iron clusters inside dust grains is $\phi_{\rm sp} = 0.1, 0.2, 0.3$, from left to right, respectively. By increasing the magnetic susceptibility, more VLGs can be aligned with $\B$, and more micron-sized grains can have MRAT alignment inside the disk scale. However, it cannot help to enhance the internal alignment of dust grains, the major origin of which causes low polarization degree within several hundred au around the protostar.}
     \label{fig:alignment_disk_phisp}
\end{figure*}
 
One can see that for both metallic iron, $\rm Fe_{\rm 3}\rm O_{\rm 4}$, and $\gamma- \rm Fe_{\rm 2}\rm O_{\rm 3}$, adopting $N_{\rm cl} = 10^{4}$ in the rest of the paper still be under the allowance range for SPM grains regardless of their position inside YSOs (except the case of $\rm Fe_{\rm 3}\rm O_{\rm 4}$ inside the envelope with $T_{\rm d} < 15$ K-the case of $L_{\rm star} = 5L_{\odot}$). Theoretically, we can increase $N_{\rm cl}$ to nearly $10^{5}$ to enhance the grain magnetic alignment of SPM grains inside the disk. However, given the unclear form of iron inclusions and mechanisms producing the big iron cluster inside large grains, it is very risky to adopt $N_{\rm cl} \sim 10^{5}$ to study the grain alignment dynamic inside protostellar/protoplanetary disk environments. Another related issue of $N_{\rm cr}$ is that if iron clusters are in form of $\rm Fe_{\rm 3} \rm O_{\rm 4}$ or $\gamma - \rm Fe_{\rm 2} \rm O_{\rm 3}$, dust grains which only hold big iron clusters with size $N_{\rm cl} > 5\times 10^{3}$ may become ferromagnetic material inside the envelope due to the low dust temperature there (see column 5 of Table \ref{tab:Ncr}). However, ferromagnetic grains still respond with magnetic fields stronger than PM grains. Since micron-sized PM grains still can have the magnetic alignment by RATs there (Figure \ref{fig:distribution_alignment}), ferro grains can also couple with $\B$ and produce polarized dust emission in the envelope scale. Indeed, as discussed in Section \ref{sec:discuss_increase_Ncl}, all dust grains in the envelope do not only hold a single iron inclusion size as our assumption, and the grain can contain both big iron clusters (with $N_{\rm cl}$ may exceed $N_{\rm cr}$) and smaller ones (\citealt{Yang_2021}). Even big iron clusters cannot have a well-defined magnetic moment direction; the large amount of remaining small iron inclusions embedded inside dust grains do, which allows grains to still behave as SPM. As shown in Figure \ref{fig:CASA_Ncl}, SPM grains with  $N_{\rm cl} \sim 5-10^{3}$ can already reproduce $p \sim 5-30\%$ of dust polarization inside the envelope. Consequently,  even grains hold big iron clusters with $N_{\rm cl} > 5\times 10^{3}$ as grains inside the disk, they can still aligned with magnetic fields, and their dust emission still helps to trace magnetic fields inside the envelope (\citealt{Giang_2023a}).

Besides the uncertainty of the upper limit of iron cluster sizes, the magnetic moment of iron atom $p$ also affects the alignment efficiency of dust grains and the observed polarization fraction. In our study, we adopt $p = 5.5$ for silicate of $\rm Mg Fe Si O_{\rm 4}$ grains (\citealt{Draine_1996}, \citealt{Hoang_Lazarian_2016_mrat}) for calculations of SPM grains. The exact values of $p$ for iron inclusion inside protostellar grains are still unclear, that \cite{Draine_1996} suggested it to be $p = 2.2-3$ (for bulk iron), \cite{Coey_2010} and \cite{Yang_2021} suggested $ \sim 1.25-1.5$ (for $\rm Fe_{\rm 3}\rm O_{\rm 4}$), and $\sim 1.18-1.33$ (for $\gamma -\rm Fe_{\rm 2} \rm O_{\rm 3}$). We provide Figure \ref{fig:magnetic_moment} the $p-N_{\rm H}/N_{\rm H,max}$ relation got from different values of $p = 1, 2, 3$, and 5.5 (our adopted value), considering model RATA$-$INELASTIC of SPM grains with $N_{\rm cl} = 10^{3}$. The results are not filtered with CASA. One can see that by reducing $p$ by 5.5 times, the polarization degree is reduced $\sim 2$ times, similar as when we reduce $N_{\rm cl}$ by 10 times from $N_{\rm cl} = 10^{3}$ to $10^{2}$ (Figure \ref{fig:CASA_Ncl}). The uncertainty of $p$ does not affect the dust polarization properties in the envelope, but if iron inclusions inside protostellar grains have lower $p \sim 1$, the required range of $N_{\rm cl}$ to reproduce ALMA dust polarization must be larger, i.e., the mean $N_{\rm cl}$ starting from $\geq 5\times 10^{2}$ instead of $\geq 10^{2}$ (Section \ref{sec:discuss_inner_envelope}) for grains in the inner envelope. Lower $p$ also further challenges micron-sized and VLGs to have magnetic alignment and produce $\sim 1\%$ of polarized dust emission inside the disk scale. As the form and the magnetic moment of iron inclusions locked inside protostellar grains remain unconstrained, these issues should be considered in further studies aiming to digest the alignment dynamic of dust grains within 100 au scale around the protostar.

\section{Grain alignment state inside the disk}\label{sec:appen_disk}

Figure \ref{fig:alignment_disk} shows the spatial distribution of $a_{\rm max,JB}^{\rm Lar}$ (first row), $a_{\rm max, JB}^{\rm DG,0.5}$ (second row), $a_{\rm min,aJ}^{\rm Bar, highJ}$ (third row), and $a_{\rm min,aJ}^{\rm iNER, highJ}$ (fourth row) inside the disk midplane within 400 au around the protostar. For the three first rows showing the effect of iron inclusions on grain alignment, the first column shows results for PM grains, and the second to fifth columns correspond to results of SPM grains with $N_{\rm cl} = 5, 10^{2}, 10^{3}$ and $10^{4}$. For the fourth row showing the impact of inelastic relaxation on grain alignment, two first panels show results of inelastic material with low $\mu Q = 3\times 10^{9}$ and $3\times 10^{10}\erg\cm^{-3}$, and the last one corresponds to highly elastic grains with high $\mu Q = 3\times 10^{12}\erg\cm^{-3}$.

Similar as finding in Figures \ref{fig:distribution_alignment}, \ref{fig:Barnett}, and \ref{fig:Inelastic}, iron inclusions and inelastic relaxation help to maintain the MRAT alignment and efficient IA for micron-sized and VLGs beyond 100 au to the protostar. However, inside the disk scale, the high number of $N_{\rm cl} = 10^{4}$ (fifth column) cannot allow VLGs to have the magnetic alignment, i.e., $a_{\rm max,JB}^{\rm Lar} \sim 10\mum$ (first row), and maintain MRAT alignment for micron-sized grains, i.e., $a_{\rm max,JB}^{\rm DG,0.5} \sim 5\mum$ (second row). Furthermore, all aligned dust grains will have inefficient IA at high-\textit{J} attractors regardless of the joint action from Barnett relaxation and inelastic relaxation owing to efficient gaseous damping there, i.e., $a_{\rm min, aJ}^{\rm Bar, highJ} = a_{\rm min,aJ}^{\rm iNER, highJ} = 50\mum$. That explains why the depolarization effect caused by the reduction of the grain alignment efficiency always appears in our dust models regardless of values of $N_{\rm cl}$ and $\mu Q$.

\subsection{Whether higher $\phi_{\rm sp}$ helps to enhance the grain alignment inside the disk?}\label{sec:alignment_disk_phisp}
 
We show in Figure \ref{fig:alignment_disk_phisp} the spatial distribution of $a_{\rm max,JB}^{\rm Lar}$, $a_{\rm max,JB}^{\rm DG,0.5}$, and $a_{\rm min,aJ}^{\rm Bar,highJ}$ within 400 au on the disk midplane, assuming the model of SPM grains with $N_{\rm cl} = 10^{4}$, with $\phi_{\rm sp} = 0.1, 0.2, 0.3$ (corresponds to $30 - 90\%$ of iron inside dust grains under the cluster form). By increasing the volume factor of iron clusters inside SPM grains, more VLGs can have the magnetic alignment by RATs and more micron-sized grains can experience MRAT alignment inside $\sim 100$ au disk scale. However, it cannot help to increase the net internal alignment degree of dust grains - the major origin causing very low polarization degree around the protostar.

\section{Effect of stellar luminosity on the grain alignment state }\label{sec:appen_Lstar} 
\begin{figure*}
    \includegraphics[width=\textwidth,height=\textheight,keepaspectratio]{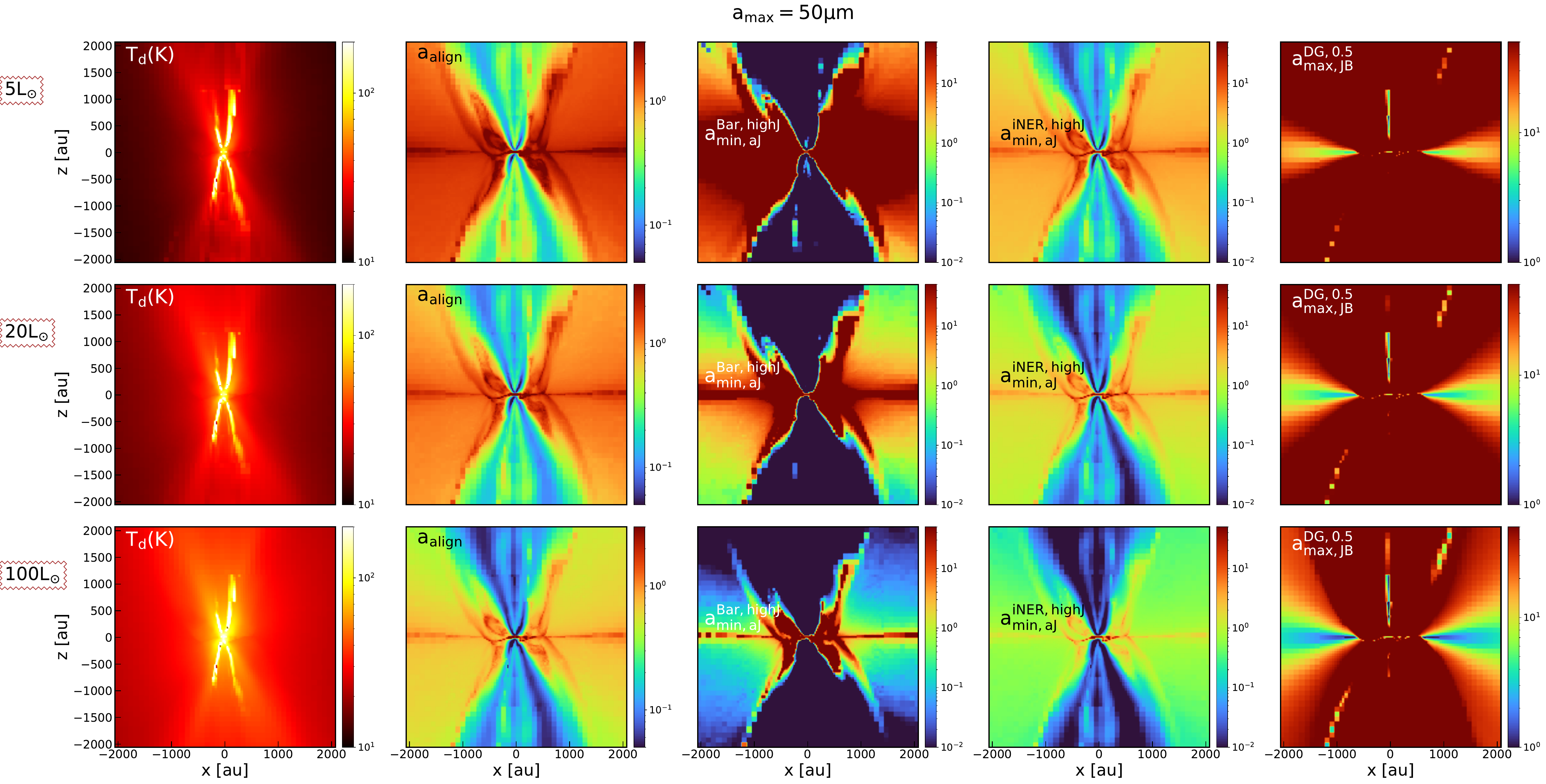}
    \caption{Effect of stellar luminosity, with $L_{\rm star} = 5L_{\odot}$ (upper row), $20L_{\odot}$ (middle row), and $100L_{\odot}$ (lower row) on the critical size of grain alignment, assuming SPM grains with $N_{\rm cl} = 10^{3}$, $\mu Q = 3\times 10^{9}\erg\cm^{-3}$, and $a_{\rm max} = 50\mu m$. The first column shows the distribution of dust temperature $T_{\rm d}$, and the second column shows the distribution of minimum alignment size $a_{\rm align}$. The third and fourth columns illustrate the minimum size having fast super-Barnett relaxation $a_{\rm min,aJ}^{\rm Bar,highJ}$ and fast inelastic relaxation $a_{\rm min,aJ}^{\rm iNER,highJ}$ at high-\textit{J} attractors. And the fifth column displays the maximum size having $f_{\rm high-J} = 0.5$ by MRAT mechanism $a_{\rm max,JB}^{\rm DG,50}$. Except for the narrow alignment range around fainter protostars, almost aligned dust grains still can achieve efficient IA by either super-Barnett or inelastic relaxation regardless of $L_{\rm center}$. Furthermore, VLGs around fainter protostars tend to experience MRAT mechanism, i.e., higher $a_{\rm max,JB}^{\rm DG,50}$, owing to the enhanced magnetic relaxation by reducing dust temperature there.} 
     \label{fig:alignment_Lstar50}
\centering     \includegraphics[width=\textwidth,height=\textheight,keepaspectratio]{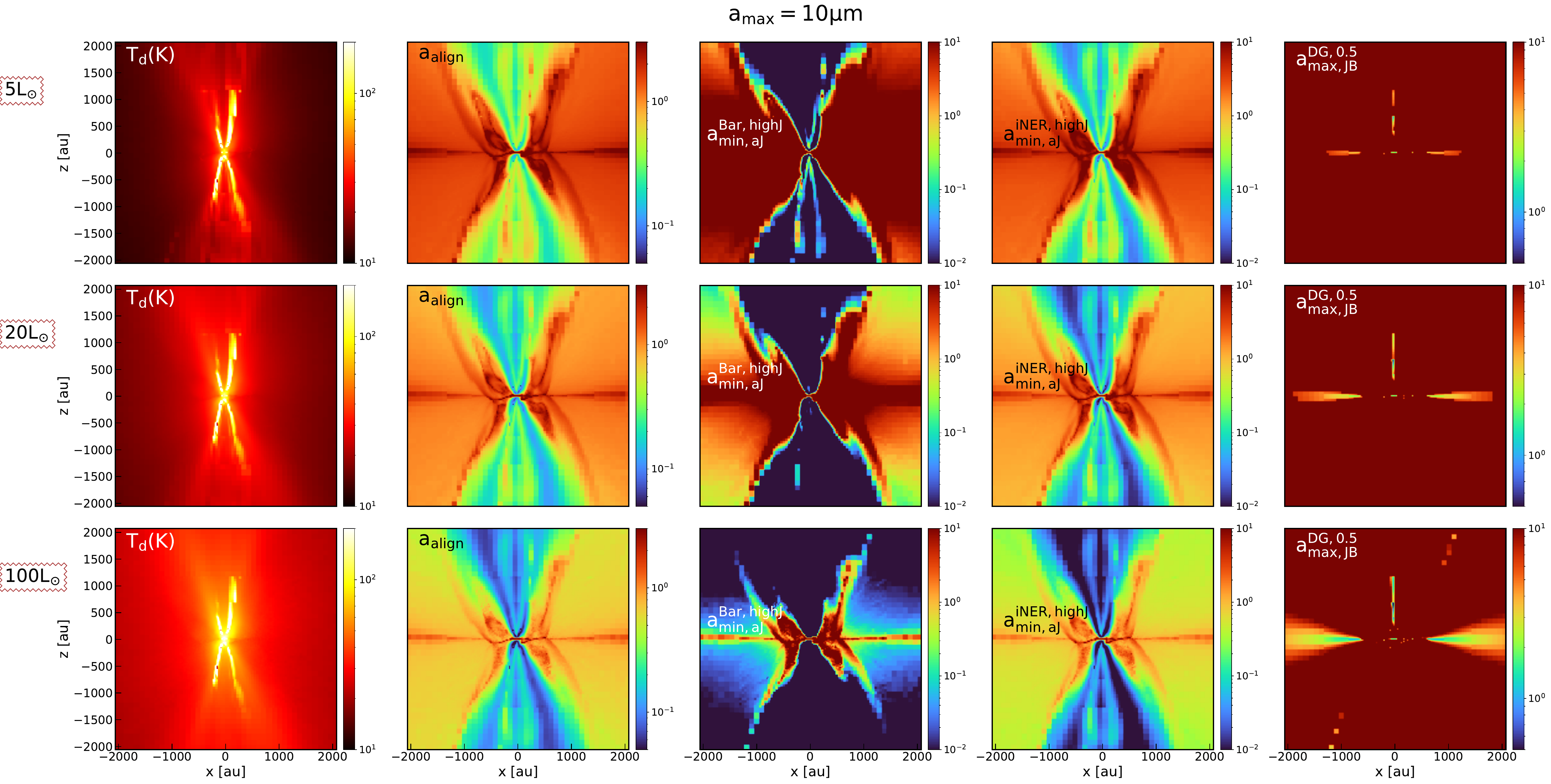}
    \caption{Similar results as Figure \ref{fig:alignment_Lstar50} but assuming $a_{\rm max} = 10\mu m$. The tendency of $T_{\rm d}$ and another critical size for alignment are similar to Figure \ref{fig:alignment_Lstar50}. But owing to the lower maximum grain size, the reduction of the alignment range (i.e., increasing $a_{\rm align}$) affects dust polarization fraction stronger, producing lower $p(\%)$ in YSOs having lower $L_{\rm star}$.}
     \label{fig:alignment_Lstar10}
\end{figure*}

Figure \ref{fig:alignment_Lstar50} shows the effect of central luminosity $L_{\rm center}$ on the distribution of dust temperature (first column) and alignment state inside the 2000 au box. The second column illustrates the minimum alignment size $a_{\rm align}$, the third and fourth columns show the distribution of the minimum size having fast super-Barnett and Inelastic relaxation at high-\textit{J} attractors $a_{\rm min,aJ}^{\rm Bar,highJ}$ and $a_{\rm min,aJ}^{\rm iNER,highJ}$. The fifth column shows the maximum size having $f_{\rm high-J} = 0.5$ by MRAT mechanism. The upper, middle, and lower rows display results obtained with $L_{\rm center} = 5L_{\odot}, 20L_{\odot}$, and $100L_{\odot}$, respectively. We consider SPM grains with $N_{\rm cl} = 10^{3}$ and $\phi_{\rm sp}=0.1$.

Generally, the overall dust temperature will decrease with decreasing $L_{\rm center}$. Less sub-micron grains inside the outflow cavity and less small micron-sized grains inside the outflow cavity wall, inner envelope, and envelope can be aligned with $\B$ by receiving weaker RAT efficiency, i.e., larger $a_{\rm align}$. The minimum sizes with fast super-Barnett and inelastic relaxation also increase with decreasing $L_{\rm center}$. However, given the weak dependence of inelastic relaxation timescale on grain sizes ($\tau_{\rm iNER} \sim a^{2}$ while $\tau_{\rm Bar} \sim a^{7}$), almost micron-sized grains and VLGs which can be aligned with $\B$ still can have fast inelastic relaxation regardless of $L_{\rm center}$ (second and fourth columns). Furthermore, more VLGs inside the envelope of protostars with low $L_{\rm center}$ can be aligned with $\B$ by MRAT mechanism, i.e., larger $a_{\rm max,JB}^{\rm DG,50}$, owing to the enhanced magnetic relaxation by decreasing the grain temperature there (first column). As aligned dust grains still can maintain their efficient magnetic alignment in fainter protostars, and given $a_{\rm max} = 50\mum$, the slight increase of $a_{\rm align}$ (from $\sim 0.6\mum$ if $L_{\rm center} = 100L_{\odot}$ to $\sim 3\mum$ if $L_{\rm center} = 5L_{\odot}$) is negligible in causing the clear change in the net alignment degree of protostellar grains. It explains why  $L_{\rm center}$ does not have a strong impact on $p(\%)$ if $a_{\rm max} = 50\mum$ (Figure \ref{fig:CASA_Lstar}, left panel).

We provide the distribution map of grain alignment size in case $a_{\rm max} = 10\mum$ in Figure \ref{fig:alignment_Lstar10}. Similar to Figure \ref{fig:alignment_Lstar50}, almost aligned dust grains inside protostars with low $L_{\rm center} = 5L_{\odot}$ still can achieve efficient IA and magnetic alignment as grains inside objects with high $L_{\rm center} = 100L_{\odot}$. But given smaller $a_{\rm max} = 10\mum$, the narrower of the alignment size range with decreasing $L_{\rm center}$ (from $[\sim 0.8,10\mum]$ for $L_{\rm center} = 100L_{\odot}$ to $[\sim 3-10\mum]$ for $L_{\rm center} = 5L_{\odot}$) becomes clearer, inducing lower net observed polarization fraction shown in Figure \ref{fig:CASA_Lstar}, right panel).

\section{Fraction of grains at high-\textit{J} inside the protostellar core}\label{sec:appen_align_fhighJ}
 \begin{figure*}
    \includegraphics[width=\textwidth,height=\textheight,keepaspectratio]{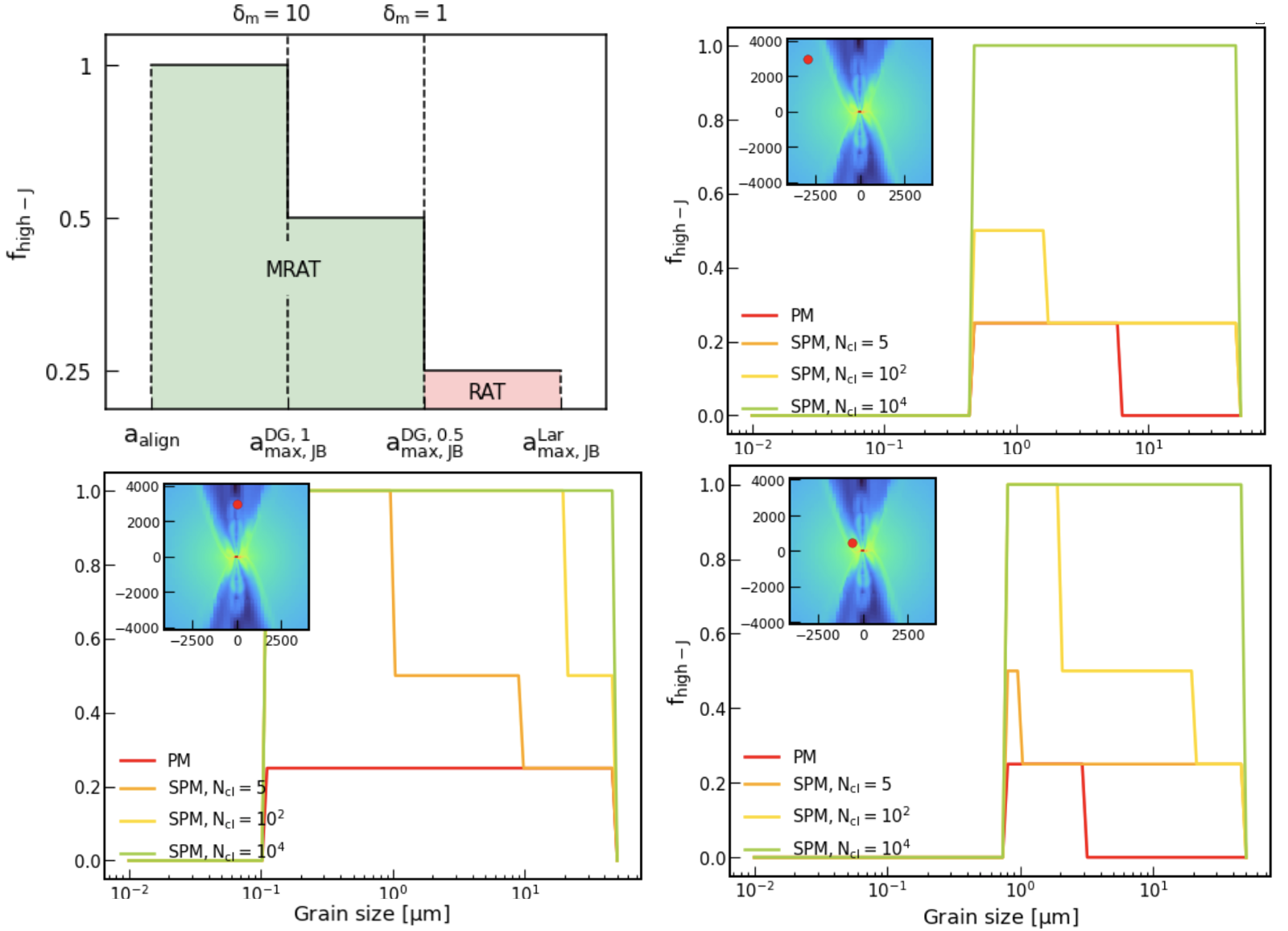}
    \caption{Upper left panel: Illustration of the variation of $f_{\rm high-J}$ and their external alignment with grain sizes within the alignment range $[a_{\rm align} - a_{\rm max,JB}^{\rm Lar}]$. The green area determines the size range that grains are aligned with $\B$ by MRAT mechanism, with $f_{\rm high-J} = 1$ for grains having high magnetic relaxation ratio $\delta_{\rm m}\geq 10$, and $f_{\rm high-J} = 0.5$ for grains having $1 \geq \delta_{\rm m} \leq 10$. The red area determines the size range for RAT alignment with typical $f_{\rm high-J} = 0.25$. The maximum size for grains having $f_{\rm high-J} = 1$ and $f_{\rm high-J} = 0.5$ by MRAT alignment is denoted by $a_{\rm max,JB}^{\rm DG,1}$ and $a_{\rm max,JB}^{\rm DG,0.5}$, respectively. The original picture is in \citet{Giang_2023a}. Upper right, lower left, lower right panel: Distribution of $f_{\rm high-J}(a)$ for dust grains inside the envelope, outflow cavity, and outflow cavity wall, respectively. The position of grains used in each panel is marked on the density map placed in the upper left corner. Generally, SPM grains containing larger iron inclusions have higher possibility to be aligned with $\B$ by MRAT with high $f_{\rm high-J} \sim 0.5 - 1$ regardless of their position inside the core. In contrast, MRAT tends to work with sub-micron and micron-sized grains, while VLGs tend to experience RATs with lower $f_{\rm high-J} \sim 0.25$.} 
     \label{fig:fhighJ}
\end{figure*}

The upper left panel of Figure \ref{fig:fhighJ} illustrates the variation of $f_{\rm high-J}$ on grain sizes based on the magnetic relaxation ratio $\delta_{m}$ (the original figure is in \citealt{Giang_2023a}). The detailed picture of $f_{\rm high-J}(a)$ for PM and SPM grains with different $N_{\rm cl}$ in the envelope, outflow cavity, and inner envelope is shown the upper right and lower panels of Figure \ref{fig:fhighJ}. The position of dust grains is marked in the gas density map placed in the upper left corner of each panel. Dust grains beyond the alignment range $a_{\rm align} -a_{\rm max,JB}^{\rm Lar}$ are denoted to have $f_{\rm high-J} = 0$. One can see that PM grains are majorly aligned with $\B$ by RATs with $f_{\rm high-J} = 0.25$ in the protostellar core owing to their weak magnetic relaxation efficiency. For SPM grains with $N_{\rm cl} \leq 10^{2}$, VLGs still tend to be aligned with $\B$ by RATs with low $f_{\rm high-J} \sim 0.25$ because of their high inertia moment that reduces the magnetic relaxation strength (see numerical calculations in \citealt{Hoang+2022} and \citealt{Giang_2023a}). Micron-sized grains below $10\mum$ can have the magnetic alignment by MRAT mechanism with $f_{\rm high-J} = 0.5$. And sub-micron grains tend to have higher $f_{\rm high-J} \sim 1$ owing to their smaller inertia moment. The fraction of large grains being aligned with $\B$ at high-$J$ attractors increases as increasing the grain magnetic susceptibility. For example, SPM grains with $N_{\rm cl} = 10^{4}$ can have $f_{\rm high-J} \sim 1$, regardless of their location inside the core.
     
\section{Polarization degree map after filtering}\label{sec:appen_filter}
 \begin{figure*}
\centering
 \includegraphics[width=\textwidth,height=\textheight,keepaspectratio]{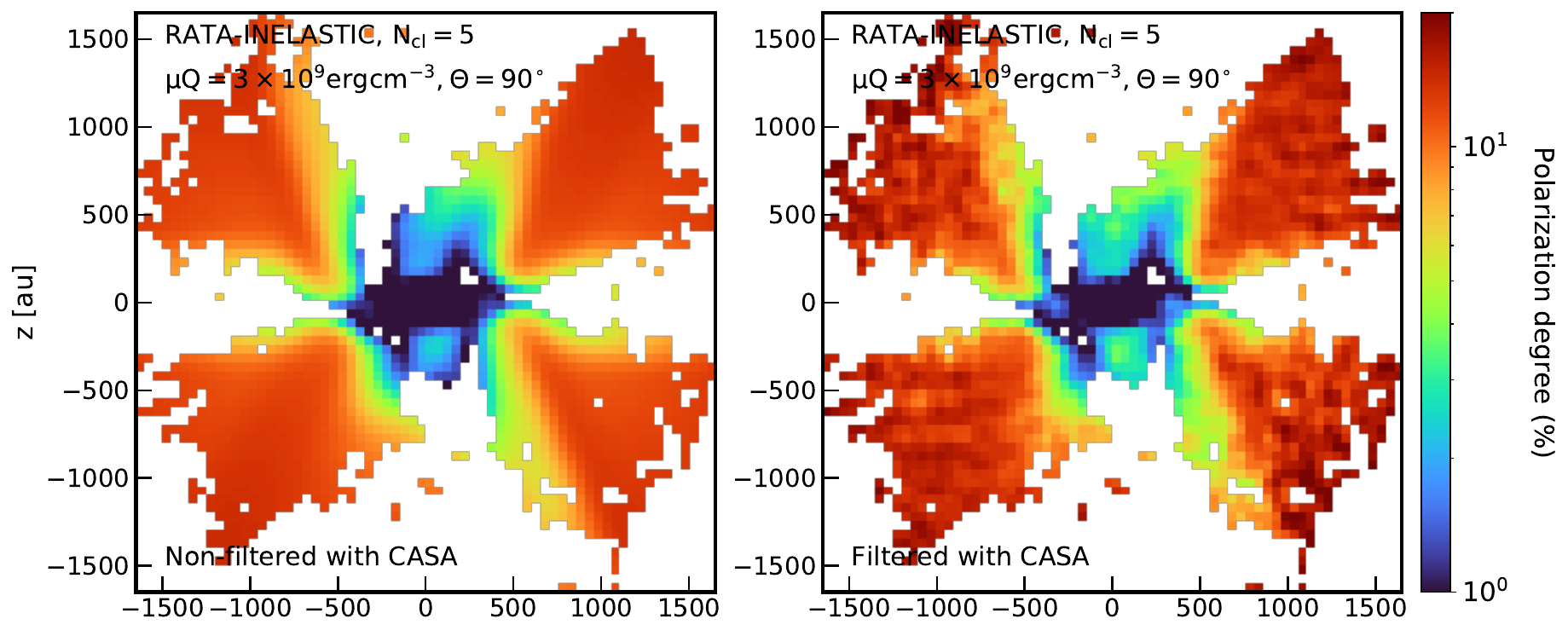}   
\includegraphics[width=\textwidth,height=\textheight,keepaspectratio]{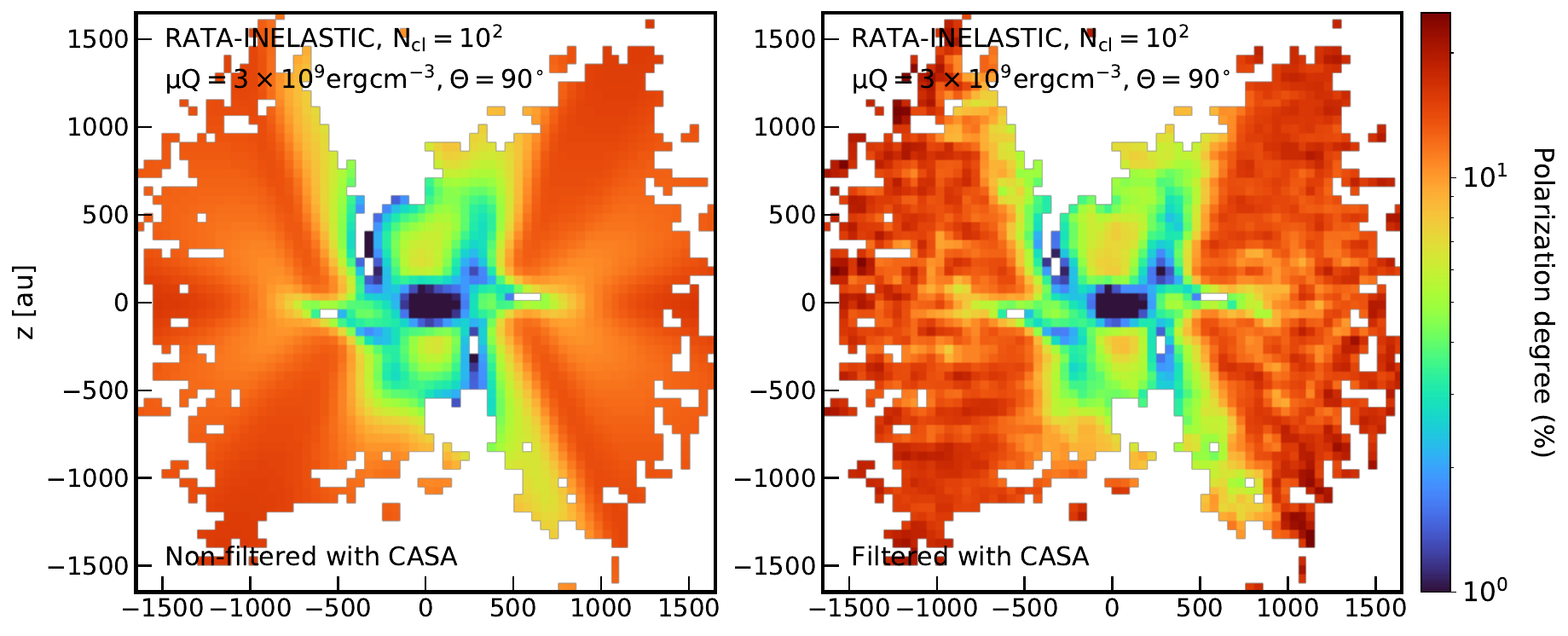}   \includegraphics[width=\textwidth,height=\textheight,keepaspectratio]{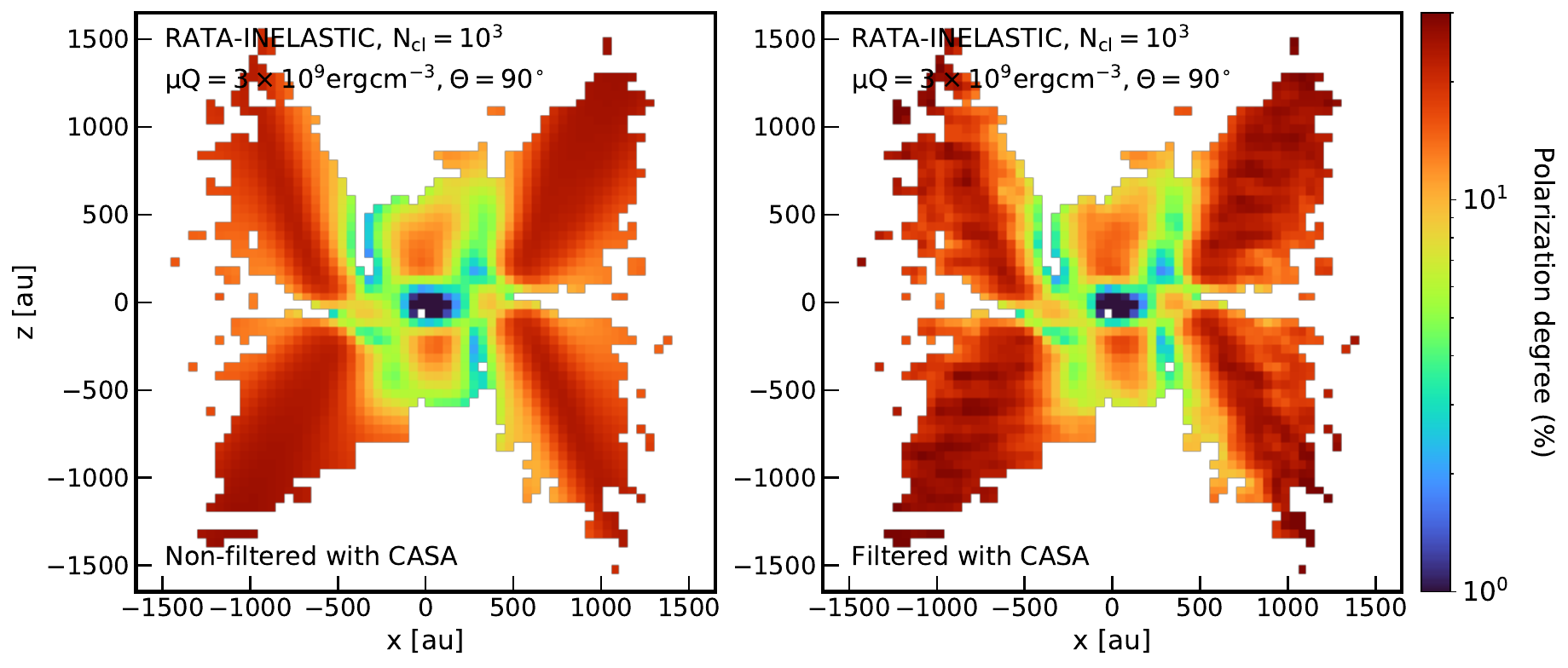}   
    \caption{Polarization degree observed at 1.3mm before (left column) and after (right column) filtering with CASA, considering model RATA$-$INELASTIC with SPM grains with $N_{\rm cl} =5, 10^{2}, 10^{3}$, $\mu Q = 3\times 10^{9}\erg\cm^{-3}$, $a_{\rm max} = 50\mum$, and $\Theta = 90^{\circ}$. The map is shown when $I \geq 3\sigma_{\rm I}$ and $I_{\rm pol} \geq 5\sigma_{\rm P}$. Generally, ALMA observation can recover the structure and detail of the intrinsic polarization degree map produced by aligned dust grains inside intermediate Class 0 YSOs. The filtering tends to amplify the polarization degree in the outermost, low-intensity envelope region by a few percent owing to the leakage of Stokes I emission. However, the difference is not significant enough to totally change the intrinsic properties of polarized emission radiating from the core.}
     \label{fig:filtering_map}
\end{figure*}

We show in Figure \ref{fig:filtering_map} the polarization degree map obtained from POLARIS (left column) and from the filtering with CASA (right column), considering results from model RATA$-$INELASTIC with $N_{\rm cl} = 5$ (first row), $10^{2}$ (second row), and $10^{3}$ (third row). We show the cell which satisfies $I \geq 3\sigma_{\rm I}$ and $I_{\rm pol}\geq 5\sigma_{\rm P}$, the values of $\sigma_{\rm I}$ and $\sigma_{\rm P}$ are denoted in each panel. Generally, the polarization degree map after filtering with CASA is nearly similar to the synthetic map obtained from POLARIS, except for the amplification of $p(\%)$ in the boundary of the recovered field due to the leakage of Stokes I emission. However, as shown in Figure \ref{fig:filtering_Ncl_theta}, the interferometric filtering only increases the intrinsic polarization fraction to $\sim 5-10\%$, which verifies the reliability of using observational data in inferring behind physics of grain alignment and tracing $\B$ fields in this region.

 
\end{document}